\documentclass[a4paper, 11pt]{article}
\usepackage{aapreprint}
\makeatletter
\def\@fpheader{}
\makeatother

\usepackage{graphicx,wrapfig,float,slashed}
\usepackage{amsmath,amssymb,dsfont,epsfig,graphicx,xcolor}
\usepackage{mathtools} % loads amsmath
\usepackage{multirow}
\usepackage{blkarray}
\usepackage{epstopdf}
\usepackage{ragged2e}
\usepackage[utf8]{inputenc} 
\usepackage{cancel}
\usepackage[table]{xcolor}
\usepackage[toc]{appendix}
\usepackage[normalem]{ulem}
\usepackage{enumitem}
\usepackage[font=small,skip=1pt]{caption}
\allowdisplaybreaks
\usepackage{pifont}
\usepackage{subcaption}
\usepackage{footnote}

\usepackage{cleveref}

\newcommand{\nc}{\newcommand}
\nc{\non}{\nonumber}
\nc{\hc}{\hbox {h.c.}}
\nc{\noi}{\noindent}
\nc{\barx}{\bar{x}}
\nc{\pbarn}{\;\hbox {pb}}
\nc{\fbarn}{\;\hbox {fb}}

\nc{\hsp}{\hspace{0.5cm}} 
\nc{\lsp}{\hspace{1cm}}
\nc{\Lsp}{\hspace{2cm}}
\nc{\LLsp}{\lsp\lsp}
\nc{\lra}{\longrightarrow}
\nc{\p}{\prime}
\nc{\sgn}{\text{sgn}}
\nc{\tr}{\text{Tr}}
\nc{\ph}{\varphi}
\nc{\op}{{\cal O}}

\nc{\beq}{\begin{equation}}  \nc{\eeq}{\end{equation}}
\nc{\bea}{\begin{eqnarray}}  \nc{\eea}{\end{eqnarray}}
\nc{\baa}{\begin{array}}     \nc{\eaa}{\end{array}}
\nc{\bit}{\begin{itemize}}   \nc{\eit}{\end{itemize}}
\nc{\ben}{\begin{enumerate}} \nc{\een}{\end{enumerate}}
\nc{\bce}{\begin{center}}    \nc{\ece}{\end{center}}
\nc{\bpm}{\begin{pmatrix}}   \nc{\epm}{\end{pmatrix}}
\nc{\bvt}{\begin{verbatim}}  \nc{\evt}{\end{verbatim}}

\def\lsim{\mathrel{\raise.3ex\hbox{$<$\kern-.75em\lower1ex\hbox{$\sim$}}}}
\def\gsim{\mathrel{\raise.3ex\hbox{$>$\kern-.75em\lower1ex\hbox{$\sim$}}}}

\def\udots{\mathinner{\mkern1mu\raise1pt\vbox{\kern7pt\hbox{.}}\mkern2mu\raise4pt\hbox{.}\mkern2mu\raise7pt\hbox{.}\mkern1mu}}

\def\mev{\;\hbox{MeV}}
\def\gev{\;\hbox{GeV}}
\def\tev{\;\hbox{TeV}}

\def\mpl{M_{\rm Pl}}

\def\ads#1{\text{AdS}$_{#1}$}

\def\eq#1{Eq.~(\ref{#1})}
\def\fig#1{Fig.~\ref{#1}}
\def\sec#1{Sec.~\ref{#1}}
\def\tab#1{Table~\ref{#1}}
\def\rcite#1{Ref.~\cite{#1}}

\begin{document}
%%%%%%%%%%%%%%%%%%%%%%%%%

%%%%%%%%%%%%%%%%%%%%%%%%%%%%%%%%%%%%%%%%%%%%%%%%%%%%%%
\title{Composite Dark Matter and Neutrino Masses from a Light Hidden Sector}
\author[1]{Aqeel Ahmed,}
\emailAdd{aqeel.ahmed@mpi-hd.mpg.de}
\author[2]{Zackaria Chacko,}
\emailAdd{zchacko@umd.edu}
\author[3]{Niral Desai,}
\emailAdd{npd393@utexas.edu}
\author[2]{Sanket Doshi,}
\emailAdd{sdoshi1@umd.edu}
\author[3]{\\ Can Kilic,}
\emailAdd{kilic@physics.utexas.edu}
\author[4]{and Saereh Najjari}
\emailAdd{snajjari@uni-mainz.de}
\affiliation[1]{Max-Planck-Institut f??r Kernphysik,\\ Saupfercheckweg 1, 69117 Heidelberg, Germany}
\affiliation[2]{Maryland Center for Fundamental Physics, Department of Physics,\\ University of Maryland, College Park, MD 20742-4111 USA}
\affiliation[3]{Theory Center, Weinberg Institute for Theoretical Physics,\\ Department of Physics, University of Texas at Austin, Austin, TX 78712, USA}
\affiliation[4]{PRISMA+ Cluster of Excellence \& Mainz Institute for Theoretical Physics,\\ Johannes Gutenberg University, 55099 Mainz, Germany}

%%%%%%%%%%%%

\abstract{\noindent 
We study a class of models in which the particle that constitutes dark matter arises as a composite state of a strongly coupled hidden sector. 
The hidden sector interacts with the Standard Model through the neutrino 
portal, allowing the relic abundance of dark matter to be set by 
annihilation into final states containing neutrinos. The coupling to the 
hidden sector also leads to the generation of neutrino masses through 
the inverse seesaw mechanism, with composite hidden sector states 
playing the role of the singlet neutrinos. We focus on the scenario in 
which the hidden sector is conformal in the ultraviolet, and the 
compositeness scale lies at or below the weak scale. We construct a 
holographic realization of this framework based on the Randall-Sundrum 
setup and explore the implications for experiments. We determine the 
current constraints on this scenario from direct and indirect detection, 
lepton flavor violation and collider experiments and explore the reach 
of future searches. We show that in the near future, direct detection 
experiments and searches for $\mu \rightarrow e$ conversion will be able 
to probe new parameter space.  At colliders, dark matter can be produced 
in association with composite singlet neutrinos via Drell Yan processes 
or in weak decays of hadrons. We show that current searches at the Large 
Hadron Collider have only limited sensitivity to this new production 
channel and we comment on how the reconstruction of the singlet 
neutrinos can potentially expand the reach. 
}

%%%%%%%%%%%%

\preprint{\begin{flushright}MITP-23-016\\ UTWI-12-2023 \end{flushright}}
\arxivnumber{2305.09719}

\maketitle
\flushbottom
%%%%%%%%%%%%

%%%%%%%%%%%%%%%%%%%%%%%%%%%%%%%%%%%
\section{Introduction	\label{s.intro}}
%%%%%%%%%%%%%%%%%%%%%%%%%%%%%%%%%%%

Despite the remarkable success of the Standard Model (SM) in explaining 
the interactions of the elementary particles, there is now indisputable 
evidence that it is incomplete. In particular, cosmological 
and astrophysical observations have established that more than 80\% of 
the matter in the universe is composed of some form of non-luminous, 
cold dark matter (DM), but there is no particle in the SM that can play 
this role. Furthermore, although the SM predicts that the neutrinos are 
massless, over the last few decades various oscillation experiments have 
established that the masses of the neutrinos, although tiny relative to 
those of the other SM fermions, are non-vanishing. Any explanation of 
these puzzles require new physics beyond the SM.

At present, the nature of the particles of which DM is composed remains 
completely unknown. One interesting possibility is that the particles 
that constitute DM are the composites of a strongly coupled hidden 
sector. Within this class of theories, the DM candidates that have been 
studied include dark glueballs \cite{Faraggi:2000pv,Soni:2016gzf}, dark 
pions 
\cite{Bai:2010qg,Bhattacharya:2013kma,Cline:2013zca,Hochberg:2014kqa,Beylin:2019gtw} 
and dark baryons \cite{Hodges:1993yb,Berezhiani:2000gw,Foot:2004pa}. For 
a clear review of composite DM with many additional references, 
see~\cite{Cline:2021itd}. Another intriguing possibility is that there 
is a close connection between the particles that constitute DM and the 
SM neutrinos. Examples of such theories include sneutrino DM in 
supersymmetric extensions of the SM 
\cite{Ibanez:1983kw,Hagelin:1984wv,Hall:1997ah}, sterile neutrino DM 
\cite{Dodelson:1993je,Shi:1998km,Abazajian:2001nj}, and scotogenic 
models of neutrino masses \cite{Ma:2006km,Boehm:2006mi,Hambye:2009pw}.

In this paper, we propose a new class of composite DM models that can 
account for both the observed abundance of DM and the origin of neutrino 
masses. We consider a scenario in which the particle that constitutes 
the observed DM arises as a composite state of a strongly coupled hidden 
sector. The stability of DM is ensured by a discrete symmetry of the 
hidden sector. The SM is assumed to couple to the hidden sector through 
the neutrino portal.\,{\footnote{For earlier work on models in which DM couples 
to the SM through the neutrino portal see, for 
example, Refs.~\cite{Escudero:2016ksa,Batell:2017cmf,Lamprea:2019qet}.}} This 
allows the relic abundance of DM to be set by annihilation to a final 
state consisting of a SM neutrino and antineutrino, or alternatively, to 
a final state consisting of a SM (anti)neutrino and a composite hidden 
sector particle. The neutrino portal interaction leads to mixing of the 
SM neutrinos with composite fermions of the hidden sector. The neutrinos 
thereby acquire tiny Majorana masses through the inverse seesaw 
mechanism~\cite{Mohapatra:1986bd,Mohapatra:1986aw}. In this framework, 
the neutrinos are partially composite.\,{\footnote{For earlier work on 
models of composite neutrinos see, for 
example, Refs.~(\cite{Arkani-Hamed:1998wff,vonGersdorff:2008is,Grossman:2010iq,Okui:2004xn}). 
Neutrino compositeness has been linked to DM 
in~\cite{Robinson:2012wu,Robinson:2014bma,Hundi:2011et,Davoudiasl:2017zws} and to the 
origin of the baryon asymmetry of the universe 
in~\cite{Grossman:2008xb}.}} We focus on a scenario in which the 
compositeness scale lies at or below the weak scale, leading to rich 
experimental signals.

The strong dynamics of the hidden sector is taken to be approximately 
conformal in the ultraviolet. This can allow the small parameters 
necessary to explain the observed neutrino masses within the framework of 
a low-scale seesaw model to naturally arise from the scaling dimensions of 
operators in the conformal field theory (CFT)~\cite{Chacko:2020zze}. 
These small parameters also play a role in realizing the correct
relic density of DM when the DM mass lies below the weak scale.  
To explore the dynamics of this class of models, we construct a holographic 
realization of this scenario based on the AdS/CFT correspondence 
\cite{Maldacena:1997re,Gubser:1998bc,Witten:1998qj,Klebanov:1999tb}. The 
correspondence relates large-$\mathcal{N}$ CFTs in four dimensions to 
theories of gravity in warped space in a higher number of dimensions. Our 
realization takes the form of a five-dimensional (5D) Randall-Sundrum (RS) 
model with two branes~\cite{Randall:1999ee}. Operators in the hidden 
sector CFT are dual to fields in the bulk whereas the SM fields, being 
elementary, are mapped to states localized on the ultraviolet brane. 
Note that in this framework we are not attempting to address the 
hierarchy problem of the SM, and so the Higgs boson is also an elementary 
field localized on the UV brane. Within this framework, we can calculate 
the relic abundance of DM and explore the phenomenology associated with 
this class of models.

Within the framework of the RS solution to the hierarchy problem, 
several authors have addressed the generation of neutrino masses, for 
example, 
\cite{Grossman:1999ra,Huber:2002gp,Huber:2003sf,Agashe:2015izu,Gherghetta:2003hf}, 
and the origin of DM, for example, 
\cite{Agashe:2004ci,Agashe:2004bm,Belanger:2007dx,Panico:2008bx}. 
However, an important difference is that because these models are built 
around the RS solution to the hierarchy problem, all the SM quarks and 
leptons are necessarily partially or entirely composite. The 
compositeness scale is then constrained to lie above a TeV, and so the 
implications for experiments are very different. A holographic model of 
neutrino masses with a lower compositeness scale that shares some 
features with our construction was considered in~\cite{McDonald:2010jm} 
(see also~\cite{Gripaios:2006dc}).

This class of DM models gives rise to signals in direct and indirect 
detection experiments and at colliders. Since the couplings of the 
hidden sector to the SM through the neutrino portal will in general 
violate flavor, we also expect signals in experiments searching for 
lepton flavor violating processes, such as $\mu \rightarrow e \gamma$ 
and $\mu \rightarrow e$ conversion. We determine the constraints from 
existing searches and explore the reach of future experiments. In this 
framework, although the hidden sector is neutral under the weak 
interactions, the DM acquires a coupling to the $Z$-boson at loop level 
through the neutrino portal interaction and can therefore be searched 
for in direct detection experiments. We find that for some range of DM 
masses, future direct detection experiments such as 
LZ~\cite{LUX-ZEPLIN:2018poe} and XENONnT~\cite{XENONnT} will have 
sensitivity to this scenario. In some regions of parameter space, in 
addition to neutrinos, other SM particles are also produced as the 
result of DM annihilation. This can be used to constrain the model in 
indirect detection experiments, both from precision observations of the 
CMB and from searches for gamma rays and positrons that are the products 
of DM annihilation.

The spectrum of composite states includes singlet neutrinos that carry a 
small charge under the weak interactions through their mixing with the 
SM neutrinos. These particles fall into the category of heavy neutral 
leptons (HNLs), which are being searched for at the LHC and at beam 
dumps. However, the composite singlet neutrinos differ from conventional 
HNLs in that, in some regions of parameter space, their primary decay 
mode is completely invisible. For the case when the dominant decays of 
the composite singlet neutrinos are visible, we use the current limits 
on HNLs to place bounds on this scenario and study the reach of future 
searches. Since the hidden sector couples to the SM through the neutrino 
portal, DM particles can also be produced in association with a 
composite singlet neutrino. The challenge in detecting DM is therefore 
to identify the effects of additional invisible particles on top of 
standard HNL signatures. We study the sensitivity of existing experimental 
HNL searches for events produced via this new channel, and we also explore a strategy 
involving the reconstruction of the HNL that would extend the reach for 
DM particles. We find that, although the parameter space of the model is 
highly constrained by non-collider observations, there is nevertheless 
a limited region where these searches have sensitivity.

 The outline of the paper is as follows. We discuss the general 
framework in \sec{s.4d_model}. In \sec{s.holographic}, we lay out the 
extra-dimensional model and determine the mass spectra, couplings and 
mixing angles. In \sec{s.updm}, we present a comprehensive 
phenomenological analysis of DM with a focus on its production, direct 
and indirect detection. The collider aspects of the phenomenology are 
presented in \sec{s.colliders}. We conclude in \sec{s.conclusion}.

%%%%%%%%%%%%%%%%%%%%%%%%%%%%%%%%%%%
\section{A Framework for Composite Dark Matter and Neutrino Masses} 	
 \label{s.4d_model}
%%%%%%%%%%%%%%%%%%%%%%%%%%%%%%%%%%%

In this section, we outline the general features of the scenario we are 
exploring. We consider a framework in which the particle that 
constitutes DM arises as a composite state of a strongly coupled hidden 
sector that is approximately conformal in the ultraviolet. We show 
that couplings between the hidden sector and the SM through the neutrino 
portal can give rise to the observed abundance of DM, while also 
generating the neutrino masses. We discuss the current constraints on 
this class of models and outline the possibilities for discovering the 
DM candidate in direct and indirect detection experiments and collider 
searches.

\subsection{Composite Dark Metter through the Neutrino Portal}

Consider a hidden sector composed of a strongly coupled CFT, to which we 
add a relevant deformation $\op_{\rm def}$,
 \beq
 {\cal L}_{\rm UV}\supset {\cal L}_{\rm CFT} + 
 \lambda_{\rm def}\op_{\rm def} \;.
 \eeq
 When the deformation grows large in the infrared (IR), it causes the 
breaking of the conformal dynamics. This occurs at a scale that we 
denote by $\Lambda$. 

We assume that the spectrum of light hidden sector states includes three 
composite Dirac fermions $N^{\alpha}$, which play the role of composite 
singlet neutrinos. Here $\alpha = 1,2,3$ represents a flavor index. The 
low energy effective Lagrangian contains kinetic and mass terms for the 
singlet neutrinos,
 \begin{align}
{\cal L}_{\rm IR}&\supset  
i\bar{N} \gamma^{\mu} \partial_{\mu} N - m_{N} \bar N N	, 
\label{eq:lag_ir}
 \end{align}
 where we have suppressed the flavor indices. Here $m_N$ is the singlet 
neutrino mass, which is expected to be of the order of the conformal 
symmetry breaking scale $\Lambda$. We can decompose $N$ into components 
with left- and right-handed chiralities, $N = (N_L,N_R)$.

The hidden sector interacts with the SM through the neutrino portal,
 \beq
{\cal L}_{\rm UV}\supset -\frac{\hat{\lambda}}{M_{\mathrm{UV}}^{\Delta_{\widehat{N}}-3 / 2}} \bar L \widetilde H \mathcal{\widehat{O}}_{\!N}+\mathrm{h.c.},
\label{eq:portal_int_uv}
 \eeq
 where $L$ is the SM left-handed lepton doublet, $\widetilde H \equiv 
i\sigma_2 H^\ast$ where $H$ is the SM Higgs doublet, and 
$\mathcal{\widehat{O}}_{\!N}$ represents a primary operator of scaling 
dimension $\Delta_{\widehat{N}}$ that transforms as a right-handed Weyl 
fermion. Here $\hat{\lambda}$ is a dimensionless coupling constant and 
$M_{\rm UV}$ denotes the ultraviolet (UV) cutoff of the theory. At the 
conformal breaking scale~$\Lambda$, this interaction gives rise to the 
following term in the low-energy Lagrangian,
 \begin{align}
{\cal L}_{\rm IR}&\supset -\lambda\, \bar L \widetilde H N_{\!R}+{\rm h.c.},	
\label{eq:portal_int_ir}
 \end{align}
 where the dimensionless coupling $\lambda$ scales as
 \beq
\lambda \sim \hat{\lambda}\left(\frac{\Lambda}{M_{\mathrm{UV}}}\right)^{\Delta_{\widehat{N}}-3 / 2}.	\label{eq:lambda}
 \eeq
  This represents a coupling of the SM to the composite singlet 
neutrinos $N_{\!R}$ through the neutrino portal. Once Higgs acquires 
a vacuum expectation value (VEV), this interaction and the mass term in 
Eq.~(\ref{eq:lag_ir}) lead to mixing between the SM neutrinos and the 
composite fermions $N_{\!L}$. Therefore the light neutrinos contain an 
admixture of hidden sector states, while the composite singlet neutrinos 
acquire an admixture of the SM neutrino. In this way the composite 
singlet neutrinos acquire a small coupling to the weak gauge bosons of 
the SM.

The scaling dimension of the primary fermionic operator 
$\mathcal{\widehat{O}}_{\!N}$ is bounded from below by unitarity, 
$\Delta_{\widehat{N}} \geq 3/2$, where the limiting case of 
$\Delta_{\widehat{N}}=3/2$ corresponds to the case of a free fermion. On 
the other hand, for scaling dimensions $\Delta_{\widehat{N}} \geq 5/2$ 
the interaction in \eq{eq:portal_int_uv} leads to the theory becoming 
ultraviolet sensitive, which requires the addition of new counterterms 
involving the SM fields for consistency~\cite{Chacko:2020zze}. 
Therefore, in this work, we limit our analysis to values of the scaling 
dimension of $\mathcal{\widehat{O}}_{\!N}$ in the range $3/2 <
\Delta_{\widehat{N}} < 5/2$. With this choice of $\Delta_{\widehat{N}}$, 
the coupling $\lambda$ in \eq{eq:lambda} is hierarchically small for 
$\Lambda \ll M_{\rm UV}$, so that the mixing between the SM neutrinos and 
their singlet counterparts is suppressed. As we explain below, this 
feature of our model can help explain both the smallness of the neutrino 
masses and the observed abundance of DM. In this work, we focus on low 
values of the compositeness scale, corresponding to values of $\Lambda$ 
at or below the electroweak scale.

We now assume that at the scale $\Lambda$, in addition to the composite singlet
neutrinos $N$, the spectrum of hidden sector states also 
includes a composite Dirac fermion $\chi$, 
which plays the role of DM. Then the low energy Lagrangian at scales of 
order $\Lambda$ includes the terms,
 \begin{align}
{\cal L}_{\rm IR}\supset &\,
i\bar{\chi} \gamma^{\mu} \partial_{\mu} \chi  - m_{\chi} \bar\chi \chi   \; .
\label{eq:lag_{ir}}
 \end{align}
 Here $m_\chi$ is the DM mass, which we again take to be of the order of 
$\Lambda$. To ensure the stability of DM we assume that the hidden 
sector respects a discrete $Z_2$ symmetry under which $\chi$ is odd, but 
the singlet neutrinos $N$ as well as the SM fields are even. We further 
assume that there are no Nambu-Goldstone bosons or other light states, 
so $\chi$ is the lightest state in the hidden sector.

In our framework, the neutrino portal interaction keeps the hidden 
sector in equilibrium with the SM in the early universe. Because of the 
composite nature of the fermions $\chi$ and $N$, the 
low energy theory at the scale $\Lambda$ contains nonrenormalizable 
interactions between the DM particle and the singlet neutrinos of the 
schematic form,
 \begin{align}
 {\cal L}_{\rm IR}&\supset -\frac{\tilde{\kappa}}{\Lambda^2}(\bar \chi N)(\bar N \chi) \;,
 \label{eq:chiN_eff}
 \end{align}
 where $\tilde{\kappa}$ is of order $16 \pi^2/\mathcal{N}^2$ in the 
large-$\mathcal{N}$ limit. Once the temperature falls below $m_{\chi}$, 
these interactions allow the DM particles to annihilate away through 
processes such as $\chi \bar{\chi} \rightarrow (\nu \bar{N}, N 
\bar{\nu})$ and $\chi \bar{\chi} \rightarrow \nu \bar{\nu}$. Naively, 
the annihilation rate would be expected to be enhanced compared to the 
freeze-out of DM of weak scale mass because of the low scale $\Lambda$ 
that sets the mass of $\chi$ and the strength of its interactions, 
resulting in a too low abundance of DM. However, this can be 
compensated for by the small mixing between the SM neutrinos and the 
composite singlet neutrinos. This class of theories can therefore easily 
accommodate the observed abundance of DM.

%%%%%%%%%%%%%%%%%%%%%%%%%%%%%%%%
\subsection{Neutrino masses via the Inverse Seesaw Mechanism 
\label{s.neutrino_masses}}
%%%%%%%%%%%%%%%%%%%%%%%%%%%%%%%%%%%

In this subsection, we outline how this framework can naturally 
incorporate the generation of neutrino masses through the inverse seesaw 
mechanism. Our discussion is based on the analysis 
in~\cite{Chacko:2020zze}. We now assume that the hidden sector possesses 
a global symmetry under which the operator $\mathcal{\widehat{O}}_{\!N}$ 
is charged. The charges under this global symmetry can be normalized 
such that $\mathcal{\widehat{O}}_{\!N}$, and therefore $N_{\!R}$, 
carries charge $+1$. Then, we see from the coupling 
Eq.~(\ref{eq:portal_int_ir}) that this symmetry can be subsumed into an 
overall lepton number symmetry under which both $N_{\!R}$ and $N_{\!L}$ 
carry charge $+1$.

In order to employ the inverse seesaw mechanism to generate the SM 
neutrino masses, we require a source of lepton number violation in the 
model. Accordingly, we add to the theory a lepton number violating 
deformation arising from an operator $\op_{2N}$, which has scaling 
dimension $\Delta_{2N}$,
 \beq
\mathcal{L}_{\rm UV} \supset -\frac{\hat{\mu}}{M_{\mathrm{UV}}^{\Delta_{2 N}-4}} \op_{2 N}+\mathrm{h.c.}
 \label{lnv_deformation}
 \eeq
 Here $\hat \mu$ is a dimensionless constant that parametrizes the 
strength of the deformation. We assume that $\op_{2N}$ carries a 
charge of $+2$ under the global symmetry of the hidden sector, so that 
this deformation violates lepton number by two units. In the low-energy 
effective theory at the scale $\Lambda$, this gives rise to terms in the 
Lagrangian of the form,
 \begin{align}
{\cal L}_{\rm IR} &\supset -(\mu N_{\!L} N_{\!L}+\mu' N_{\!R} N_{\!R}) + \textrm{h.c.}\,,
 \label{eq:lag_ir_lnv}
 \end{align}
 where the Majorana masses $\mu$ and $\mu'$ parameterize the strength of lepton 
number violation. Their values scale with the parameters of the 
theory as
 \beq
\mu \sim \mu' \sim \hat{\mu} \Lambda\left(\frac{\Lambda}{M_{\mathrm{UV}}}\right)^{\Delta_{2 N}-4}.	\label{eq:muc}
 \eeq
 The scaling dimension $\Delta_{2N}$ of the lepton number violating 
scalar operator $\op_{2N}$ is constrained by unitarity to satisfy 
$\Delta_{2N}\geq 1$, where the limiting case $\Delta_{2N} = 1$ 
corresponds to the case of a free scalar. For $\Delta_{2N} > 4$, the 
Majorana mass terms $\mu$ and $\mu'$ are hierarchically smaller than 
$\Lambda$.

With the inclusion of the lepton number violating terms in 
\eq{eq:lag_ir_lnv} the low-energy effective theory now possesses all the 
ingredients required to realize inverse seesaw mechanism,
 \begin{align}
{\cal L}_{\rm IR} \supset &i\bar{N} \gamma^{\mu} \partial_{\mu} N-m_{N} \bar N N- \Big[\mu N_{\!L} N_{\!L}+\lambda \bar L \widetilde H N_{\!R}+{\rm h.c.}\Big].	\label{eq:lag_ir_2}
 \end{align}
By integrating out the composite singlet neutrinos $N$ we obtain a contribution
to the masses of the light neutrinos, 
 \beq
m_{\nu}= \mu\Big(\frac{\lambda v_{\rm EW}}{m_{N}}\Big)^2 \;.	
 \label{eq:neutrino_mass}
 \eeq
 Here $v_{\rm EW}\equiv\langle H\rangle\simeq 174 \gev$. When the 
effects of higher resonances are included, this relation is only 
approximate, so that
 \beq
m_{\nu} \sim \mu\Big(\frac{\lambda v_{\rm EW}}{m_{N}}\Big)^2 \;.
 \label{eq:neutrino_mass'}
 \eeq
 Note that the neutrino masses depend on both the parameter $\lambda$, 
which controls the mixing with the composite states as seen 
in~\eq{eq:lambda}, and the parameter $\mu$, which controls the extent 
of lepton number violation as seen in~\eq{eq:muc}. Then the smallness of 
the SM neutrino masses can naturally be explained by either the small 
parameter $\lambda$ that sets the size of the neutrino mixing or the 
small lepton number violating coupling $\mu$. Since the small values 
of these parameters admit a simple explanation in terms of the scaling 
dimensions of the operators $\mathcal{\widehat{O}}_{\!N}$ and $\op_{2N}$, this class of 
models can provide a natural explanation for the smallness of neutrino 
masses.

 Note that in this construction both the Dirac mass term $\lambda v_{\rm 
EW}$ and the Majorana mass term $\mu$ need to be smaller than the 
compositeness scale $\Lambda$. This leads to the following range 
for the coupling $\lambda$,
 \begin{align}
 \label{lambdarange}
\frac{\sqrt{m_{\nu}m_{N}}}{{v_{\rm EW}}}&\lesssim\lambda \lesssim \frac{m_{N}}{v_{\rm EW}} \;,
 \end{align}
 where we have employed \eq{eq:neutrino_mass'} in obtaining the lower 
bound, after setting $m_N\sim \Lambda$.

The neutrino mixing angle to the physical mass eigenstates in the limit 
$m_{N} \gg \lambda v_{\rm EW}$ is defined as,
 \beq
U_{N\ell} \equiv \frac{\lambda v_{\rm EW}}{m_{N}}.	\label{eq:mixing}
 \eeq
 It follows from Eqs.~(\ref{lambdarange}) and (\ref{eq:mixing}) that the 
mixing angle $U_{N\ell}$ lies in the range
 \begin{align}
 \label{mixingrange}
 \sqrt{\frac{m_{\nu}}{m_{N}}} & \lesssim U_{N\ell} \lesssim 1 \; .
 \end{align}

%%%%%%%%%%%%%%%%%%%%%%%%%%%%%%%%
\subsection{Abundance of Dark Matter 
\label{s.darkmatter}}
%%%%%%%%%%%%%%%%%%%%%%%%%%%%%%%%%%%

In this subsection, we outline how this class of models can reproduce 
the observed abundance of DM. At high temperatures in the early 
universe, the hidden sector was in thermal contact with SM through the 
portal operator in~\eq{eq:portal_int_uv}. This interaction populates the 
hidden sector, bringing it into thermal equilibrium with the SM. Once 
the temperature falls below their masses, the hidden sector states begin 
to exit the bath. The observed DM today is composed of the lightest 
$Z_2$ odd Dirac fermion $\chi$ that survives as a thermal relic.

We first show that the hidden sector is in thermal equilibrium with the 
SM at temperatures of order the compositeness scale. The hidden sector 
states can be produced from SM neutrinos via processes such as $\nu \nu 
\to {\cal U}_N$, $\nu \nu \to \nu {\cal U}_N$, $\nu {\cal U}_N \to \nu 
{\cal U}_N$, where the label ${\cal U}_N$ denotes hidden sector states. 
To see this, note that the strongly coupled nature of the hidden sector 
implies large self-interactions between the composite singlet neutrino 
states, 
 \beq
\mathcal{L}_{\mathrm{IR}} \supset
-\frac{\kappa}{\Lambda^{2}}\left(\bar{N} \gamma^{\mu} N\right)^{2} +
\cdots,
 \label{neutrinoselfinteractions}
 \eeq
 where the size of the coupling $\kappa$ is of the order of 
$\sim(4\pi)^2/\mathcal{N}^2$ in the large-$\mathcal{N}$ limit and the 
ellipses denote other Lorentz contractions. These interactions are 
characteristic of the composite nature of the singlet neutrinos. To see 
that the hidden sector is in equilibrium with the SM at temperatures $T$ 
of order $\Lambda$, we estimate the rate for the $\nu \nu \to {\cal 
U}_N$ process. When $T \sim \Lambda$, this rate is expected to be 
parametrically of the same order as the $\nu \nu \to N N$ rate. From 
Eq.~(\ref{neutrinoselfinteractions}) we can estimate the cross section 
for this process as,
 \beq
\sigma_{\nu \nu \rightarrow \mathcal{U}_N} \sim \sigma_{\nu \nu \rightarrow N N} \sim \frac{1}{4 \pi}\bigg(\frac{\kappa}{\Lambda^{2}}\bigg)^{2} \big|U_{N\ell}\big|^{4} \, T^{2}.
 \eeq
From this, we can estimate the thermally averaged 
interaction rate as 
 \beq
\bar n_\nu \langle\sigma_{\nu\nu\to NN} \, v\rangle \sim \frac{\kappa^2}{2 \pi^3} \big|U_{N\ell}\big|^{4} \, \frac{T^{5}}{\Lambda^4}\,,
 \eeq 
 where $\bar n_\nu\simeq 2\,T^3/\pi^2$ represents the equilibrium 
number density of SM neutrinos. The SM and hidden sector will be in
thermal and chemical equilibrium at the compositeness scale if the 
interaction rate is larger than the Hubble rate ${\cal H}\!=\! 
\sqrt{\pi^2 g_\star/90}\, T^2/\mpl$ at temperatures $T\sim \Lambda$. Here 
$g_\star$ represents the effective number of relativistic degrees of 
freedom 
contributing to 
the energy density at temperature $T$, while $\mpl$ denotes the reduced 
Planck mass. Taking $\kappa \sim (4\pi)^2$ and $g_\star \sim 100$, the 
condition for thermal equilibrium at temperatures $T\sim \Lambda$ can be 
translated into a lower bound on the neutrino mixing angle $U_{N\ell}$ as,
 \beq
|U_{N\ell}|^2\gtrsim 0.1\sqrt{\Lambda/\mpl}\,.
\label{therm_eqbm}
 \eeq 
 Recall that the condition that the theory generate a realistic spectrum 
of neutrino masses already places a lower bound on $|U_{N\ell}|^2$, shown 
in \eq{mixingrange}. Comparing the two conditions, we find that the hidden 
sector is in thermal equilibrium with the SM at temperatures $T\!\sim\! 
\Lambda$, while successfully generating light neutrino masses of the order 
of $0.1\,{\rm eV}$, for values of the compositeness scale 
$\Lambda \lesssim 1\,{\rm GeV}$.

However, we note that thermal equilibrium can also arise 
from processes such as $\nu\, {\cal U}_N \to {\cal U}_N$. Although such 
processes require a small initial abundance of hidden sector states ${\cal 
U}_N$, 
these could have been produced earlier from other processes such as $\nu 
\nu \to 
{\cal U}_N$.
At temperatures of order the compositeness scale, the rate for $\nu\, 
{\cal U}_N \to {\cal U}_N$ is 
expected to be 
parametrically of the same order as the rate for $\nu N \to N N$. 
From 
\eq{neutrinoselfinteractions}, we can estimate the thermally averaged 
interaction rate at temperatures $T\sim \Lambda$ as,
 \beq
\bar n_\nu \langle\sigma_{\nu N\to NN} \, v\rangle \sim \frac{\kappa^2}{2 \pi^3} \big|U_{N\ell}\big|^{2} \, \Lambda\,.
 \eeq 
  Comparing this against the Hubble expansion rate for $\kappa\sim 
(4\pi)^2$ 
and $g_\star\sim 100$, the condition for thermal equilibrium at 
temperatures $T\sim \Lambda$ translates to a lower bound on 
$U_{N\ell}$,
 \beq
|U_{N\ell}|^2\gtrsim 0.01\, \Lambda/\mpl\,.
 \eeq
  This condition, when taken together with the condition for successful 
generation of neutrino masses in \eq{mixingrange}, implies that the hidden 
sector is in thermal equilibrium with the SM at temperatures of order the 
compositeness scale provided $\Lambda\lesssim 10\sqrt{m_\nu\mpl}$. For 
$m_\nu\sim 0.1\,{\rm eV}$ it follows that the two sectors are in thermal 
equlibrium for values of the compositeness scale $\Lambda\lesssim 
10^5\,{\rm GeV}$. This encompasses the entire parameter space of 
interest.
 
\begin{figure} [t!]
\centering
\includegraphics[width=0.7\textwidth]{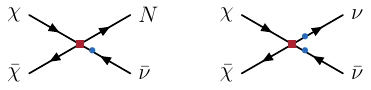}
\caption{Feynman diagrams for DM annihilation are shown above where the composite DM--neutrino effective vertex $\kappa/\Lambda^2$ is denoted by a red-square and neutrino mixing $U_{N\ell}$ is shown as a blue circle. For the left diagram, there is an analogous process for the final state $\bar N \nu$.  Flavor indices on the neutrinos have been suppressed. }
\label{fig:chichi_ann_eff}
\end{figure}

It follows from this discussion that, in this scenario, 
the abundance of DM 
is set by freeze-out. Once the temperature falls below the compositeness 
scale, the hidden sector states begin to exit the bath. The heavier hidden 
sector states annihilate away into lighter hidden sector states through 
processes such as $\bar{N} N \rightarrow \bar{\chi} \chi$.  Since the 
hidden sector is strongly coupled, these processes are very efficient and 
so the abundance of the heavier hidden sector states at temperatures well 
below the compositeness scale is extremely small. The DM particle $\chi$, 
being the lightest state in that sector, cannot annihilate into other 
hidden sector states. Instead, it annihilates to the visible sector 
through the neutrino portal and eventually freezes out as a thermal 
relic. The dominant DM annihilation channels to the visible sector are,
 \begin{align}
\lsp \chi \bar \chi \to \left(N \bar \nu, \nu \bar{N} \right)
\lsp \chi \bar \chi \to \nu \bar \nu.
 \end{align}
 The Feynman diagrams for the above DM annihilation processes are shown 
in \fig{fig:chichi_ann_eff}, where the vertex shown as a red-square 
corresponds to the interaction between DM and singlet neutrinos given in 
\eq{eq:chiN_eff} and a blue-circle denotes the neutrino mixing angle 
$U_{N\ell}$.

 The cross sections for the above DM annihilation processes at 
temperatures $T$ of order $\Lambda$ can be estimated as,
 \beq
\sigma_{\chi\bar\chi\to N\bar \nu}\sim \frac{\tilde{\kappa}^2\, U_{N\ell}^2}{4\pi }\frac{T^2}{\Lambda^4}, \lsp \sigma_{\chi\bar\chi\to \nu\bar \nu}\sim  \frac{\tilde{\kappa}^2\, U_{N\ell}^4}{4\pi}\frac{T^2}{\Lambda^4}.
 \eeq 
 DM freeze-out happens when its thermally averaged interaction 
rate becomes comparable to the Hubble rate, $\bar n_\chi 
\langle\sigma v\rangle\sim {\cal H}\sim T^2/\mpl$, where $\bar n_\chi \sim 
(m_\chi T)^{3/2}\,e^{-m_\chi/T}$ is the non-relativistic equilibrium 
number density for DM. The freeze-out happens at temperatures $T$ 
of order $m_\chi/20$.

The thermally averaged DM annihilation cross sections at DM freeze-out, 
i.e. for $T=T_{\rm fo}\sim m_\chi/20$, can be estimated as,
 \begin{align}
\langle\sigma_{\chi\bar \chi\to N\bar \nu}  v\rangle_{\rm fo}\!\sim\!\frac{\tilde{\kappa}^2\, U_{N\ell}^2}{32\pi\,\Lambda^2} , \qquad 
\langle\sigma_{\chi\bar \chi\to \nu\bar \nu}  v\rangle_{\rm fo}\!\sim\!\frac{\tilde{\kappa}^2\, U_{N\ell}^4}{32\pi\,\Lambda^2},
 \end{align}
 where we have made the simplification $m_N\sim m_\chi\sim\Lambda$.
 The observed DM relic abundance can be obtained when $\langle\sigma v\rangle_{\rm fo}\sim 2\times10^{-9}/\gev^{2}$. 

Provided that the $\chi \bar{\chi} \rightarrow (N \bar{\nu}, \bar{N} \nu)$ 
channel is kinematically open, this will be the dominant annihilation 
mode. For values of the composite scale $\Lambda\sim1\gev$ and $ 
\tilde{\kappa} \sim 4\pi$, we obtain the correct relic abundance from the 
$\chi\bar \chi\to (N\bar \nu, \nu \bar{N})$ process with 
elementary-composite neutrino mixing $|U_{N\ell}|^2\sim 10^{-8}$. For 
values of the DM mass $m_\chi< m_N/2$, this annihilation channel is 
kinematically forbidden. In this case the relic abundance of DM is set by 
the $\chi\bar \chi\to \nu\bar \nu$ channel. However, in this case, larger 
values of the mixing angle $U_{N\ell}$ are required. For instance, for a 
compositeness scale $\Lambda\sim1\gev$ and $\tilde{\kappa}\sim4\pi$, 
we require $|U_{N\ell}|^2\sim 10^{-4}$ to obtain the correct relic 
abundance. The discussion above implies that the neutrino mixing angle 
$U_{N\ell}$ plays a crucial role in setting the DM relic abundance within 
this framework. Although this analysis has been based on rough estimates, 
in \sec{s.updm} we perform detailed relic abundance calculations within 
the holographic realization of this model. After solving the full set of 
Boltzmann equations, we find that these conclusions are robust.

It is worth noting that, if the DM mass is larger than the composite 
neutrino mass, i.e. $m_\chi \gtrsim m_N$, the DM can directly annihilate 
to composite neutrinos, $\bar{\chi} \chi \rightarrow \bar{N} N$. In this 
scenario the annihilation cross-section taking $m_\chi\sim \Lambda$ can be 
estimated as $\langle\sigma_{\chi\bar \chi\to N\bar N} 
v\rangle\!\simeq\!\tilde{\kappa}^2/(32\pi\,\Lambda^2)$. In a strongly 
coupled theory we have $\tilde{\kappa}\sim4\pi$. We then require the DM 
mass $m_\chi\sim \Lambda\sim 10\tev$ to obtain the observed DM relic 
abundance. However, in this paper, we do not explore this high-mass 
region. Instead, as mentioned above, our focus here is on the low-mass 
region corresponding to composite scales at or below the electroweak 
scale.

 A characteristic feature of composite DM within this class of theories 
is that the DM particles have sizable self-interaction cross sections of 
order,
 \begin{equation}
\sigma_{\rm self}\sim \frac{\kappa_{\chi}^2}{8\pi}\frac{m_\chi^2}{\Lambda^4} \;.		\label{eq:sigma_self}
 \end{equation}
 These self-interactions arise from terms in the Lagrangian of the form, 
 \beq
\mathcal{L}_{\mathrm{IR}} \supset
-\frac{\kappa_\chi}{\Lambda^{2}}\left(\bar{\chi} \gamma^{\mu} \chi\right)^{2} +
\cdots,
 \label{DMselfinteractions}
 \eeq
 where the size of the coupling $\kappa_\chi\sim(4\pi)^2/\mathcal{N}^2$ 
in the large-$\mathcal{N}$ limit and the ellipses denote alternative 
Lorentz contractions.
 
There are several constraints on the DM self-interactions. The most 
stringent are based on observations of the Bullet Cluster and lead to 
$\sigma_{\rm self}/m_\chi\lesssim 0.7 \,{\rm cm^2/g}\sim3000/{\rm 
GeV^3}$, (for a review see~\cite{Tulin:2017ara}). It is straightforward 
to convert this to a constraint on the DM mass as
 \begin{equation}
m_\chi\gtrsim \frac{2}{3}\Big(\frac{\kappa_\chi}{16\pi^2}\Big)^{\!2/3}\Big(\frac{m_\chi}{\Lambda}\Big)^{\!4/3} \gev.		\label{eq:mdm_selfint}
 \end{equation}
 Going forward, we will consider the above result as a rough lower bound 
on the DM mass. 
 
There could also be constraints on this class of models from effects 
arising from neutrino self-interactions inside 
supernovae~\cite{Chang:2022aas}. However, the region of parameter space 
where these effects are likely to be important is already disfavored by 
the constraints on DM self-interactions.

%%%%%%%%%%%%%%%%%%%%%%%%%%%%%%%%%%%
 \subsection{Signals}
%%%%%%%%%%%%%%%%%%%%%%%%%%%%%%%%%%%
 Before we close this section, we would like to briefly remark on the 
implications of this class of models for direct and indirect detection 
experiments and for collider searches. 
 \bit
 \item {\bf Indirect detection:}
 Since the DM particle is a thermal relic, its annihilation cross 
section is of order $\langle\sigma v\rangle \sim 2\times 10^{-9}\gev^{-2}\sim 
2\times 10^{-26} \,{\rm cm^3/s}$. The dominant annihilation channels are either 
$\chi\bar \chi\to (N\bar \nu, \bar{N} \nu) $ or $\chi\bar \chi\to 
\nu\bar \nu$. When the dominant annihilation channel is $\chi \bar{\chi} 
\to (N \bar \nu, \bar{N} \nu)$, the visible end products such as 
electrons and photons produced in the decay of composite singlet neutrinos 
are constrained by 
indirect detection experiments and precision observations of the CMB. 
When the dominant annihilation channel is $\chi\bar \chi\to \bar\nu\nu$, 
the constraints from indirect detection are much weaker. All the 
annihilation channels give rise to monochromatic neutrinos and 
antineutrinos in the final state. Currently, the most stringent 
constraints on such a signal are provided by Super-Kamiokande 
(SuperK)~\cite{Super-Kamiokande:2002weg}, and the reach will be further 
expanded by Hyper-Kamiokande (HyperK)~\cite{HyperK} , DUNE~\cite{DUNE1, 
DUNE2, DUNE3} and JUNO~\cite{JUNO:2015zny}. Unfortunately, as we show in 
\sec{s.updm}, for both the $\chi\bar \chi\to (N\bar \nu, \bar{N} \nu) $ 
and $\chi\bar \chi\to \nu\bar \nu$ annihilation channels the region of 
parameter space that can be probed by these future searches is already 
disfavored by other considerations.

\begin{figure}
\centering
\includegraphics[width=0.35\textwidth]{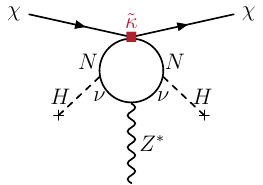}
 \caption{The figure shows the Feynman diagram for the loop-induced 
coupling of DM to the $Z$-boson once the Higgs boson acquires a VEV.}
 \label{fig:feyn_dd_comp}
\end{figure}

 \item {\bf Direct detection:} 
 The dominant contribution to DM scattering off nuclei arises from 
$Z$-boson exchange. 
The coupling of the $Z$ boson to the fermionic 
DM particles $\chi$ arises from a dimension-six effective 
operator,
 \begin{align}
\mathcal{L}_{\text{eff}} &\supset i\frac{C_{H\chi}}{\Lambda^2}H^{\dagger} 
\overleftrightarrow{D}_\mu H \bar{\chi} \gamma^{\mu} \chi\supset 
g_{Z\bar\chi\chi}Z_\mu \bar{\chi} \gamma^{\mu} \chi \;. \end{align}
 Here $H$ is the SM Higgs doublet and $H^{\dagger} 
\overleftrightarrow{D}_\mu H \equiv 
H^{\dagger}(\overrightarrow{D}_\mu-\overleftarrow{D}_\mu) H$, with $D_\mu$ 
being the covariant derivative. This operator is generated from the 
one-loop 
diagram shown in \fig{fig:feyn_dd_comp},
which has the heavy composite neutrinos and SM leptons 
running in the 
loop. 
The coupling of 
DM to the $Z$ boson is generated once the Higgs boson acquires a VEV. The 
Wilson 
coefficient above can be estimated as,
 \beq
C_{H\chi} \sim \frac{\tilde{\kappa} \,\lambda^2}{16\pi^2},
 \eeq
 where $\tilde{\kappa}$ and $\lambda$ are the DM and Higgs couplings to 
the composite singlet neutrinos respectively, while the factor of 
$16\pi^2$ arises from the loop. The scale $\Lambda$ arises from
the composite neutrino states that run in the loop, which have masses 
$m_N\sim \Lambda$. 
Therefore the DM interaction with the $Z$ boson 
is given by
 \beq
g_{Z\bar\chi\chi}=\frac{g\, C_{H\chi} 
}{2\cos\theta_W}\frac{v_{\text{EW}}^2}{\Lambda^2}\sim 
\frac{g}{2\cos\theta_W}\frac{\tilde{\kappa}}{(4\pi)^2}  U_{N\ell}^2 \;,
 \eeq
  where $g$ and $\theta_W$ are the weak coupling and the Weinberg angle, 
respectively. Since the DM coupling to the $Z$-boson is suppressed by the 
square of the mixing angle $U_{N\ell}$, the cross section for direct 
detection is expected to be small.

The spin-independent DM-nucleon cross 
section arising from this coupling can be estimated as,
 \beq
\sigma_{\chi n}\sim \frac{g^4 \,\tilde\kappa^2\, U_{N\ell}^4}{64\pi \cos^2\theta_W (4\pi)^4}\frac{\mu_{\chi n}^2}{m_Z^4},
 \eeq
where $g$ is the $SU(2)_L$ gauge coupling constant and $\mu_{\chi n}$ 
is the reduced mass of the DM-nucleon system. The factor of $(4\pi)^4$ 
in the denominator arises from the loop-suppressed coupling of the DM 
particle to the $Z$-boson. Since $\tilde\kappa$ is big and can be as large as 
$(4 \pi)^2$, this can compensate, at least partially, for the loop 
suppression of the cross section. Hence, in this class of models, the 
DM-nucleon cross section is primarily determined by the DM mass and 
mixing angle $U_{N\ell}$. Currently the most stringent limits on the 
DM-nucleon cross section have been set by the first results from LZ 
experiment~\cite{LUX-ZEPLIN:2022qhg}. In the near future, 
LZ~\cite{LUX-ZEPLIN:2018poe} and XENONnT~\cite{XENONnT} will be able to 
explore some part of the unconstrained parameter space.

 \item {\bf Collider Searches:}
 The composite singlet neutrinos $N$ in this class of models fall into 
the category of HNLs, for which the collider 
signals have been well studied in the literature. At colliders and beam 
dumps, these particles can be produced through Drell-Yan processes and in weak decays of hadrons. In 
the region of parameter space where $m_{\chi} > m_N/2$, corresponding to 
the $\chi \bar{\chi} \to (\bar{\nu} N, \bar{N} \nu)$ annihilation 
channel, the dominant decay mode of the $N$ is the same as that of a 
conventional HNL, and therefore the bounds from HNL searches in beam 
dumps and the LHC are directly applicable. However, in the region of 
parameter space for which $m_{\chi} < m_N /2$, where the relic abundance 
is set by $\chi \bar{\chi} \to \nu \bar{\nu}$, the dominant decay 
channel of $N$ is into $\nu\chi\bar{\chi}$. This decay is completely 
invisible, and so many of the standard collider and beam dump searches for 
HNLs do not apply.

 At colliders and beam dumps, DM can be pair produced in association 
with one or more composite singlet neutrinos. In our analysis we focus 
on the regime $m_{\chi} > m_N/2$, where the resulting signals are 
similar to those from the production of a conventional HNL, but come with 
additional missing energy. Bounds on this scenario are less well 
studied, as the event rates and kinematics are model-dependent. In this 
scenario, to discover the DM candidate, it is necessary to first 
discover the composite singlet neutrinos. Singlet neutrinos are 
typically produced in association with a charged lepton. In this regime 
they then decay to another charged lepton (or neutrino) and two SM 
fermions. In the case of a Majorana $N$, the smoking gun signature is a 
pair of same-sign leptons. When $m_N$ is heavier than several GeV, the 
strongest limits come from Drell-Yan production, while for lower values 
of $m_N$ the strongest limits are from meson decays. Searches for $N$ 
are broadly divided based on whether $N$ decays promptly, via displaced 
vertices, or whether it is long-lived. In \sec{s.colliders}, we map the 
existing bounds from ATLAS and CMS on HNLs onto the parameter space of 
our model and project the reach of the HL-LHC and the proposed MATHUSLA 
experiment~\cite{Mathusla}. We then turn our attention to the production 
of DM in association with $N$. For both the prompt and displaced 
searches, we describe how optimizing the cuts can increase the 
efficiency for events in the DM+$N$ signal. We also describe how reconstructing 
the $N$ (in a fully visible decay channel such as $N \to \ell q 
\bar{q}$) would offer the most promising avenue to detect the 
additional signal. Unfortunately, most of the parameter space of 
interest for future collider searches is in tension with the existing 
constraints from indirect detection. However, the allowed parameter space 
can be expanded if $\chi$ constitutes only a subcomponent of DM. 

\eit

%%%%%%%%%%%%%%%%%%%%%%%%%%%%%%%%
\section{Holographic Realization \label{s.holographic}}
%%%%%%%%%%%%%%%%%%%%%%%%%%%%%%%%%%%

In this section we present a holographic realization of our framework 
for composite DM via the neutrino portal. Theories in which the 
strong conformal dynamics is spontaneously broken are 
dual~\cite{ArkaniHamed:2000ds,Rattazzi:2000hs} to the two-brane 
Randall-Sundrum (RS) construction~\cite{Randall:1999ee}. Accordingly we 
consider a slice of 5D anti-de Sitter (\ads5) space bounded by two 
3-branes. The metric in the \ads5 slice is given by
 \begin{equation}
    ds^2=\left(\frac{R}{z} \right)^2 \eta_{MN}^{}\,dx^M dx^N,	\label{e.metric}
\end{equation}
 where $x^M\!=\!(x^\mu,z)$ with $\mu\!=\!0,1,2,3$ represent the familiar 
four-dimensional (4D) coordinates and the fifth coordinate $z$ is confined to the interval 
between the two branes. The branes are located at $z\!\equiv\!R$ and 
$z\!\equiv\!R^\prime$, i.e. $R\leq z \leq R'$. The AdS/CFT 
correspondence relates the location in the fifth dimension in the AdS 
space to the energy scale in the dual CFT. The locations of the two 
branes correspond to the UV and IR scales, 
$M_{\rm UV}\sim 1/R$ and $\Lambda_{\rm IR} \sim 1/R' \equiv \Lambda$, and so the two 
branes will be referred to as the UV-brane and the IR-brane. The 
presence of the UV-brane is associated with the 4D theory 
being defined with a cutoff, while the presence of the IR-brane is 
associated with the spontaneous breaking of conformal dynamics. The 
singlet neutrinos and the states that constitute DM arise as 
composites of the hidden sector, therefore arise from bulk fields in 
the higher-dimensional construction. On the other hand, since the SM 
fields are elementary, they are localized on the UV-brane.

 We introduce Dirac fermions $\widehat{\Psi}_{N}$ and $\Psi_{N}$ in the 
bulk of the extra dimension. The Dirac fermion $\widehat{\Psi}_{N}$ is 
the holographic dual of the operator $\mathcal{\widehat{O}}_{\!N}$ in 
the CFT, and will give rise to the right-handed composite singlet 
neutrino $N_R$ in the low-energy theory. Similarly, $N_L$ arises from 
the bulk Dirac fermion $\Psi_{N}$, which is assumed to be dual to an 
operator $\mathcal{O}_{\!N}$ of dimension $\Delta_N$ in the CFT. In 
addition, we introduce bulk Dirac fermions $\widehat{\Psi}_{\chi}$ and 
$\Psi_{\chi}$, which will give rise to the right- and left-handed 
chiralities of the Dirac fermion DM particle $\chi$. These are assumed 
to be dual to operators $\mathcal{\widehat{O}}_\chi$ and 
$\mathcal{O}_{\chi}$ of dimension $\Delta_{\widehat{\chi}}$ and 
$\Delta_{\chi}$ in the CFT. These bulk fields can be written out in 
terms of two-component spinors that transform as Weyl fermions under the 
Lorentz group in four spacetime dimensions,
 \begin{align}
 \Psi_{N}&=\left(\begin{array}{l}
N_{L} \\
N_{R}^{\prime}
 \end{array}\right), 
 &\widehat\Psi_{N}&=\left(\begin{array}{c}
N_{L}^{\prime} \\
N_{R}
 \end{array}\right),
 &\Psi_{\chi}&=\left(\begin{array}{l}
\chi_{L} \\
\chi_{R}^{\prime}
 \end{array}\right),  
 &\widehat\Psi_{\chi}&=\left(\begin{array}{c}
\chi_{L}^{\prime} \\
\chi_{R}
 \end{array}\right).
 \end{align}
 The CFT operators $\mathcal{O}_N$ and $\widehat{\mathcal{O}}_N$ 
transform as two-component Weyl fermions under the Lorentz group. In 
contrast, the bulk fermions $\Psi_{N}$ and $\widehat\Psi_{N}$ transform 
as four-component Dirac fermions. Therefore, to realize the duality we 
must impose boundary conditions on the UV-brane such that only one of 
the two Weyl fermions contained in each bulk Dirac fermion is sourced by 
the fields on that brane. Accordingly, on the UV-brane we impose the 
boundary conditions
  \begin{equation}
\left.\psi_{L}^{\prime},\,\psi_{R}^{\prime}\right|_{\rm UV}=0 \;, 
 \end{equation}
 where we have employed the notation $(\psi=N,\chi)$ to denote the bulk 
fermions. 
Furthermore, since we wish to consider a theory without any light 
states below the compositeness scale, on the IR-brane we impose the
boundary conditions,
 \begin{equation}
\left.\psi_{L},\,\psi_{R}\right|_{\rm IR}=0 \;.
 \end{equation}
 The action for the bulk fermions includes kinetic terms and mass terms, 
 \begin{align}
S_{\rm bulk}& \supset \int \!\!d^4 x\!\int\!\! dz \sqrt{g}\bigg[
    \frac{i}{2}\Big(\bar\Psi_{N}e_a^M \gamma^a \nabla_M \Psi_{N} - \nabla_M \bar\Psi_{N} e_q^M \gamma^a \Psi_{N}\Big) - \dfrac{c_{N}^{}}{R}\bar\Psi_{N}\Psi_{N} 	\notag\\
 &\qquad+\frac{i}{2}\Big(\bar\Psi_{\chi}e_a^M \gamma^a \nabla_M \Psi_{\chi} - \nabla_M \bar\Psi_{\chi} e_q^M \gamma^a \Psi_{\chi}\Big) - \dfrac{c_{\chi}}{R}\bar\Psi_{\chi}\Psi_{\chi} + \big\{\Psi\rightarrow \widehat \Psi, c\to \widehat c\big\}\bigg].
 \end{align} 
 Here $\nabla_M\equiv \partial_M + \omega_M$ where $\omega_M$ represents 
the spin connection and $e^a_M\equiv (R/z) \delta^a_M$ is the vierbein 
that relates the locally flat 5D coordinates $x^a$ to the warped 5D 
coordinates $x^M$. The bulk mass parameters $c_{\psi}$ and $\widehat 
c_{\psi}$ with $\psi=(N,\chi)$ are related to the scaling dimensions of 
the corresponding CFT operators~\cite{Contino:2004vy, 
Cacciapaglia:2008bi}
 \begin{align}
\Delta_{\psi} &= 2 + c_{\psi}  \,& \Delta_{\widehat{\psi}} &= 2 - \widehat c_{\psi} \;.
 \end{align} 
 The scaling dimensions of primary fermionic operators are bounded by 
unitarity to be $\Delta_{\psi}, \Delta_{\widehat{\psi}} \geq 3/2$, where 
the limiting case corresponds to a free fermion. Furthermore, it was 
noted earlier that fermionic scaling dimensions larger than $5/2$ render 
the theory UV sensitive. Therefore, in this work we consider scaling 
dimensions of the fermion fields in the range $3/2 < \Delta_{\psi}, 
\Delta_{\widehat{\psi}} < 5/2$. For the bulk mass parameters this 
translates to the range $-1/2<c_{\psi},\widehat{c}_{\psi}<1/2 $. Given 
that the bulk fields corresponding to $N$ and $\chi$ satisfy similar 
equations of motion and boundary conditions, the desired mass ordering 
$m_N > m_\chi$ can be obtained from an appropriate choice of the scaling 
dimensions, i.e. the bulk mass parameters.

 In the higher-dimensional framework, the SM is localized on the 
UV-brane. Then the interaction of the SM neutrinos with the hidden 
sector in the 4D theory, Eq.~(\ref{eq:portal_int_uv}), corresponds to 
the following brane-localized interaction in the higher-dimensional 
theory,
 \begin{align}
S_{\rm UV} &\supset \int d^4x\int d z\left(\frac{R}{z}\right)^{4} \delta(z-R)   \,\hat \lambda\,\sqrt{R}\, \bar L \widetilde H N_{\!R} +{\rm h.c.}
\label{e.S_UV}
 \end{align}
Here $\hat \lambda$ is a dimensionless coupling constant. To generate a 
Majorana mass term for $N$ as required by the inverse seesaw mechanism, we add a brane-localized term
 \begin{equation}
S_{\rm UV}\supset\int d^4x\int d z\left(\frac{R}{z}\right)^{4} \delta\left(z-R\right) \, \hat \mu\, N_L N_L +{\rm h.c.} \;, 
 \label{e.S_UV_maj}
 \end{equation}
 where $\hat \mu$ parametrizes the strength of lepton number violation. 
This term is the dual of Eq.~(\ref{lnv_deformation}) in the 4D theory, 
with the role of the holographic dual to the operator 
$\mathcal{O}_{2N}$ being played by $N_L N_L$. In order to generate 
the Dirac mass terms between $N_R$ and $N_L$ in the low energy theory, 
Eq.~(\ref{eq:lag_ir}), and between $\chi_L$ and $\chi_R$, 
Eq.~(\ref{eq:lag_{ir}}), we introduce Dirac mass terms for 
$\psi=(N,\chi)$ on the IR-brane,
 \begin{equation}
S_{\rm IR}\supset\int d^4x\int d z\left(\frac{R}{z}\right)^{4} \delta\!\left(z-R^{\prime}\right) \, \kappa_{\psi}\left(\bar{\psi}_{\!L} \psi_{\!R}+\bar{\psi}^{\prime}_{\!L} \psi^{\prime}_{\!R}+\hc \right),	\label{e.S_IR}
 \end{equation}
 The value of the parameter $\kappa_\psi$ determines the size of the 
resulting mass term. Note that the boundary conditions for the fermions 
$N_{L}$ and $N_{R}$ are modified on both the UV- and IR-branes because of 
the brane-localized terms~\eq{e.S_UV}, \eq{e.S_UV_maj} 
and~\eq{e.S_IR}. However, the boundary conditions for the DM fields 
$\chi_{L},\chi_{R}$ are only modified on the IR-brane due to the 
brane-localized term in \eq{e.S_IR}.

In order to mediate interactions between the DM candidate and the 
neutrinos, we introduce a complex scalar field $\Phi$ in the bulk. In 
addition to the kinetic term and mass term, the action for the scalar 
contains a Yukawa interaction,
 \begin{align}
S_{\rm bulk} \supset \int \!\!d^4 x\!\int\!\! dz &\sqrt{g}\bigg\{ g^{MN} \partial_M\Phi \partial_N\Phi^\ast - \frac{a^2}{R^2}|\Phi|^2+ \Big(\hat{y}\,\sqrt{R}\,\Phi \bar\Psi_{\chi} \widehat{\Psi}_{N}+{\rm h.c.}\Big) \bigg\} \;.	
\label{e.DM-N}
 \end{align} 
 Here $a$ is the mass parameter for the bulk scalar which is related to 
the scaling dimension $\Delta_{\Phi}$ of the corresponding primary 
operator $\mathcal{O}_{\Phi}$ as,
 \begin{equation}
\alpha \equiv \sqrt{4+a^2} = \Delta_{\Phi}-2. 
 \end{equation}
 We choose the following boundary conditions for the scalar field,
 \begin{equation}
\left.\Phi\right|_{\rm UV}=0, \lsp \text { and }	\lsp \left.\partial_{z} \Phi\right|_{\rm IR}=0.
 \end{equation}
 These boundary conditions ensure that the bulk scalar does not give rise to
any light states below the compositeness scale.

The interactions in \eq{e.S_UV} and \eq{e.DM-N} respect an overall 
lepton number symmetry under which $\widehat{\Psi}_N$ and $\Phi$ carry 
charges of $+1$ and $-1$, respectively. This symmetry is violated by the 
term in \eq{e.S_UV_maj}, giving rise to masses for the light neutrinos. 
All the interactions respect a discrete $Z_2$ symmetry under which the 
DM fields $\Psi_\chi$ and $\widehat{\Psi}_\chi$ and the scalar $\Phi$ 
are odd, while the remaining fields are even. This discrete symmetry 
ensures the stability of DM.

%%%%%%%%%%%%%%%%%%%%%%%%%%%%%%%%
\subsection{Kaluza-Klein Decomposition and Mass Spectrum \label{s.kkmass}}
%%%%%%%%%%%%%%%%%%%%%%%%%%%%%%%%%%%

In this subsection, we perform a Kaluza-Klein (KK) decomposition of the 
bulk fields and obtain the profiles of the various modes and their mass 
spectra. In what follows we will use the notation $\psi_n$ to denote the 
$n$th-KK mode ($n\!=\!1$ being the lowest mode) of the bulk fermion 
field $\psi$ ($\psi=N,\chi$). We will employ the analogous convention 
for all other bulk fields. The bulk fermion fields give rise to a tower 
of Dirac states,
 \begin{align} 
\psi_{L}(x, z) &=\sum_{n} g_{n}^{\psi}(z)\psi_{n,L}(x), & \psi_{L}^{\prime}(x, z)&=\sum_{n} g_{n}^{\psi^{\prime}}(z) \psi_{n,L}(x), \\ 
\psi_{R}(x, z) &=\sum_{n} f_{n}^{\psi}(z)\psi_{n,R}(x), & \psi_{R}^{\prime}(x, z)&=\sum_{n} f_{n}^{\psi^{\prime}}(z) \psi_{n,R}(x), 
 \end{align} 
 where the $\psi_{n}(x)=(\psi_{n,L},\psi_{n,R})$ satisfy the Dirac equation in the usual four spacetime dimensions at linear order. 
Substituting these expansions into the equations of motion for the bulk 
fermions we obtain the following equations for the bulk profiles of the 
KK modes,
 \begin{align}
\partial_z f_{n}^\psi-\frac{\widehat c_\psi+2}{z} f_{n}^\psi+m_{n}^{\psi} g_{n}^{\psi^{\prime}}-\kappa_{\psi} g_{n}^\psi \delta\!\left(z-R^{\prime}\right)&=0,	\\
\partial_z f_{n}^{\psi^{\prime}}-\frac{c_{\psi}+2}{z} f_{n}^{\psi^{\prime}}+m_{n}^{\psi} g_{n}^\psi-\kappa_{\psi} g_{n}^{\psi^{\prime}} \delta\!\left(z-R^{\prime}\right)&=0,	\\
\partial_z g_{n}^\psi+\frac{ c_\psi-2}{z} g_{n}^\psi-m_{n}^{\psi} f_{n}^{\psi^{\prime}}+\kappa_{\psi} f_{n}^\psi \delta\!\left(z-R^{\prime}\right)&=0,	\\
\partial_z g_{n}^{\psi^{\prime}}+\frac{\widehat c_{\psi}-2}{z} g_{n}^{\psi^{\prime}}-m_{n}^{\psi} f_{n}^\psi+\kappa_{\psi} f_{n}^{\psi^{\prime}} \delta\!\left(z-R^{\prime}\right)&=0 \;.
 \end{align}
 Here we have neglected the UV-localized Dirac and Majorana mass term 
contributions, assuming that $\lambda v_{\rm EW}$ and $\hat \mu$ are 
small compared to the other terms. The solutions to these equations 
for $g_n^\psi$ and $f_n^\psi$ are given by,
 \begin{align}
g_{n}^{\psi}(z) &=-\frac{z^{5 / 2}}{{\cal N}_{n}^{\psi}} Z_{L}^\psi(m_{n}^{\psi} z)\equiv -\frac{z^{5 / 2}}{{\cal N}_{n}^{\psi}}\Big[J_{-c_{\psi}-1 / 2}(m_{n}^{\psi} z)+b_{n}^{\psi} Y_{-c_{\psi}-1 / 2}(m_{n}^{\psi} z)\Big] , \\
f_{n}^{\psi}(z) &=\frac{z^{5 / 2}}{{\cal N}_{n}^{\psi}} Z_{R}^\psi(m_{n}^{\psi} z)\equiv \frac{z^{5 / 2}}{{\cal N}_{n}^{\psi}}\Big[J_{\widehat c_{\psi}-1 / 2}(m_{n}^{\psi} z)+\widehat b_{n}^{\psi} Y_{\widehat c_{\psi}-1 / 2}(m_{n}^{\psi} z)\Big] .
 \end{align}
 The solutions for $g_{n}^{\psi^{\prime}}$ and $f_{n}^{\psi^{\prime}}$ 
can be obtained by the replacements: $\psi \rightarrow \psi^\prime, 
c_{\psi} \leftrightarrow \widehat c_{\psi},$ and $ b_{n}^{\psi} 
\leftrightarrow \widehat b_{n}^{\psi}$. Imposing the above boundary 
conditions on the UV-brane, we obtain expressions for the $\widehat 
b_n^\psi$ and $b_n^\psi$,
 \begin{align}
\widehat b_{n}^{\psi}&=-\frac{J_{-\widehat c_{\psi}-1 / 2}(m_{n}^{\psi} R)}{Y_{-\widehat c_{\psi}-1 / 2}(m_{n}^{\psi} R)},
&b_{n}^{\psi}&=-\frac{J_{-c_{\psi} +1 / 2}(m_{n}^{\psi} R)}{Y_{-c_{\psi}+1 / 2}(m_{n}^{\psi} R)} \; .
 \end{align}
 Then imposing the boundary conditions on the IR-branes will determine 
the mass spectra. The normalization factors ${\cal N}_{n}^{\psi}$ can be 
obtained from the condition that the kinetic terms of the KK modes be 
canonically normalized. For the fermions, implementing the appropriate 
jump conditions in the limit of large Dirac mixing 
parameter~$\kappa_{\psi}$ leads to
 \begin{equation}
Z_{L}^{\psi}\!\big(m_{n}^{\psi} R^{\prime}\big)\, Z_{R}^{\psi}\!\big(m_{n}^{\psi} R^{\prime}\big) \simeq-Z_{R}^{\psi^{\prime}}\!\big(m_{n}^{\psi} R^{\prime}\big) \,Z_{L}^{\psi^{\prime}}\!\big(m_{n}^{\psi} R^{\prime}\big).
 \end{equation}
The mass spectrum in units of the IR scale can be obtained by 
determining the values of $m_nR^\prime$ for which this equation is 
satisfied. This is most easily done numerically. However, approximate 
analytical expressions can be obtained from the large and small argument 
expansions for the Bessel functions. These approximate forms are given by
\begin{equation}
\frac{ m_n^\psi }{ \Lambda} \approx \, \frac{n\,\pi}{2} + \Big(\Delta_\psi + \Delta_{\widehat{\psi}} - 5\Big)\frac{\pi}{4}\,.
\end{equation}

Note that the expressions for the 
spectra of $\chi$ and $N$ have the same form and hence the desired 
ordering of the masses of the lightest KK modes can be obtained by 
suitably choosing the scaling dimensions. Furthermore, we note that  
if the lepton number violating Majorana mass parameter $\hat \mu$ 
in~\eq{e.S_UV_maj} is small and the Dirac mass parameter $\kappa_N$ 
in~\eq{e.S_IR} is large, the two-component Weyl fermions $N_R$ and $N_L$ 
form a quasi-Dirac state $N=(N_L,N_R)^T$. Similarly, the two-component 
Weyl fermions $\chi_R$ and $\chi_L$ also form a Dirac state 
$\chi=(\chi_L,\chi_R)^T$.

The bulk scalar $\Phi(x,z)$ can be expanded out in KK modes as, 
 \begin{align}
\Phi(x, z) =\frac{1}{\sqrt{R}} \sum_{n} \phi_{n}(x) f_{n}^{\phi}(z),
\end{align}
 where the quadratic terms in the action for the $\phi_n(x)$ satisfy the 
Klein-Gordon equation. The bulk profile $f_{n}^{\phi}(z)$ can be obtained 
from the equation,
 \begin{align}
 \partial_z^2 f_{n}^{\phi}-\frac{3}{z}  \partial_z f_{n}^{\phi}+\Big((m_n^{\phi})^{2}-\frac{a^{2}}{z^2}\Big) f_{n}^{\phi} =0 \;.
 \end{align}
 This admits a solution of the form,
 \begin{align}
f_{n}^{\phi}(z) =\frac{z^{2}}{{\cal N}_{n}^{\phi}} Z_{\alpha}^{\phi}(m_{n}^{\phi} z)\equiv \frac{z^{2}}{{\cal N}_{n}^{\phi}}\Big[J_{\alpha}(m_{n}^{\phi} z)
+b_{n}^{\phi} Y_{\alpha}(m_{n}^{\phi} z)\Big].
 \end{align}
 Imposing the boundary condition on the UV-brane we have,
 \begin{align}
b_{n}^{\phi} = -\frac{J_{\alpha}(m_{n}^{\phi} R)}{Y_{\alpha}(m_{n}^{\phi} R)} \;.
 \end{align}
 Applying the IR boundary condition then determines the spectrum of 
scalar KK-modes, i.e. $m_{n}^{\phi}$. We find that scalar modes are 
typically heavier than the corresponding fermion modes.

%%%%%%%%%%%%%%%%%%%%%%%%%%%%%%%%
\subsection{The Effective Four-Dimensional Lagrangian 
\label{s.numixing}}
%%%%%%%%%%%%%%%%%%%%%%%%%%%%%%%%%%%

 In this subsection we determine the effective 4D Lagrangian for the 
light KK modes. We begin with a discussion of the mixing of the KK modes 
of the bulk singlet neutrinos with the SM neutrino. This mixing arises 
from the brane localized interaction in ~\eq{e.S_UV}. The resulting 
neutrino mixing term is given by,
 \begin{align}
{\cal L}_{\rm UV} &\supset \int \!d z\left(\frac{R}{z}\right)^{4} \delta(z-R) \,\hat\lambda\,\sqrt{R}\, v_{\rm EW} \,\bar{\nu}_{L} N_{R}(x,z) +{\rm h.c.}, \notag\\
& \equiv \sum_{n} \lambda_n \,v_{\rm EW}\,\bar{\nu}_{L} N_{n,R}(x) +{\rm h.c.}  	
 \label{e.S_UV_mix}
 \end{align}
 Here $\lambda_n$ is the parameter that controls the mixing between the 
$n$th-KK mode and the SM neutrino. After performing a KK decomposition 
of the $N_{R}(x,z)$ field and integrating over the $z$ coordinate we 
obtain an expression for $\lambda_n$,
 \beq
\lambda_n \equiv \hat\lambda\,\frac{R^3}{{\cal N}_{n}^N}\left|Z_{R}^{N}\left(m_{N_{n}} R\right)\right|.	
 \eeq
 The parameters in the equation above can be evaluated numerically to 
determine $\lambda_n$. The mixing angle between the SM neutrino and 
the $n$-th KK mode in the limit, $m_{N_n} \gg \lambda_n v_{\rm EW}$ is 
given by,
 \beq
U_{N_n\ell} \equiv \frac{\lambda_n v_{\rm EW}}{m_{N_n}}.	
\label{e.mixing}
 \eeq

After performing the KK decompositions of all the bulk fields,
the 4D Lagrangian takes the form, 
 \begin{align}
\mathcal{L}_{\rm IR}\supset&\,i\bar{\chi}_n \gamma^{\mu} \partial_{\mu} \chi_n+i\bar{N}_n \gamma^{\mu} \partial_{\mu} N_n+\partial_{\mu}\phi_n\partial^{\mu}\phi_n^\ast  +m_{\phi_n}^2 |\phi_n|^2 -m_{\chi_n} \bar\chi_n \chi_n - m_{N_n} \bar N_n N_n	\notag\\
&-\Big[\mu_{mn} (N_{m,L}\,N_{n,L})+\lambda_n \bar L \widetilde H N_{n,R}+y_{npq}\,\bar{\chi}_{n,L}\, N_{p,R} \,\phi_q+{\rm h.c.}\Big],	
\label{eq:lag}
 \end{align}
 where we have employed four-component Dirac notation for 
$N_n\!=\!(N_{n,L},N_{n,R})^T$ and 
$\chi_n\!=\!(\chi_{n,L},\chi_{n,R})^T$. In this expression summation 
over repeated indices is implied. The Majorana mass~$\mu_{mn}$ for 
the KK modes is defined through \eq{e.S_UV_maj} as
 \beq
\mu_{mn}\equiv \hat\mu\,R^5\bigg|\frac{Z_{L}^{N}\left(m_{N_{m}} R\right)\,Z_{L}^{N}\left(m_{N_{n}} R\right)}{{\cal N}_{m}^{N}\,{\cal N}_{n}^{N}}\bigg|.
\eeq
  The effective 4D coupling corresponding to the bulk $\Phi
\chi N$ interaction can be obtained from the overlap integral,
 \beq
y_{npq}=-\int d z \, \frac{\hat{y}\, R^{5}z^2}{{\cal N}_{n}^{N}\,{\cal N}_{p}^{\chi}{\cal N}_{q}^{\phi}} \, Z_{R}^{N}\big(m_{n}^{N} z\big)
Z_{L}^{\chi}\big(m_{p}^{\chi} z\big) Z_{\alpha}^{\phi}\big(m_{q}^{\phi} z\big) \;.
 \eeq
 We expect the effective coupling between the lightest KK modes, 
$y_{111}$, to be of the order of $\sim\! {4\pi}/\mathcal{N}$, where the 
value of $\mathcal{N}$ is set by the dual large-$\mathcal{N}$ CFT. In 
our analysis we will consider values of $y_{111}$ in the range from 1 to 
$4 \pi$.

At each KK level $n$, the scalars $\phi_n$ are heavier 
than the fermions $N_n$ and $\chi_n$. For the greater part of our 
analysis, we will restrict our attention to just the lightest KK mode of 
$\phi$ and the two lowest KK modes of $N$ and $\chi$. This turns out to 
be an excellent approximation. There are regions of parameter space 
where decays of $N_{2}$ to $\phi_{1}+\chi_{1}$ are kinematically allowed 
but decays to three $N_{1}$ are forbidden. This opens up interesting 
phenomenological prospects. Going forward, when there is no danger of 
ambiguity, we will omit the subscript $n=1$ for the lightest KK-mode of 
these fields, and only include a subscript $n\ge2$ when explicitly 
referring to one of the higher KK-modes.

We now consider the flavor structure of the model. The bulk fermions 
$\widehat{\Psi}_N$ and $\Psi_N$ and the scalar $\Phi$ each come in three 
flavors. However, we have just a single flavor of the DM fields 
$\widehat{\Psi}_\chi$ and $\Psi_\chi$. We assume that the bulk theory 
respects an $SU(3)$ flavor symmetry under which the fermions 
$\Psi_N^{\alpha}$ and $\widehat{\Psi}_N^{\alpha}$ transform in the 
fundamental representation. Here $\alpha$ is an $SU(3)$ flavor index. 
The complex scalar $\Phi_{\alpha}$ is assumed to transform in the 
antifundamental representation under this bulk flavor symmetry, while 
the DM fields $\Psi_\chi$ and $\widehat{\Psi}_\chi$ are singlets. Then 
the interaction in \eq{e.DM-N} is invariant under this symmetry. This 
flavor symmetry ensures that the different flavors of $N_{n,L}$, 
$N_{n,R}$ and $\phi_n$ are degenerate up to effects arising from the 
coupling to the SM in \eq{e.S_UV} and the lepton number violating term 
in \eq{e.S_UV_maj}. These need not respect the bulk flavor symmetry, and 
can give rise to a realistic spectrum of neutrino masses.

Restoring the flavor indices, the neutrino portal interaction in the 
Lagrangian in \eq{eq:lag} takes the form,
 \beq
\mathcal{L}_{\rm IR}\supset \lambda_n^{i \alpha} \bar L_i \widetilde H N_{n,R}^\alpha +{\rm h.c.} \; ,
 \eeq
 where $i\!=\!1,2,3$ is a SM flavor index. Similarly, the neutrino 
mixing \eq{e.mixing} can be written as,
 \beq
U_{N_n^{\alpha}\ell_i} \equiv \frac{\lambda_n^{i \alpha} v_{\rm EW}}{m_{N_n}} \; .	
\label{e.mixing_flavor}
 \eeq
 In general the neutrino mixing matrix $\lambda_n^{i\alpha}$ will give 
rise to mixing between the different flavors of the SM neutrino $\nu^i$ 
and composite neutrinos $N_n^\alpha$. However, for simplicity, in most 
of the phenomenological studies that follow we will take the portal 
coupling $\lambda_n^{i \alpha}$ to be flavor-diagonal and universal, so 
that $U_{N_n^{1}e} = U_{N_n^{2}\mu} = U_{N_n^{3}\tau} \equiv 
U_{N_n\ell}$.  We will relax this assumption when considering the 
implications of this scenario for lepton flavor violation~
\footnote{In some cases, extra dimensional models can 
provide a natural explanation for the suppression of flavor-violating 
couplings, see e.g.~\cite{Desai:2020rwz}.}. In the 
following sections we shall suppress the flavor indices unless there is 
a need to distinguish between flavors.

In Table~\ref{t.masses} we have given some representative values of the 
masses of the lightest KK modes. These correspond to the choices 
$\Delta_{N}=\Delta_{\widehat{N}}=9/4$ and $\Lambda=1$ GeV. We have set
$M_{\rm UV}=10^6$ GeV, the flavor scale. 
\begin{table}[t] 
 \centering 
 \begin{tabular}{||c|c|c|c|c|c|c||} 
 \hline 
$\Delta_{N}=\Delta_{\widehat N}$& $m_{N_{1}} / \Lambda$ & $m_{N_{2}} / \Lambda$ & $\Delta_{\chi}=\Delta_{\widehat \chi}$ & $m_{\chi_{1}} / m_{N_1}$ & $ \Delta_{\Phi}$ & 
$m_{\phi_{1}} / m_{N_1}$ \\ 
 \hline \hline 
2.25 & 1.12 & 2.75 & 1.75&0.4 & 2.0 &1.5 \\ 
2.25 & 1.12 & 2.75 & 1.91 & 0.6 & 2.0 &1.5 \\ 
2.25 & 1.12 & 2.75 & 1.91 & 0.6 & 1.5 &1.95 \\ 
2.25 & 1.12 & 2.75 & 2.0 & 0.7 & 2.0 &1.5 \\ 
 \hline 
 \end{tabular}
 \caption{In this table we present the masses of the lowest KK-modes for 
a few benchmark points. The values of $\Lambda=1 \gev$ and $M_{\rm 
UV}=10^6\gev$ are the same for all the benchmark points.}
 \label{t.masses}
\end{table}

 A comment regarding the spectrum of KK-gravitons and their potential 
effects on the dynamics under consideration is in order here. In RS-like 
warped geometries, with the 5D Ricci scalar and a negative cosmological 
constant in the bulk along with UV- and IR-brane tensions, the effective 
4D Planck mass is related to the UV scale of the 5D theory 
$M_{\rm UV}$~\cite{Randall:1999ee} as,
 \beq
\mpl^2\simeq R \, M_{\rm UV}^3 \sim M_{\rm UV}^2, 
 \eeq
 where we have employed the relation $R \sim 1/M_{\rm UV}$ and 
$R^\prime\gg R$. The above correspondence implies that in order to 
obtain 4D gravity in the low-energy effective theory, one requires 
$M_{\rm UV}\sim \mpl$. However, in our holographic setup discussed 
above, we are taking $M_{\rm UV}$, the scale of the UV-brane, as a free 
parameter. In the case when $M_{\rm UV}\ll\mpl$, the effective 4D 
gravity cannot be reproduced in this minimal gravitational setup. In 
such a scenario, in order to obtain effective 4D gravity when $M_{\rm 
UV}\ll\mpl$, it is necessary to add an Einstein-Hilbert term localized 
to the UV-brane to the action, see for 
example~\cite{Davoudiasl:2008hx,George:2011sw}. Accordingly we add to 
the action the term,
 \beq
S_{\rm UV} \supset\int d^4x\int d z \delta\left(z-R\right) \sqrt{-\hat g} \,2M_{0}^2\hat R,
 \eeq
 where $\hat g$ is the determinant of the induced metric $\hat 
g_{\mu\nu}$ and $\hat R$ is the corresponding Ricci scalar. The parameter
$M_{0}$ has the dimensions of mass. In the presence of such a brane 
localized Einstein-Hilbert term, the effective 4D Planck mass 
in the limit $R^\prime\gg R$ is given by,
 \beq
\mpl^2\simeq R \, M_{\rm UV}^3\bigg[1+\frac{M_0^2}{R \, M_{\rm UV}^3}\bigg] \simeq M_{0}^2. 
 \eeq
 Hence, taking the parameter $M_0$ to be of the order of the Planck 
mass, i.e. $M_{0}\sim \mpl$, we recover 4D gravity. In this scenario, as 
noted in~\cite{George:2011sw}, the spectrum of graviton KK modes remains 
similar to that of the RS model. In particular, the graviton KK spectrum is 
approximately given by the zeros of the Bessel function $J_1(m_n^{\rm g} 
R^\prime)$. From a more precise calculation we find that the first 
KK-graviton has a mass $m_1^{\rm g}\sim 3 \Lambda$ and its dependence on 
the UV scale $M_{\rm UV}$ is negligible. The masses of the higher graviton 
KK modes can be approximated as,
 \beq
\frac{m_n^{\rm g}}{\Lambda}\approx \Big(n+\frac{1}{4}\Big) \pi\,. 
\eeq
 Note that the first graviton KK-mode $m_1^{\rm g}$ is typically about 
twice as heavy as the first KK-mode of any other bulk particle in our 
model, as can be seen from the benchmark values in \tab{t.masses}. This 
large mass, coupled with the fact that the KK-graviton wave functions 
are highly suppressed at the location of the UV-brane where the SM 
particles reside, allows us to safely neglect the effects of the KK 
gravitons on the dynamics we are studying.

%%%%%%%%%%%%%%%%%%%%%%%%%%%%%%%%
\section{Dark Matter Phenomenology \label{s.updm}}
%%%%%%%%%%%%%%%%%%%%%%%%%%%%%%%%%%%

In this section, we study the phenomenological aspects of our model in 
detail. We determine the regions of parameter space where we reproduce 
the observed abundance of DM and explore the prospects for direct and 
indirect detection. Our analysis is based on the 4D Lagrangian obtained 
after the KK decomposition of the higher dimensional theory.

%%%%%%%%%%%%%%%%%%%%%%%%%%%%%%%%%%%
\subsection{Relic Abundance\label{s.relic}}
%%%%%%%%%%%%%%%%%%%%%%%%%%%%%%%%%%%
%
\begin{figure}
\centering
\includegraphics[width=0.77\textwidth]{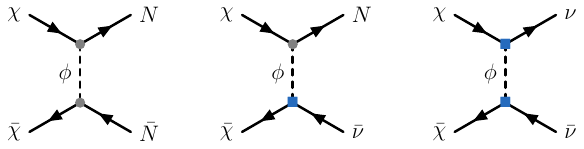}
 \caption{The dominant DM annihilation processes in the mass ranges 
$m_\chi/m_N\gtrsim 0.8$ (left), $0.5 \lesssim m_\chi/m_N\lesssim 0.8$ 
(center), and $m_\chi/m_N< 0.5$ (right). For the center diagram, there is an analogous process for the final state $\bar N \nu$. A grey-dot denotes an effective 
coupling $y_{\rm eff}$ and a blue-square denotes an effective coupling 
$y_{\rm eff} U_{N\ell}$. The flavor indices have been suppressed.  }
 \label{fig:chichi_ann}
\end{figure}

We begin our phenomenological analysis with a study of the DM relic 
abundance. Annihilation processes mediated by the scalar $\phi$ play the 
dominant role. Their rate depends sensitively on the coupling $y_{\rm 
eff}\!\equiv \!y_{111}$. In our analysis we will consider values of 
$y_{\rm eff}$ in the range from 1 to $4 \pi$. In the early universe, the 
states of the composite sector are initially in thermal equilibrium with 
the SM states through the neutrino portal via processes such as 
$N\nu\leftrightarrow N\nu$ and $NN\leftrightarrow \nu\nu$. The thermal 
freezeout of $\chi$ can occur through annihilation into three different 
final states, as shown in ~\fig{fig:chichi_ann}. The relative importance 
of these processes depends on the ratio $m_\chi/m_N$:
 \bit
 \item \underline{$m_{\chi}/m_N\geq 0.8$}: Note that $\chi$ can be 
stable even when $m_{\chi}>m_N$, as long as $m_{\chi} < m_{\phi}$. The 
dominant DM annihilation process in this mass range is $\bar \chi 
\chi\to \bar{N} N$. Even though this process cannot proceed at zero 
temperature if $m_{\chi} < m_N$, the kinetic energy of the $\chi$ 
particles means that it can be important at finite temperatures. From 
numerical calculations we find that it continues to be the dominant 
annihilation process for values of $m_{\chi}/m_N$ above about 0.8. This 
is an example of forbidden DM~\cite{DAgnolo:2015ujb}.  For this 
annihilation channel, the thermally averaged cross section is 
proportional to $y_{\rm eff}^4$, which in a strongly coupled theory is 
expected to be large. Therefore, the annihilation is extremely effective and the observed dark matter abundance is only obtained in a limited region of parameter space.
 \item \underline{$0.5 \lesssim m_{\chi}/m_N\lesssim 0.8$}: In this mass 
range the dominant annihilation process is $\bar \chi \chi \to (\bar{N} 
\nu, \bar{\nu} N)$. The thermally averaged cross section is proportional 
to $y_{\rm eff}^4 U_{N\ell}^2$, and the observed DM relic abundance can 
be obtained at sufficiently small mixing.
 \item \underline{$m_{\chi}/m_N<0.5$}: In this case the dominant 
annihilation process is $\bar \chi \chi\to \bar{\nu} \nu$ and the thermally 
averaged cross section is proportional to $y_{\rm eff}^4 U_{N\ell}^4$. 
Therefore larger mixing angles are favored compared to the mass range 
above. 
 \eit

 We used the package \texttt{micrOMEGAS-5.2}~\cite{micrOMEGAs} to 
determine the relic abundance. In our analysis we were careful to 
include the coannihilation processes involving higher KK modes. However, 
for pedagogical reasons, in the discussion below we limit ourselves to 
an approximate analytic calculation of the DM relic abundance that 
involves only the lowest KK modes. 

We first consider the case when the dominant DM annihilation channel is 
$\bar \chi\chi\to (\bar{N} \nu, \bar{\nu} N)$. In the limit $m_\phi^2\gg 
t$, we can approximate the spin-averaged cross section to a single 
flavor of the final state neutrinos ($\bar\chi \chi \to \bar{N} \nu$) as
 \beq
\sum_{\rm spins}\Big|{\cal M}(\bar \chi\chi\to \bar N \nu)\Big|^2\approx \frac{ y_{\rm eff}^4U_{N\ell}^2  (m_{\chi}^2-t) (m_{\chi}^2+m_{N}^2-t)}{4 (t-m_{\phi}^2)^2} \;,
 \eeq
 where $t$ is the Mandelstam variable.
 After summing over the different flavors and thermally averaging the 
cross section, we obtain
 \begin{align}
\langle\sigma_{\bar\chi\chi\to \bar N\nu, N \bar \nu}  v\rangle\approx \frac{3\,  y_{\rm eff}^4\, U_{N\ell}^2 \big(4 m_{\chi}^2-m_{N}^2\big)^2 \big(4 m_{\chi}^2+m_{N}^2\big)}{512 \pi  m_{\chi}^4 \big(2 m_{\chi}^2-m_{N}^2+2 m_{\phi}^2\big)^2}.
 \end{align}
 The Boltzmann equation for the yield ($Y_\chi\equiv n_\chi/s$) as a 
function of $x\equiv m_{\chi}/T$ is given by,
 \begin{align}
\frac{d Y_{\chi}}{d x}=-\frac{\lambda_\chi \left\langle\sigma_{\bar\chi\chi\to \bar N\nu, N \bar \nu} v\right\rangle}{x^2}\left[Y_{\chi}^{2}-\bar Y_{\chi}^{2}\right] \;,
 \end{align}
where $n_\chi$ is the DM number density and $s$ is the entropy density.
 In this expression the equilibrium yield $\bar Y_\chi$ and the parameter
$\lambda_\chi$ are defined as
 \begin{align}
\bar Y_\chi&\equiv \frac{\bar n_\chi}{s}=\frac{45}{4 \pi^4}\frac{g_\chi}{g_{\star S}} x^2 K_2(x), &\lambda_\chi &\equiv \frac{x\, s}{{\cal H}}=\sqrt{\frac{8 \pi^{2}}{45}} \,\frac{g_{\star S}}{\sqrt{g_{\star}}}\,\mpl \,m_\chi \;.
 \end{align}
 Here $\mpl$ is the reduced Planck mass, ${\cal H}$ is the Hubble rate and 
$K_2(x)$ is the modified Bessel function. The parameters $g_{\star}$ and 
$g_{\star S}$ represent the effective number of relativistic degrees of 
freedom for the energy and entropy densities of radiation respectively, 
while $g_\chi$ denotes the number of degrees of freedom in DM.

Equivalent expressions for the $\bar\chi\chi\to \bar\nu \nu$ 
annihilation channel are
 \beq
\sum_{\rm spins}\Big|{\cal M}(\bar \chi\chi\to \bar \nu \nu)\Big|^2\approx \frac{ y_{\rm eff}^4U_{N\ell}^4  (m_{\chi}^2-t)^2}{4 (m_{\phi}^2-t)^2} \;.
 \eeq
 and 
 \begin{align}
\langle\sigma_{\bar\chi\chi\to \bar \nu\nu}\,  v\rangle\approx \frac{3\,  y_{\rm eff}^4\, U_{N\ell}^4\, m_{\chi}^2}{32 \pi \, m_{\phi}^4} \;.
 \end{align}
The present-day DM relic density can be obtained by solving the 
Boltzmann equation,
 \beq
\Omega_\chi h^2=\frac{h^2 s_0}{\rho_{\text{cr}}}m_\chi Y_{\chi}^{(0)}=2.742\times 10^{8}\left(\frac{m_\chi}{\text{GeV}}\right)Y_{\chi}^{(0)},	\label{eq:relic}
 \eeq
 where $Y_{\chi}^{(0)}$ is the DM yield, $s_0=2970\,{\rm cm}^{-3}$ is 
the total entropy density today, and $\rho_{\text{cr}}=1.054 \times 
10^{-5} h^{2} \gev\,{\rm cm}^{-3}$ is the critical density. The observed 
DM relic abundance is $\Omega_{\rm obs} 
h^2=0.12\pm0.012$~\cite{Planck:2018vyg}.

\begin{figure}
\centering
\includegraphics[width=0.7\textwidth]{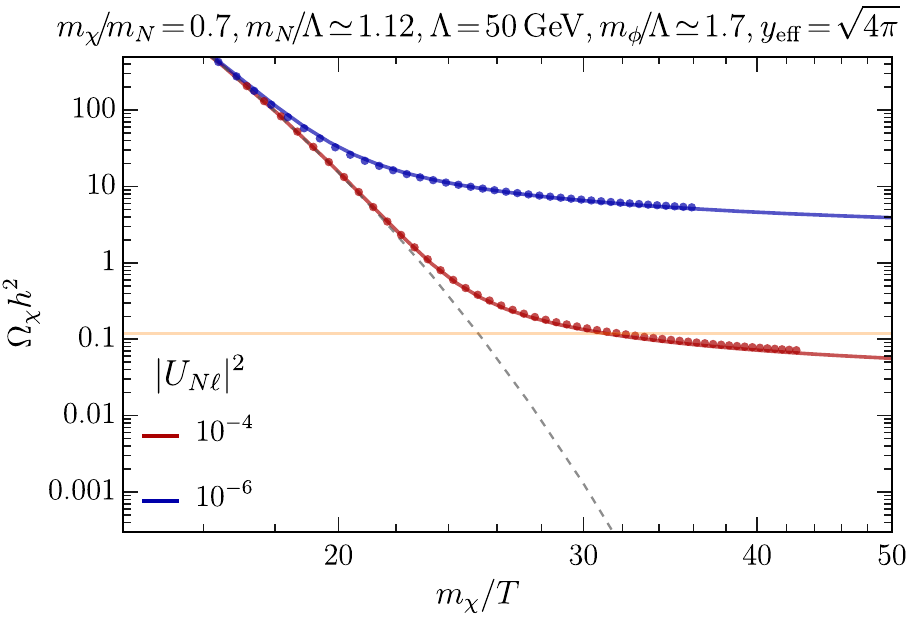}
 \caption{The DM relic abundance as a function of $x=m_\chi/T$ is shown 
for a benchmark spectrum with the parameters $\Lambda=50\gev$, $m_\chi\simeq 40\gev$, 
$m_N/\Lambda\simeq1.12$, $m_\phi/\Lambda\simeq 1.7$ and two choices of the mixing 
angle, $|U_{N\ell}|^2=10^{-4}$ (\textcolor{BrickRed}{red}) and $10^{-6}$ 
(\textcolor{Blue}{blue}). The dominant annihilation channel is $\bar 
\chi \chi\to (\bar{N} \nu, \bar{\nu} N$). The dots show the numerical 
results obtained using \texttt{micrOMEGAS}. The dashed-grey line 
represents the equilibrium density and the orange horizontal band 
corresponds to the observed DM relic density.}
 \label{fig:megah2xmix_NuN}
\end{figure}

\begin{figure}
\centering
\includegraphics[width=0.7\textwidth]{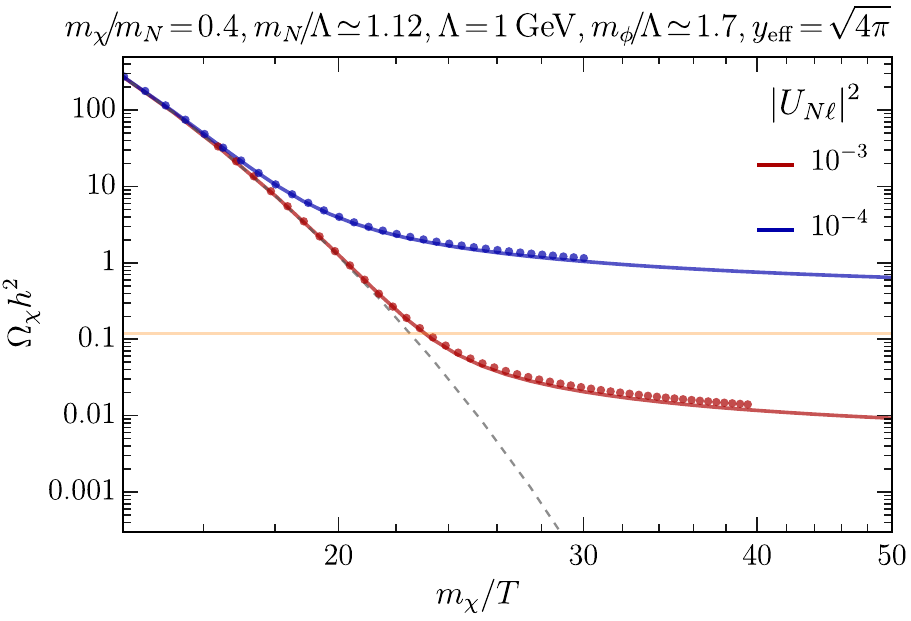}
 \caption{The DM relic abundance as a function of $x=m_\chi/T$ is shown 
for a benchmark spectrum with $m_\chi/m_N=0.4$, 
$\Lambda=m_N/1.12=1\gev$, $m_\phi=1.7\gev$ and two choices of the mixing 
angle, $|U_{N\ell}|^2=10^{-3}$ (\textcolor{BrickRed}{red}) and $10^{-4}$ 
(\textcolor{Blue}{blue}). The dominant annihilation channel is $\bar 
\chi \chi\to \bar{\nu} \nu$. The dots show the numerical results 
obtained using \texttt{micrOMEGAS}. The dashed-grey line represents the 
equilibrium density and the orange horizontal band corresponds to the 
observed DM relic density. }
 \label{fig:megah2xmix_NuNu}
\end{figure}

In \fig{fig:megah2xmix_NuN} we show the evolution of the DM relic 
abundance as a function of $x=m_\chi/T$ for a benchmark spectrum with 
$\Lambda=50\gev$, $m_N/\Lambda=1.12$, $m_\chi/m_N=0.7$, and $m_\phi/\Lambda=1.7$, the dominant 
annihilation channel being $\bar \chi \chi\to (\bar{N} \nu, \bar{\nu} 
N)$. We have taken $y_{\rm eff}=\sqrt{4\pi}$, and considered two values 
of the mixing, $|U_{N\ell}|^2=10^{-6}$ (\textcolor{BrickRed}{red}) and 
$10^{-8}$ (\textcolor{Blue}{blue}). The solid lines result from the 
approximate analytic calculation presented above, whereas the dots show 
the numerical results obtained with \texttt{micrOMEGAS}, which include 
the effects of the higher KK modes. \fig{fig:megah2xmix_NuNu} shows an 
equivalent plot for a benchmark wherein $m_\chi/m_N=0.4$, 
$\Lambda=m_N/1.12=1\gev$ and $m_\phi=1.7\gev$, the dominant annihilation 
channel being $\bar \chi \chi\to \bar \nu \nu$. With the same value for 
$y_{\rm eff}$, we consider larger values for the mixing 
$|U_{N\ell}|^2=10^{-3}$ (\textcolor{BrickRed}{red}) and $10^{-4}$ 
(\textcolor{Blue}{blue}). This is necessary because of the extra factors 
of the mixing that appear in the corresponding annihilation cross 
section. The good agreement between the analytic and numerical results 
confirms that neglecting the higher KK-modes is a good approximation.
%

%%%%%%%%%%%%%%%%%%%%%%%%%%%%%%%%%%%
\subsection{Direct Detection\label{s.direct_detection}}
%%%%%%%%%%%%%%%%%%%%%%%%

Direct detection experiments are significantly less sensitive to 
composite DM that couples to the SM through the neutrino portal than to 
conventional WIMPs. This is because the DM-nucleon interactions are 
induced only at the loop level and are further suppressed by the small 
mixing between the SM neutrinos and their singlet counterparts. While 
this has the effect of weakening the constraints on the model, it of 
course also makes the model more challenging to discover in direct 
detection experiments. In particular, we will show below that in a large 
region of parameter space, the DM-nucleon cross section lies below the 
neutrino floor.

\begin{figure}
\centering
\includegraphics[width=0.35\textwidth]{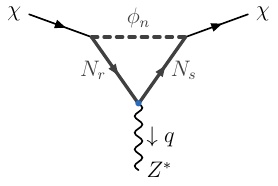}
 \caption{The figure shows the Feynman diagram for the loop-induced 
coupling of DM to the $Z$-boson. The subscripts on the fields in the 
loop denote the KK modes. The blue dot denotes the coupling of the $Z$ 
in the mass eigenbasis to $N_r$ and $N_s$, which is suppressed by mixing 
angles.}
 \label{fig:feyn_dd}
\end{figure}

The loop diagram shown in \fig{fig:feyn_dd} leads to an 
effective $Z\bar\chi\chi$ vertex of the form
 \beq
g_{Z\bar\chi\chi }\,Z_\mu\bar{\chi}\gamma^{\mu}P_L\chi \;,
 \eeq 
 where $P_L$ is the projection operator for left-handed states. This 
gives rise to DM scattering off nuclei via $ Z$ exchange. In what 
follows, we calculate the cross section for this process, working in the 
physical mass eigenbasis. Although an effective coupling to the Higgs 
boson is also induced, the Higgs exchange contribution to direct 
detection is additionally suppressed by the small coupling of the Higgs 
to nuclei, and can therefore safely be neglected. The effective coupling 
$g_{Z\bar\chi\chi }$ is given by
 \beq
 g_{Z\bar\chi\chi } = \sum_{n,r,s}\frac{U_{N_r\ell}\,U_{N_s\ell} \,y_{r1n} \,y_{s1n}}{16\pi^2}\frac{3g \,m_{N_r}m_{N_s}}{2\cos\theta_W}\,
   C_0\left(m_\chi^2,m_{\chi}^
   2,q^2;m_{N_r},m_{\phi_n},m_{N_s}\right),
 \label{eq:L_Zchichi}
 \eeq   
where the sum is over the KK modes of $N$ and $\phi$. Here $C_0$ is the 
scalar Passarino--Veltman three-point function as defined in 
\rcite{Patel:2016fam} and $q^\mu$ is the four-momentum carried by the 
$Z$. Since each term in the sum is proportional to the fermion masses in 
the loop, the contribution from the light neutrinos is negligible. All 
the terms in the sum are suppressed by the squares of the mixing 
parameters. On the other hand, since the couplings $y_{r1n}$ arise from 
strong dynamics, they are expected to be large.

The contribution to $g_{Z\bar\chi\chi }$ from just the lowest KK-modes 
of $N$ and $\phi$ in the limit $m_\phi\gg m_N$ is given by
 \beq
g_{Z\bar\chi\chi }= \frac{U_{N\ell}^2\,y_{\rm eff}^2}{8\pi^2}\frac{g}{2\cos\theta_W}\frac{m_{N}^2}{m_\phi^2}\bigg[1+\ln\bigg(\frac{m_N^2}{m_\phi^2}\bigg)\bigg].	\label{eq:g_Zchichi}
 \eeq
 However, we find that including the higher KK-modes of $N$ in the loop 
corrects this expression by an order one factor. We therefore provide 
formulas for the general terms appearing in the sum. In particular, in 
the limit $q^2\ll m_\chi^2$, one can simplify the above 
Passarino--Veltman $C_0$~function. For $r=s$, the $C_0$ function 
simplifies to
 \begin{align}
C_0(m_\chi^2,m_{\chi}^2,0;m_{N_r},m_{\phi_n},m_{N_r})&\!=\! -\frac{(m_{\chi }^2-m_{N_r}^2+m_{\phi _n}^2)
   \Lambda\big(m_{\chi }^2,m_{N_r},m_{\phi
   _n}\big)}{\lambda\big(m_{\chi }^2,m_{N_r}^2,m_{\phi
   _n}^2\big)}-\frac{1}{2 m_{\chi }^2}
   \ln\!\!\bigg(\frac{m_{N_r}^2}{m_{\phi _n}^2}\bigg).
\end{align}
 For $r\neq s$, we get instead,
 \beq
\frac{1}{m_{N_r}^2-m_{N_s}^2}\bigg[ \bigg( \Lambda \left(m_{\chi }^2,m_{\phi_n},m_{N_r}\right)- \frac{(m_{\chi }^2+m_{N_r}^2-m_{\phi _n}^2)}{2 m_{\chi }^2}
   \ln\!\Big(\frac{m_{N_r}^2}{m_{\phi _n}^2}\Big)\bigg) - \left(r \leftrightarrow s\right)\bigg].
 \eeq
 The K??ll??n kinematic triangular polynomial $\lambda(a,b,c)$ and the 
function $\Lambda(m_0^2, m_i, m_j)$ appearing in the formulae above are 
defined as
 \begin{align}
\lambda\big(a,b,c\big)&\equiv a^2 +b^2 +c^2 -2ab -2bc -2ca\,,	\\
\Lambda(m_0^2, m_i, m_j)&\equiv\frac{1}{m_0^2}\sqrt{\lambda\big(m_0^2,m_i^2,m_j^2\big)} \ln\!\bigg(\frac{m_i^2+m_j^2-m_0^2+\sqrt{\lambda\big(m_0^2,m_i^2,m_j^2\big)}}{2 m_i m_j}\bigg).
 \end{align}

\begin{figure}
\centering
\includegraphics[width=0.7\textwidth]{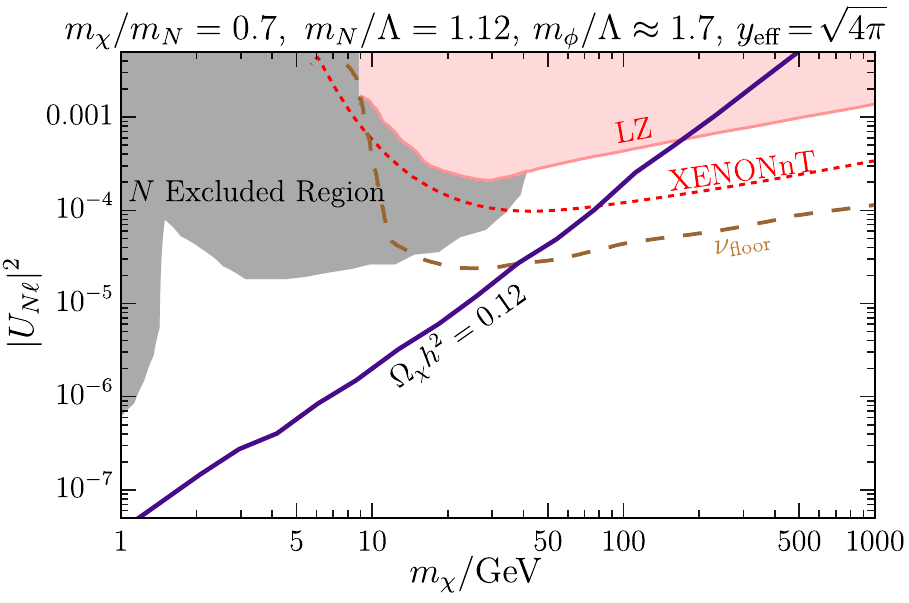}
 \caption{We plot the current exclusion from the LZ 
experiment (2022)~\cite{LUX-ZEPLIN:2022qhg}, along with the expected 
sensitivity of the XENONnT experiment~\cite{XENONnT} and the mixing 
angle corresponding to the neutrino floor~\cite{XENONnT}, for the 
benchmark model parameters listed at the top of the figure. The gray-shaded region is ruled out by a variety of collider and beam dump 
searches. The solid blue curve represents the contour of $y_{\rm eff}=\sqrt{4\pi}$ which 
produces the observed DM relic abundance.}
 \label{fig:dd_NuN}
\end{figure}

\begin{figure}
\centering
\includegraphics[width=0.7\textwidth]{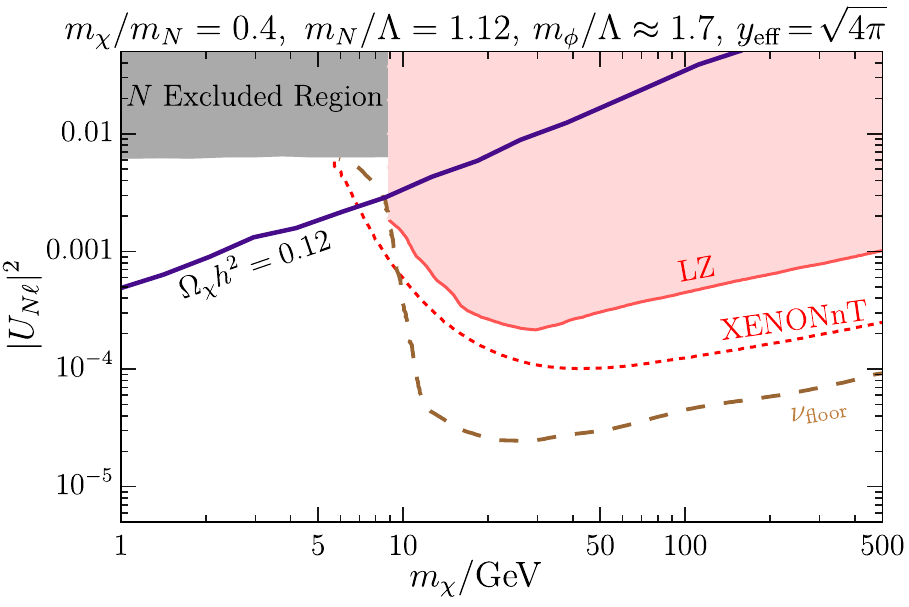}
 \caption{We plot the current exclusion from the LZ 
experiment~\cite{LUX-ZEPLIN:2022qhg}, along with the expected 
sensitivity of the XENONnT experiment~\cite{XENONnT} and the mixing 
angle corresponding to the neutrino floor~\cite{XENONnT}, for the 
benchmark model parameters listed at the top of the figure. The 
gray-shaded region is ruled out by a variety of collider searches. The 
solid blue curve produces the observed DM relic abundance for $y_{\rm 
eff}=\sqrt{4\pi}$.}
 \label{fig:dd_NuNu}
\end{figure}

 The DM--nucleon spin-independent cross section $\sigma_{\chi 
n}^{\rm SI}$ mediated via $Z$ exchange is given by,
 \beq
\sigma_{\chi n}^{\rm SI}=\frac{g^2g_{Z\chi\bar\chi}^2}{64\pi \cos^2\theta_W m_Z^4}\bigg(\frac{m_\chi m_n}{m_\chi+m_n}\bigg)^2 \left[\left(1+\frac{Z}{A}\right) V_u + \left(2-\frac{Z}{A}\right) V_d \right]^2 ,
 \eeq
 where $m_n$ is the nucleon mass, $Z/A$ is the ratio of the atomic and 
mass numbers of the target nucleus, and $V_f\!=\! (2 T^3_f-4 Q_f 
\sin^2\theta_W)$ for a fermion $f$ with electric charge $Q_f$ and 
isospin number $T^3_f$. Using this cross section, we plot in 
Figs.~\ref{fig:dd_NuN} and~\ref{fig:dd_NuNu}, for a set of benchmark 
model parameters, the mixing angles as a function of $m_\chi$ that 
correspond to the current exclusion from the first results of LZ 
experiment~\cite{LUX-ZEPLIN:2022qhg} and the expected sensitivity of the 
XENONnT experiment~\cite{XENONnT}. Also shown is the mixing angle that 
would result in a cross section equal to the neutrino 
floor~\cite{XENONnT}.

%%%%%%%%%%%%%%%%%%%%%%%%%%%%%%%%%%%
\subsection{Indirect Detection\label{s.indirect}}
%%%%%%%%%%%%%%%%%%%%%%%%%%%%%%%%%%%

The indirect detection signals of this class of models are very 
different for the $(\bar{N} \nu, \bar{\nu} N)$ and $\bar{\nu} \nu$ 
annihilation channels. In the 
case when the primary annihilation channel is $\bar \chi \chi\to 
(\bar{N} \nu, \bar{\nu} N)$, 
the singlet neutrino in the final state decays to leptons, neutrinos, and hadrons 
with ${\mathcal O}(1)$ branching ratios. When all the unstable particles 
have decayed, a continuum spectrum of electrons, positrons, photons, and 
neutrinos is produced. On average, their energies do not significantly 
exceed $m_N / 3$, since the $N$-decay is 3-body. There are strong 
astrophysical and cosmological constraints on these final states from 
indirect detection experiments. In particular, cosmic microwave 
background (CMB) measurements provide a robust probe of DM annihilations 
to energetic electrons and photons around the recombination epoch. The 
energy injected into the plasma by annihilations of DM particles 
modifies the ionization history as well as temperature and polarization 
anisotropies. The measurements of the CMB by the Planck 
collaboration~\cite{Planck:2018vyg} set stringent constraints on DM 
annihilation for GeV-scale or lighter DM. There are also constraints on 
this class of theories from present-day observations of gamma rays from 
the galactic center and from dwarf galaxies. In the case of annihilation 
to $\bar{\nu} \nu$, none of these constraints apply.

Since the DM particles are nonrelativistic, it follows that for both the 
$(\bar{N} \nu, \bar{\nu} N)$ and $\bar{\nu} \nu$ channels, the SM 
neutrinos produced in the annihilation process are monochromatic. The 
neutrino energy is given by $E_{\nu}=m_{\chi}(1-m_{N}^2/4m^{2}_{\chi})$ 
for the $(\bar{N} \nu, \bar{\nu} N)$ annihilation mode and 
$E_{\nu}=m_{\chi}$ for the $\bar{\nu} \nu$ annihilation mode. A 
monochromatic neutrino line is a striking signature for indirect 
detection experiments. After identifying the regions in parameter space 
that are consistent with the CMB and gamma-ray constraints, we will 
estimate the reach for such a signature in \sec{sec:neu-line}.

%%%%%%%%%%%%%%%%%
\subsubsection{CMB constraints}
\label{sec:cmb}

As mentioned above, CMB observations constrain the $(\bar{N} \nu, 
\bar{\nu} N)$ annihilation channel since the subsequent decays of 
composite singlet neutrinos inject energy into the intergalactic medium 
(IGM) at the recombination epoch. The constraints from the Planck 
collaboration~\cite{Planck:2018vyg} are expressed as channel-dependent 
upper-bounds on the thermally averaged annihilation cross section 
$\langle \sigma v\rangle$ at 95\% C.L.,
 \beq
f_{\!\rm eff}(m_\chi) \frac{\langle \sigma v\rangle}{m_\chi} < 3.2\times 10^{-28}\,{\rm cm^3\,s^{-1}}\gev^{-1}, 		\label{e.planckcmb}
 \eeq
 where $f_{\!\rm eff}(m_\chi)$ is the effective fraction of energy 
transferred to the IGM from DM annihilation at a redshift $z\sim 600$ 
where the CMB anisotropy data is most sensitive. In our model, $\langle 
\sigma v\rangle\sim {\rm few}\times 10^{-26}\,{\rm cm^3\,s^{-1}}$ in 
order to produce the observed DM relic abundance. Therefore, the CMB 
constraint can be translated directly into a limit on $f_{\!\rm 
eff}(m_\chi)$, which is a function of the DM mass and the final state 
particles that result from the DM annihilation process.

\begin{figure}[t]
\centering
\includegraphics[width=0.47\textwidth]{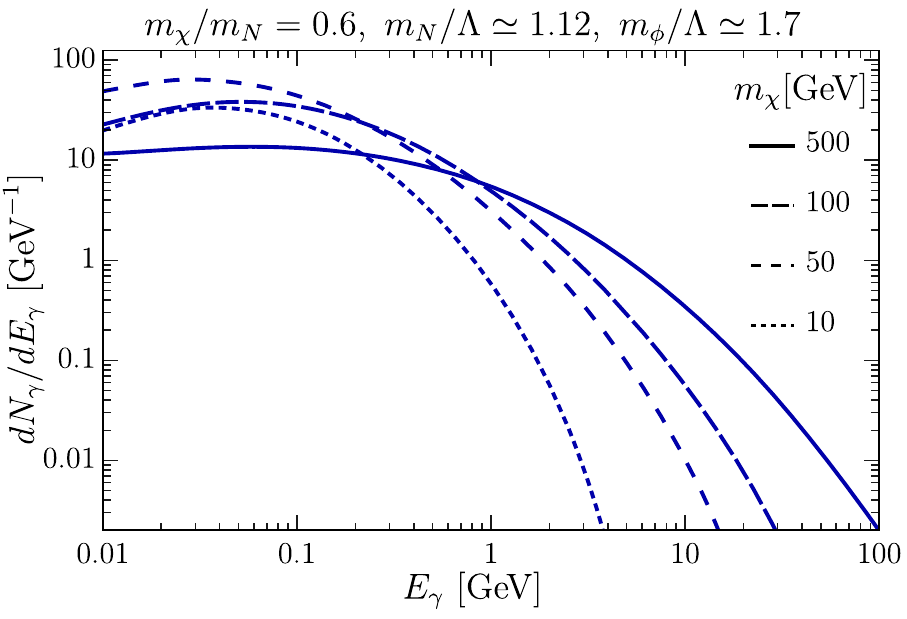}
\includegraphics[width=0.47\textwidth]{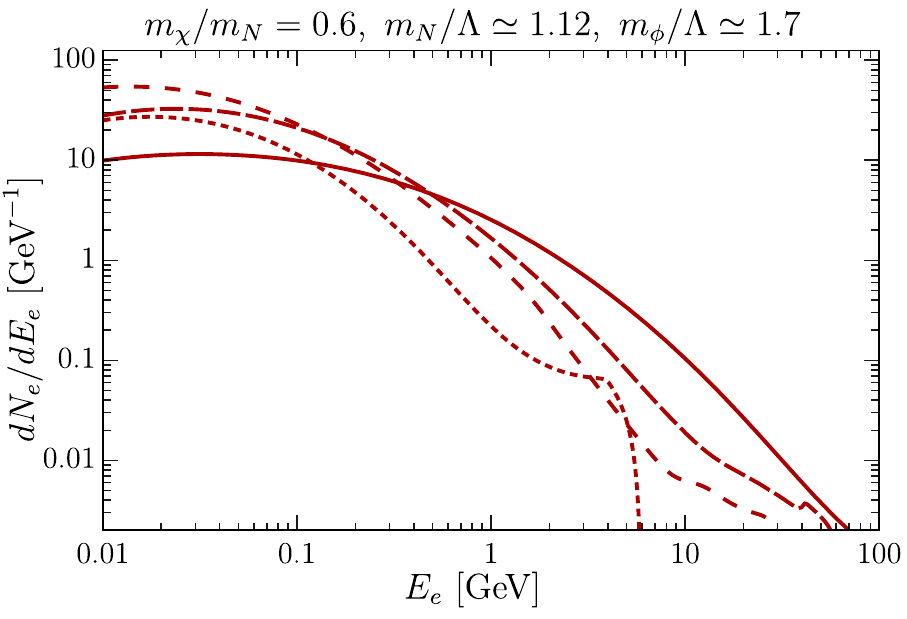}
\caption{The left (right) panel of this figure shows the photon (electron) energy spectrum arising from the annihilation channel $\bar \chi \chi\to N \nu$ and the subsequent decay of $N$.}
\label{fig:spect}
\end{figure}

 Since the Planck limits are most sensitive to electrons and photons in 
the final state, we calculate $f_{\!\rm eff}(m_\chi)$ as a weighted 
average of the electron and photon spectra $(dN_{e^-,\gamma}/dE)$ predicted 
by our model for the DM annihilation process $\bar \chi \chi\to N \bar{\nu}$ 
as,
 \beq
f_{\!\rm eff}(m_\chi)=\frac{1}{4 m_{\chi}} \int_{0}^{m_{\chi}(1+m_N^2/4m_\chi^2)} d E E\bigg[2 f_{\!\rm eff}^{e^{-}}(E)\frac{d N_{e^{-}}}{d E}+f_{\!\rm eff}^{\gamma}(E)\frac{d N_{\gamma}}{d E}\bigg] \; .
 \eeq
 To compute this we employ the results of \rcite{Slatyer:2015jla}, which 
provides data on $f_{\!\rm eff}^{e^{\!-},\gamma}(E)$, the fraction of 
energy transferred to the IGM for energies in the range [keV--TeV]. We 
use {\tt pythia8}~\cite{pythia} to calculate the photon and electron 
spectra arising from the DM annihilation $\bar{\chi}\chi\to (\bar{N} 
\nu, \bar{\nu} N)$. This incorporates the effects of showering and 
hadronization in the decays of $N$ to SM states. We sum over all the 
flavors of $N$ in the final state.  In \fig{fig:spect} we show the 
photon (left-panel) and electron (right-panel) energy spectra for a few 
benchmark values of the DM mass.

\begin{figure}[t]
\centering
\includegraphics[width=0.47\textwidth]{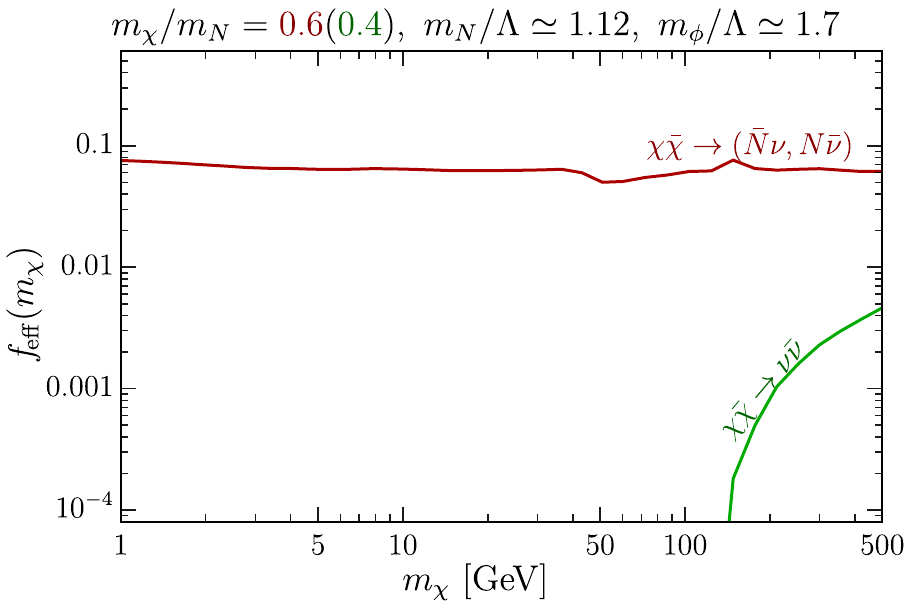}\quad
\includegraphics[width=0.47\textwidth]{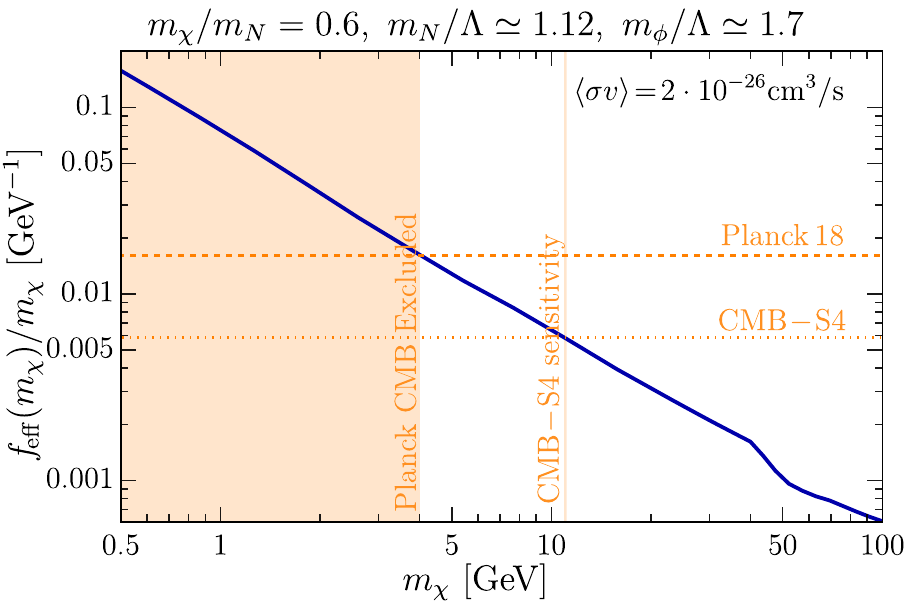}
 \caption{In the left panel we show $f_{\!\rm 
eff}(m_\chi)$, the fraction of energy transferred to the IGM due to DM 
annihilation around the time of recombination, as a function of the DM 
mass $m_\chi$ for the DM annihilation processes $\bar \chi \chi\to (\bar N\nu, 
N \bar{\nu})$ (red curve) and $\bar \chi \chi\to \nu \bar{\nu}$ (green 
curve). The blue curve in the right panel shows $f_{\!\rm 
eff}(m_\chi)/m_\chi$ as a function of $m_\chi$ for the benchmark spectrum 
listed above the figure. The dominant annihilation channel is $\chi 
\bar{\chi} \to (\bar{N} \nu, \bar{\nu} N)$. The horizontal dashed line 
represents the 95\% constraint on this quantity from the Planck 
collaboration~\cite{Planck:2018vyg} for $\langle\sigma v\rangle=2.2\times 
10^{-26}\,{\rm cm^3\,s^{-1}}$. As a result, this annihilation mode is 
ruled out for DM masses in the shaded region. The horizontal dotted line 
shows the projected sensitivity of CMB-S4 at 95\%\,C.L., which implies a 
sensitivity to DM annihilation in this channel up to 
$m_\chi\sim\!10\gev$.}
 \label{fig:feffomchi}
\end{figure}

In the left panel of \fig{fig:feffomchi} we show $f_{\!\rm 
eff}(m_\chi)$ as a function of $m_\chi$ for the DM annihilation processes 
$\bar \chi \chi\to (\bar N\nu, N \bar{\nu})$ (red curve) and $\bar \chi 
\chi\to \nu \bar{\nu}$ (green curve). Note that the fraction of energy 
injection into the IGM is smaller for the $\bar \chi \chi\to (\bar N\nu, N 
\bar{\nu})$ channel than for DM annihilation into most other SM 
states~\rcite{Slatyer:2015jla}. Moreover, for the case of DM annihilation 
directly to SM 
neutrinos, i.e. $\bar \chi \chi\to \bar{\nu} \nu$, $f_{\!\rm 
eff}(m_\chi)$ is even smaller and becomes negligible for 
$m_\chi\lesssim150\gev$. This is due to the fact that for 
this channel, energy injection into the IGM only arises from 
weak 
radiative processes, which are suppressed for $m_\chi\lesssim\op 
(100)\gev$~\rcite{Slatyer:2015jla}. In the right panel of \fig{fig:feffomchi} we plot our result for the fractional annihilation 
energy transferred per unit DM mass, $f_{\!\rm eff}(m_\chi)/m_\chi$, as 
function of $m_\chi$. The horizontal dashed orange line corresponds to the 
Planck constraint from \eq{e.planckcmb}. As can be seen in this figure, 
the Planck CMB constraint excludes DM annihilating to the $(\bar{N} \nu, 
\bar{\nu} N)$ final states for $m_\chi\lesssim 4\gev$. Therefore for DM 
masses lower than this value, only the annihilation channel to 
$\bar{\nu}\nu$, where no energy is injected into the IGM, is consistent 
with CMB bounds. We also show the sensitivity of future CMB-S4 experiments 
as the horizontal dotted orange line. This projected sensitivity is based 
on the analysis of \rcite{Wu:2014hta}, where it is shown that the most 
optimistic configuration of the CMB-S4 experiment could be sensitive to 
$f_{\!\rm eff}(m_\chi) \langle \sigma v\rangle/m_\chi\sim1.17\times 
10^{-28}\,{\rm cm^3\,s^{-1}}\gev^{-1}$ at 95\%\,C.L. Hence the resulting 
projected exclusion reach from CMB-S4 on the DM mass is $m_\chi\sim 
10\gev$ for the $(\bar{N} \nu, \bar{\nu} N)$ annihilation 
channel\,\footnote{Similar studies were perform for the annihilation 
channel $\bar{\chi}\chi\to \bar{N} N$ 
in~\cite{Batell:2017rol,Folgado:2018qlv} and for $\bar{\chi}\chi\to 
\bar{\nu} \nu$ in~\cite{Blennow:2019fhy}.}.

%%%%%%%%%%%%%%%%%
\subsubsection{Gamma Ray Constraints}
\label{s.gamma_rays}

In this subsection, we consider constraints on this class of models 
based on measurements of gamma rays from the galactic center (GC) of the 
Milky Way and from dwarf spheroidal satellite (dSphs) galaxies. In 
recent years these measurements have provided powerful bounds on DM 
models. GC searches have the advantage of a potentially large signal 
component, especially if the DM profile is cuspy, but suffer from large 
astrophysical backgrounds, while dSphs have much lower backgrounds but 
also contain a smaller signal component.

\paragraph{Galactic Center (GC):}
 We employ the Fermi-LAT Fourth Source Catalog 
(4FGL)~\cite{Fermi-LAT:2019yla} with 8~years of data for our analysis of 
the GC and follow the procedure adopted in \rcite{Abazajian:2020tww}. 
Assuming universality among the different flavors of composite singlet 
neutrinos, the expected photon flux per unit energy from DM annihilation is given by,
 \beq
\frac{d \Phi_{\gamma}}{d E_{\gamma}}=\frac{1}{4 \pi} \frac{\langle\sigma v\rangle}{4m_{\chi}^{2}} \frac{d N_{\gamma}}{d E_{\gamma}} J,	\label{eq:nu_flux}
 \eeq
 where $\langle\sigma v\rangle$ is the total thermally averaged cross 
section. The calculation of the photon energy spectrum using {\tt 
pythia8} has been described in the previous subsection, with the result 
shown in the left panel of \fig{fig:spect}. The $J$-factor is given by
 \beq
J\equiv \int\! d\Omega  \int_{0}^{\ell_{\max}}d \ell \,\rho_{\chi}^{2}(\ell),	
\label{eq:J_factor}
 \eeq
 where $d\Omega$ is integrated over the region of interest and the $d 
\ell$ integral is over the line of sight. We consider two different DM 
density profiles, the standard Navarro-Frenk-White (NFW) 
profile~\cite{Navarro:1995iw} and a cored DM halo profile proposed by Read et 
al.~\cite{Read:2015sta}. The standard NFW profile is given by
 \beq
\rho_{\chi}(r)=4\,\rho_{s}\bigg(\frac{r_s}{r}\bigg)\bigg(1+\frac{r}{r_{s}}\bigg)^{-2},
 \eeq 
 where $r$ is the distance from the GC and $\rho_{s}$ is the DM density 
at the scale radius $r_s=20\,{\rm kpc}$. We take $\rho_{s}=0.065\,{\rm 
GeV/cm^3}$, which gives the local DM density as $\rho(r_0)=0.3\,{\rm 
GeV/cm^3}$ for $r_0=8.5\,{\rm kpc}$, the distance of the sun from the 
center of the Milky Way. The line of sight $\ell$ is related to $r$ by
 \beq
r=\sqrt{r_0^2 -2 \ell r_0 \cos \theta +\ell^2},
 \eeq
 where $\theta$ is the angle between the GC and the line of sight. We 
take the integration limit $\ell_{\rm max}$ in \eq{eq:J_factor} to 
satisfy
 \beq
\ell_{\max }=\sqrt{r_{\text{MW}}^{2}-r_{0}^{2}\,\sin^{2}\theta }+r_{0} \cos \theta,
 \eeq
 where $r_{\text{MW}}\sim 40\,{\rm kpc}$ is the size of the Milky Way halo.
 
\begin{figure}[t]
\centering
\includegraphics[width=0.67\textwidth]{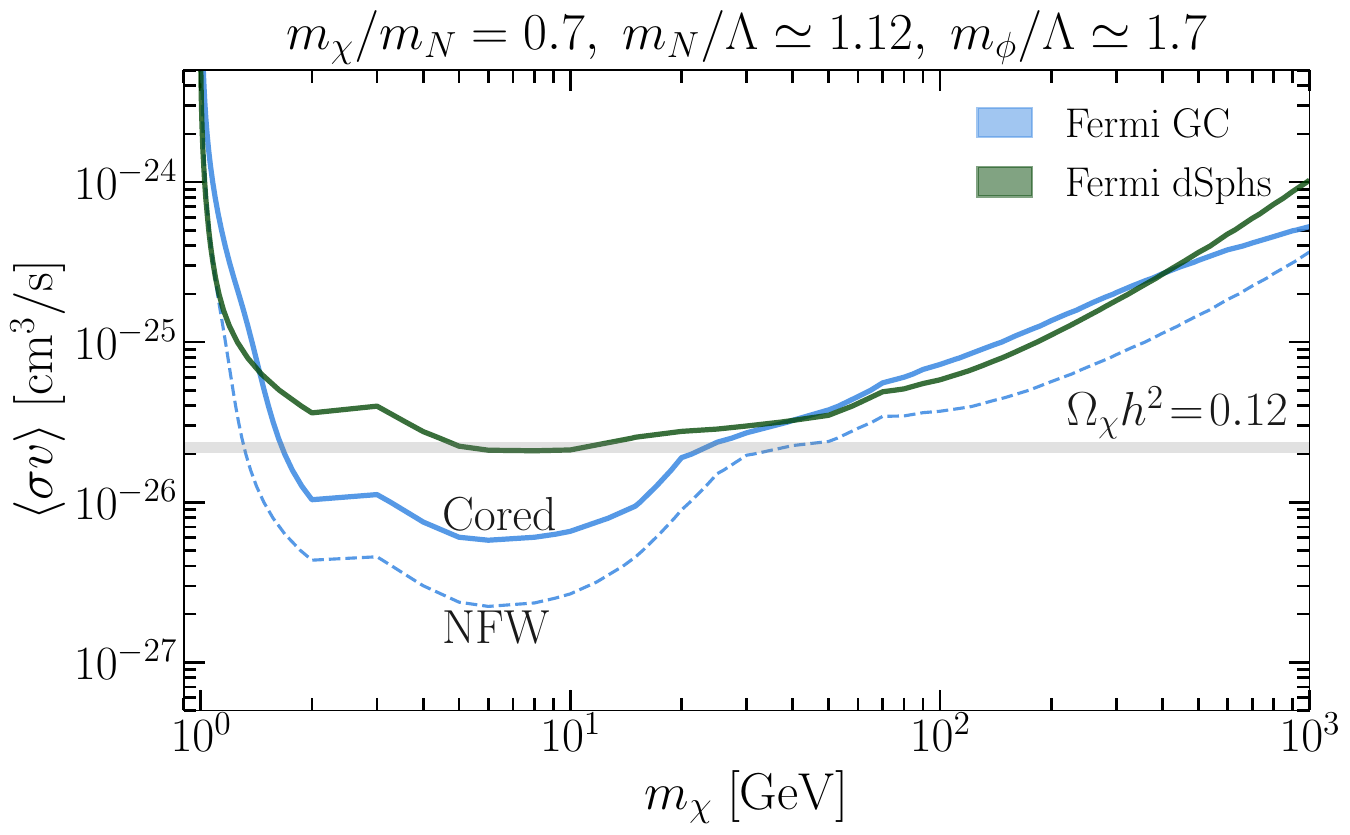}
 \caption{We show the 95\% CL limit on $\langle \sigma v\rangle$ from 
Fermi-LAT GC (blue curves) and dSphs (green curve) as a function of the 
DM mass $m_\chi$ for the benchmark value $m_\chi/m_N=0.7$, the dominant 
annihilation channel being $\bar{\chi}\chi \to (\bar{N} \nu, \bar{\nu}N)$. The horizontal gray line denotes the value required for the observed DM relic abundance $\Omega_\chi h^2\simeq 0.12$. The solid 
and dashed blue curves correspond to cored and NFW DM halo profiles respectively, based on data from Fermi-LAT fourth source catalogs.}
 \label{fig:fermi_limit}
\end{figure}

For the cored halo profile, we employ a core radius $r_c=1\,{\rm kpc}$. 
The mass of the cored profile $M_{\rm core}(r)$ asymptotically 
approaches that of the NFW profile $M_{\rm NFW}(r)$ in the outer 
regions as~\cite{Abazajian:2020tww,Read:2015sta},
 \beq
M_{\rm core}(r)=M_{\rm NFW}(r) \tanh(r/r_c).
 \eeq
 In \fig{fig:fermi_limit} we present the results of our analysis. For 
the cored DM halo profile, shown as the solid blue curve, we find that 
the range of $m_\chi$ between 2 GeV and 20 $\gev$ is excluded at 95\% 
C.L. for the benchmark case $m_\chi/m_N=0.7$. For the NFW DM halo 
profile, shown as the dashed-blue curve, the exclusion range is found to 
be between $m_\chi = 1\gev$ and $m_\chi = 30\gev$. As expected, the 
limits from the NFW profile are somewhat stronger.

\paragraph{Dwarf Spheroidal Satellite Galaxies:} 
 We also calculate the constraints from a set of dSphs galaxies with 
well-determined $J$-factors. We employ the log-likelihood profiles for 
the dSphs from Fermi-LAT data~\cite{Fermi-LAT:2015att, 
Fermi-LAT:2016uux} and we take the uncertainties in the $J$-factors 
from~\cite{Geringer-Sameth:2014yza}. These uncertainties are calculated 
from fits to the stellar kinematic data using generalized NFW profiles. 
From the analysis, we find an exclusion at 95\% CL for DM in the mass 
range $m_\chi\!\sim\!(5-10)\gev$ for $m_{\chi}/m_N=0.7$. This is shown 
in \fig{fig:fermi_limit} as the green curve. This limit is weaker than the 
bound on gamma rays from the GC.

%%%%%%%%%%%%%%%%%
\subsubsection{Neutrino Line Signal}
\label{sec:neu-line}

 Several experiments including SuperK~\cite{Super-Kamiokande:2020sgt}, 
IceCube~\cite{IceCube:2021kuw}, and ANTARES~\cite{Albert:2016emp} have 
placed limits on a neutrino line signal from DM annihilations. Their 
data has also been reanalyzed by independent groups seeking to extend 
the constraints to lower values of the DM 
mass~\cite{Olivares-DelCampo:2017feq, Klop:2018ltd,Arguelles:2019ouk, 
Bell:2020rkw}. Recently, the KamLAND experiment~\cite{Abe:2021tkw} also 
reported a bound on DM annihilation to neutrinos in the low mass range 
$m_\chi\sim[8-30]\mev$. In the top half of \tab{tab:indirect}, we 
summarize a variety of experimental constraints on the thermally 
averaged cross section $\langle\sigma v\rangle$, along with the mass 
range for which they are applicable.

For an annihilation cross section compatible with obtaining the correct 
DM relic abundance, these experiments are only sensitive when the DM 
mass is below a GeV, because then the DM number density is high and 
therefore the flux in the neutrino line is large. Since we have 
concluded that the $(\bar{N} \nu, \bar{\nu} N)$ annihilation channel is 
ruled out for light DM masses by CMB constraints, in this section we 
limit our attention to the $\bar{\nu}\nu$ annihilation channel.

The existing bounds only go down to values of $\langle\sigma v\rangle$ 
of order $10^{-25}\,{\rm cm^3/s}$, which is still significantly above 
the value required to obtain the observed DM density. However, future 
experiments such as HyperK~\cite{Hyper-Kamiokande:2018ofw}, 
JUNO~\cite{JUNO:2015zny} and DUNE~\cite{DUNE:2015lol} are projected to 
have the necessary level of sensitivity to detect thermal relic DM in 
the mass range $15\mev$ - $90\mev$. In the lower half of 
\tab{tab:indirect} we report recent 90\% projections on the reach of 
these future experiments from independent 
analyses~\cite{Olivares-DelCampo:2018pdl,Klop:2018ltd, 
Arguelles:2019ouk, Bell:2020rkw}. In our numerical analysis below, we 
make use of the most optimistic projections for these experiments.
\begin{table}[t]
\centering
\begin{tabular}{||p{3.7cm}|p{3.7cm}|p{3.7cm}||}
\hline
Experiment/analysis & DM mass range & Best upper-limit $\langle\sigma v\rangle$\\ \hline
SuperK~\cite{Super-Kamiokande:2020sgt} & \centering $1-10^4\gev$ & $ 1.2 \times 10^{-24} \, {\rm cm^3/s}$\\
IceCube~\cite{IceCube:2021kuw} & \centering $5-200\gev$ & $3.5 \times 10^{-25} \, {\rm cm^3/s}$	\\
ANTARES~\cite{Albert:2016emp} & \centering $50-10^5\gev$ & $1.5 \times 10^{-24} \, {\rm cm^3/s}$	\\ 
KamLAND~\cite{Abe:2021tkw} & \centering $8-21\mev$ & $5 \times 10^{-25} \, {\rm cm^3/s}$	\\ \hline
HyperK~\cite{Arguelles:2019ouk, Bell:2020rkw} & \centering $17\mev - 50\gev$ & $3 \times 10^{-26} \, {\rm cm^3/s}$\\
JUNO~\cite{Klop:2018ltd, Arguelles:2019ouk} & \centering $10 - 80\mev$ & $4 \times 10^{-26} \, {\rm cm^3/s}$\\
DUNE~\cite{Klop:2018ltd, Arguelles:2019ouk} & \centering $10 - 80\mev$ & $3 \times 10^{-26} \, {\rm cm^3/s}$\\
\hline
\end{tabular}
 \caption{Existing (top half) and projected future (bottom half) 
experimental limits (90\%\,C.L.) on DM annihilating directly to 
neutrinos, along with the corresponding DM mass range.}
 \label{tab:indirect}
\end{table}

In what follows, we map the sensitivity of these future experiments to 
the parameter space of our model. Since the sensitivity projections for 
different experiments use somewhat different assumptions (about the DM 
density profile, etc.), we will also need to make all dependencies on 
these assumptions explicit, so that the sensitivity of different 
experiments can be directly compared.

\begin{figure}
\centering
\includegraphics[width=0.67\textwidth]{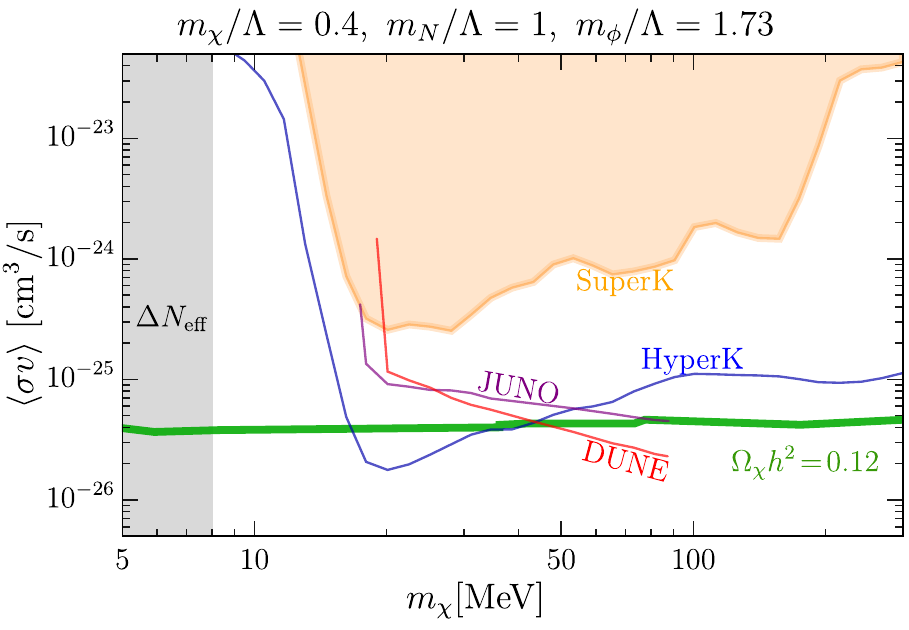}
 \caption{We plot the projected indirect detection constraint from 
the SuperK experiment (shaded orange region), along with future 
experiments (colored curves). Note that the sensitivities of SuperK and 
HyperK in this low mass range are based on 
Refs.~\cite{Arguelles:2019ouk, Bell:2020rkw}, not on the official 
results from the collaborations on which the numbers listed in 
Table~\ref{tab:indirect} are based. All the constraints correspond to a standard 
NFW DM density profile. The green line denotes the annihilation cross 
section that reproduces the observed relic abundance. At the lower end 
of the DM mass range, $\Delta N_{\rm eff}$ bounds from CMB and BBN 
disfavor values of $m_\chi$ below $8\mev$~\cite{Boehm:2013jpa, 
Nollett:2014lwa, Escudero:2018mvt}.
}
\label{fig:sigmav}
\end{figure}

Note that most experimental searches are performed for a specific 
neutrino flavor. Assuming universality among neutrino flavors, the 
expected flux on the Earth for each neutrino and antineutrino flavor from 
DM annihilation is given by,
 \beq
\frac{d \Phi_{\nu}}{d E_{\nu}}=\frac{1}{4 \pi} \frac{\langle\sigma v\rangle}{4m_{\chi}^{2}} \frac{1}{3} \frac{d N_{\nu}}{d E_{\nu}} J,	\label{eq:nu_flux}
 \eeq
 where the neutrino spectrum is mono-energetic, i.e.
 \beq
\frac{d N_\nu}{d E_\nu}= \kappa_\nu\frac{m_\chi}{E_\nu^2}\,\delta\big(1-E_\nu/m_\chi\big) \;.
 \eeq
 Here $\kappa_\nu=2$ for the annihilation channel $\bar{\chi} \chi\to 
\bar{\nu} \nu $, and the $J$-factor is given in \eq{eq:J_factor}. Note 
that using a more cuspy DM profile results in a larger $J$-factor, and 
therefore a stronger DM signal. For an NFW DM halo profile, we obtain an 
all-sky $J$-factor of $J\sim\!9\times 10^{22}\,{\rm GeV^2/cm^{5}}$. In 
\fig{fig:sigmav} we present the resulting experimental constraints and 
projections~\cite{Olivares-DelCampo:2018pdl,Klop:2018ltd, 
Arguelles:2019ouk, Bell:2020rkw}. As shown in this figure, the 
sensitivity of these experiments can reach the thermal relic cross 
section in the DM mass range 10--100~MeV. Unfortunately, this mass range 
is disfavored by the existing constraints from beam dumps and the 
bounds on DM self-interactions.

%%%%%%%%%%%%%%%%%
\subsection{Results \label{s.results}}

\begin{figure}
\centering
\includegraphics[width=0.9\textwidth]{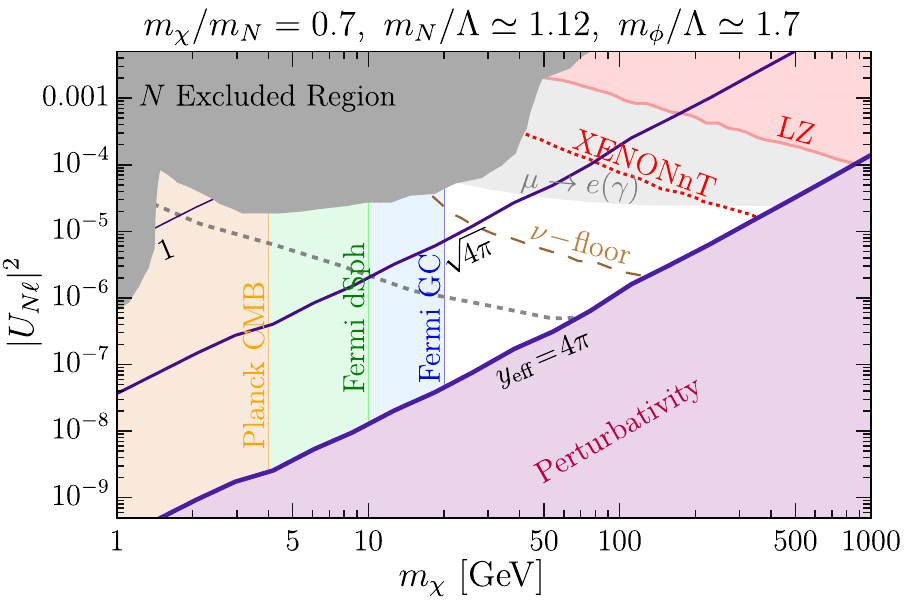}
 \caption{We show the constraints on our model in the parameter space of 
DM mass and the $N$-$\nu$ mixing angle for the benchmark parameters 
listed above the figure. The dominant annihilation channel is $\bar \chi 
\chi\to (N \bar\nu, \bar N\nu)$, and $y_{\rm eff}$ at each point has been chosen to 
reproduce the correct DM relic abundance. Contours of constant $y_{\rm 
eff}$ are shown as purple lines. The grey-shaded region is excluded by 
collider, beam dump, and astrophysical constraints~\cite{Chacko:2020zze}. 
The light-gray region corresponds to the current exclusion from the 
lepton flavor violating processes $\mu\to e$ conversion and $\mu\to e 
\gamma$ under the assumption of maximal lepton mixing, whereas the 
dotted gray curve represents the projected future sensitivity from these 
processes. The orange-shaded region on the left is excluded at 95\%~C.L. 
by Planck CMB data as discussed in \sec{sec:cmb}. The red-shaded region 
in the top-right corner is excluded by the direct detection constraint 
from the LZ (2022) experiment as discussed in \sec{s.direct_detection}. 
The red-dotted curve shows the projected sensitivity of the LZ/XENONnT 
experiment and the brown-dashed curve corresponds to the mixing angle 
below which the signal in direct detection experiments would fall below 
the neutrino floor. The gamma-ray constraints of \fig{fig:fermi_limit} 
are depicted by the green and blue shaded regions. The purple shaded region corresponds to $y_{\rm eff}>4\pi$ 
and is accordingly disfavored by unitarity considerations.
}
 \label{fig:mchilam_fdm07}
\end{figure}

We are now ready to combine the conclusions of the different parts of 
this section and identify viable regions in the parameter space of our 
model. For the case when the dominant annihilation channel is $\bar \chi 
\chi\to (\bar{N} \nu, \bar{\nu} N)$, we present all the relevant 
constraints in \fig{fig:mchilam_fdm07}, for the choices of 
$\Delta_N=\Delta_{\widehat{N}}=9/4$ (corresponding to $m_N\simeq 1.12\, 
\Lambda$) and $\Delta_\chi=\Delta_{\widehat{\chi}}\simeq 2$ 
(corresponding to $m_\chi=0.7\, m_N$). At each point, $y_{\rm eff}$ has 
been chosen such that the correct DM relic abundance is obtained. The 
inequalities in Eq.~(\ref{mixingrange}) are satisfied throughout this 
parameter region so that realistic neutrino masses can be obtained. 
The various shaded regions in the plot are excluded (see the figure 
caption for more detailed information on each constraint). The DM 
self-interaction constraint from~\eq{eq:mdm_selfint} for this benchmark 
point is only relevant for $m_\chi\lesssim 0.5\gev$, and has therefore 
not been shown. The gray-shaded region corresponds to the bounds on HNLs 
from colliders and beam dumps. The reason that this constraint weakens 
for larger masses is that the $N$ particles become too heavy to be 
produced from on-shell $W$ decays. Therefore the signal cross section 
drops precipitously while the background cross section falls more 
gradually, resulting in greatly reduced experimental sensitivity. We see 
from the plot that near-future direct detection experiments will have 
sensitivity for values of the DM mass above about 40 GeV. In the longer 
term, DM masses as low as 20 GeV may be accessible to direct detection. 
In the next section, we will evaluate the reach of future collider 
searches in this parameter space.

In this plot we have included the constraints (light-gray region) from 
the lepton flavor violating processes $\mu\to e$ conversion and $\mu\to 
e\gamma$, after relaxing the assumption that the couplings of the 
composite singlet neutrinos to the SM are flavor diagonal. Then, at the 
one-loop level, they contribute to these lepton flavor violating 
processes~\cite{Chacko:2020zze}. The light-gray region is the current 
constraint in the limit that we have maximal mixing between the $\mu$ 
and $e$ flavors, i.e. $\big\vert\!\sum_{n} U_{N_n \mu}^\ast U_{N_n 
e}\!\big\vert=\big\vert U_{N \ell}\big\vert^2$, where the 
sum over $n$ is over the KK modes in the loop. We adopt the strongest 
constraint from the MEG experiment~\cite{MEG:2016leq} for the $\mu\to 
e\gamma$ process and from the SINDRUM II 
experiment~\cite{SINDRUMII:2006dvw} for $\mu\to e$ conversion. In the 
near future the Mu2e~\cite{Mu2e:2014fns} and COMET~\cite{COMET:2018auw} 
experiments will be searching for $\mu\to e$ conversion. The future 
constraints in the absence of a signal are shown in 
\fig{fig:mchilam_fdm07} as the dotted-gray curve. Note that relaxing the 
assumption of maximal lepton mixing would lead to a weakening of the 
corresponding constraints.

\begin{figure}
\centering
\includegraphics[width=0.9\textwidth]{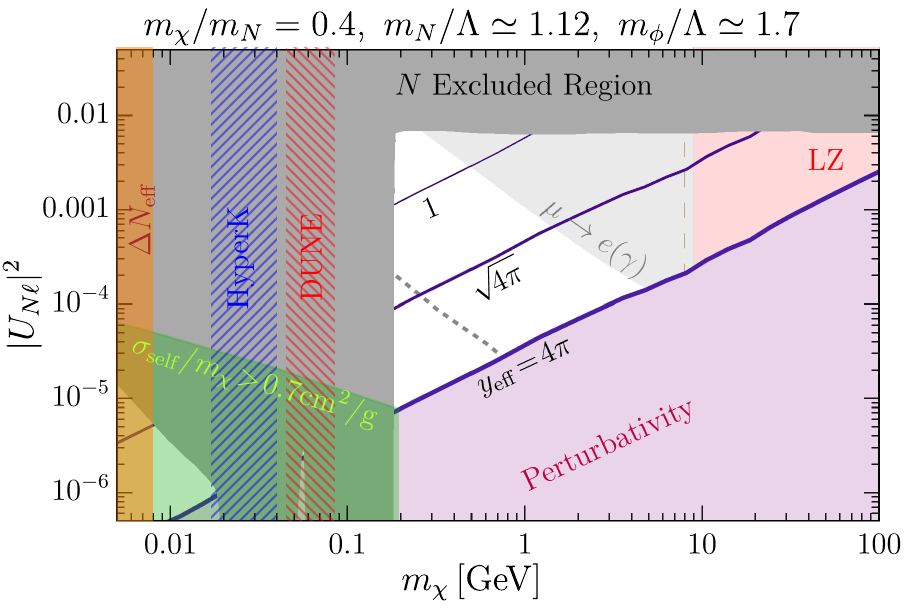}
 \caption{This plot shows the constraints on our model in the parameter space of DM mass and the $N$-$\nu$ mixing angle for the benchmark parameters listed above the figure. The dominant annihilation channel is $\bar \chi \chi\to \bar{\nu} \nu$, and $y_{\rm eff}$ at each point has been chosen to reproduce the correct DM relic abundance. Contours of constant $y_{\rm eff}$ are shown as purple lines. The grey-shaded region is excluded by 
beam dump and electroweak precision constraints~\cite{deGouvea:2015euy}. The light-gray region corresponds to the current exclusion from the 
lepton flavor violating processes $\mu\to e$ conversion and $\mu\to e 
\gamma$ under the assumption of maximal lepton mixing, whereas the 
dotted gray curve represents the projected future sensitivity from these 
processes. The red-shaded region 
in the top-right corner is excluded by the direct detection constraint 
from the LZ (2022) experiment as discussed in \sec{s.direct_detection} and to the left of the brown-dashed verticle line the direct detection experiments would fall below 
the neutrino floor. The orange-shaded region on the 
left is excluded by the CMB and BBN constraints on the effective number 
of relativistic degrees of freedom $\Delta N_{\rm eff}$. The blue and red vertical hatched bands are 
projected regions where future HyperK and DUNE experiments, 
respectively, have sensitivity for the detection of a neutrino-line. The 
green-shaded region is excluded due to DM self-interaction constraints discussed in \sec{s.darkmatter}. The purple shaded region corresponds to $y_{\rm eff}>4\pi$ 
and is disfavored by unitarity considerations.}
 \label{fig:mchilam_fdm04}
\end{figure}

For the case when the primary annihilation channel is $\bar{\chi} 
\chi\to \bar{\nu} \nu$, we present the relevant constraints in 
\fig{fig:mchilam_fdm04}, for the choices of 
$\Delta_N=\Delta_{\widehat{N}}=9/4$ (corresponding to $m_N\simeq 1.12\, 
\Lambda$) and $m_{\chi}/m_N = 0.4$ (corresponding to $\Delta_\chi\simeq 
7/4$). As before, $y_{\rm eff}$ at each point has been chosen to obtain 
the correct DM relic abundance, and the entire parameter region in the 
plot is compatible with Eq.~(\ref{mixingrange}). The shaded regions 
correspond to exclusions, except for the red and blue vertical 
hatched bands denoting the regions of sensitivity to future 
neutrino-line searches, as discussed in \sec{sec:neu-line}. 
Unfortunately, these regions are already excluded by the existing 
collider and beam dump bounds. Moreover, the DM self-interaction 
constraint in~\eq{eq:mdm_selfint}, i.e. $\sigma_{\rm 
self}/m_\chi\!\lesssim\!0.7\,{\rm cm^2/g}$, leads to a lower bound on 
the DM masses shown as the green solid band. In obtaining this bound we 
have taken $\kappa_\chi \simeq y_{\rm eff}^2$ to be consistent with 
large-$\mathcal{N}$ counting. We see that the DM self-interaction 
constraint also disfavors a part of the region where future HyperK and 
DUNE searches are sensitive to a neutrino line signal.

%%%%%%%%%%%%%%%%%%%%%%%%%%%%%%%%%%%
\section{Collider Phenomenology \label{s.colliders}}
%%%%%%%%%%%%%%%%%%%%%%%%%%%%%%%%%%%

Let us now turn our attention to the collider signatures of this class 
of models. As described earlier, composite singlet neutrinos can be 
produced at colliders via the neutrino portal. When $m_\chi < M_N / 2$ 
(i.e. when the annihilation channel is $\chi \bar\chi \to \nu \bar\nu$), 
$N$ can decay fully invisibly into $\chi \bar\chi \nu$. This decay 
occurs via the $N \bar{N} \chi \bar\chi$ interaction in the hidden 
sector, with an insertion of $N$-$\nu$ mixing. Since any $N$ 
decay channels into SM final states also require at least one mixing 
angle (in the form of the portal coupling), and are further suppressed 
by $G^2_F m^4_N$ due to the off-shell $W$ or $Z$ bosons that mediate the 
process, the decays to $\chi \bar\chi \nu$ will completely dominate over 
the SM decays. Therefore, in this region of parameter space, $N$ decays 
are completely invisible, making discovery at colliders extremely 
challenging. As a result our analysis below will focus solely on the 
region where $m_\chi > M_N / 2$, i.e. the annihilation channel is $\chi 
\bar\chi \to (N \bar\nu, \bar N \nu)$.

As we have seen in the previous section, when the annihilation channel 
is $\chi \bar\chi \to (N \bar\nu, \bar N \nu)$, only heavier ($\gsim 
20~$GeV) DM masses are consistent with the existing constraints. 
However, in this section we will explore the collider signatures across 
the entire parameter space for this annihilation channel, even 
for regions that may not be compatible with the constraints outlined in 
the previous section. There are two reasons for this. Firstly, it is 
worth considering the possibility that $\chi$ constitutes only a 
fraction of DM, in which case the indirect detection constraints on DM 
can be weakened. Secondly, similar signals may arise in the larger class 
of DM models where a dark sector couples to the SM through the neutrino 
portal. Accordingly, we will proceed with our analysis assuming only 
that we are in the regime $m_\chi > M_N / 2$.

The lowest energy state that can be probed via the neutrino portal is 
single $N$, and therefore single-$N$ production will generally have the 
highest production cross section. In collider physics contexts, neutral 
fermionic particles such as $N$ are typically categorized as HNLs (see, 
for example,~\cite{Abdullahi:2022jlv}). Depending on the mass of the 
$N$, it can be produced in Drell-Yan processes or from the decays of 
heavy mesons. The latter will have a significantly larger production 
cross section (and also significantly larger backgrounds). However, 
those channels will only be present when the $N$ is lighter than the 
heavy meson, while for $m_N\gsim 5$~GeV, only Drell-Yan production is 
available. Since charged leptons are preferable to neutrinos in collider 
searches, searches for HNLs generally focus on Drell-Yan production via 
$W$ bosons as opposed to $Z$ bosons. This results in a richer set of 
possible charge and flavor combinations of final states, allowing 
backgrounds to be better controlled. 

Once an HNL has been produced, the standard searches assume that it 
decays into a charged lepton and an off-shell $W$ boson (which can 
produce another charged lepton and a neutrino, or hadrons), or a 
neutrino and an off-shell $Z$ boson (which can produce an opposite sign 
same flavor pair of leptons, a pair of neutrinos, or hadrons). We can 
therefore expect that the beyond-the-SM channel that may be the easiest 
to observe might be single-$N$ production, followed by a leptonic decay 
of the off-shell $W$. This channel is the most commonly searched-for 
channel at the LHC for HNLs, with only null results thus far.

While LHC searches assume the HNL to be a weakly coupled particle, the 
searches are parameterized in terms of $m_N$ and the small mixing angle 
between $N$ and the SM neutrinos, so the bounds can be applied to our 
model as well. However, there is one important caveat. Even in the 
region $m_\chi > M_N / 2$, it is possible for the dominant decay mode of 
$N$ to be invisible. This is because interactions of the form $N \bar{N} 
N \bar{N}$, which are characteristic of the composite nature of the 
singlet neutrinos, can give rise to decays such as $N \to \nu {\nu} 
\bar{\nu}$ ("$N \to 3\nu$") through the mixing with the SM neutrinos. 
The corresponding width scales as $\Gamma_{N\to 3\nu} \sim |U_{N_\ell 
\ell}|^6$. This decay mode is highly suppressed for the range of masses 
and mixing angles for which $\chi$ can constitute all of DM. However, it 
can play a role for larger values of the mixing angle, corresponding to 
a reduced abundance of $\chi$. When the mixing angle is sufficiently 
large, the presence of this channel weakens the bounds on HNL-like 
searches. For nearly all of the parameter space we are interested in, 
the effect of this $N\to 3\nu$ channel remains negligible. Hence, we 
will treat $N$ as though it decays as a conventional HNL for our 
analysis, but in the figures we will show the region in which the $N\to 
3\nu$ channel becomes competitive or dominates over more typical HNL 
decay channels. As we shall see, very little of the parameter space we 
consider is affected by this channel.

\begin{figure}
	\centering
	\includegraphics{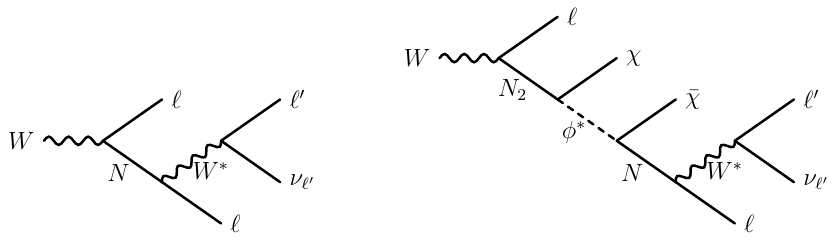}
	\caption{Collider signal processes of interest for this work. The initial $W$ may be on or off-shell. Note that the leptons labeled $\ell$ in each diagram may have the same sign and flavor if $N$ is a Majorana particle.}
	\label{feynN}
\end{figure}

Let us now imagine what the timeline of collider searches may look like. 
Most likely, the first signal will be seen at a traditional HNL search, 
though, as we will describe below, this may still correspond to prompt 
$N$ decays, displaced $N$ decays or very long-lived $N$ decays. The 
question will then become whether the discovered particle is a single 
weakly coupled particle, or whether it is the harbinger of a new sector 
with many states, as in our model. Once the $N$ discovery is firmly 
established, therefore, the focus will shift to searching for additional 
particles that are produced along with the $N$ in subleading channels. 
If the dark sector is a strongly coupled one, as in our model, and has a 
connection to DM, one may expect to see multiple $N$ production (such as 
3$N$ in our setup) which would most easily be identified in multi-lepton 
channels, or $N$ production along with other particles such as the DM 
particle itself, which would manifest itself as a presence of additional 
missing energy in channels that naively look like single $N$ production. 
In our analysis, we will first determine the region of parameter space 
where $N$ can be detected in future searches at the LHC, and then 
explore the possibility of subsequently detecting DM in the 
$N$-$\chi$-$\bar{\chi}$ (hereafter labelled $N\chi \bar\chi$) channel. The 
$N\chi \bar\chi$ final state is most easily accessed from decays of 
$N_2$. Since the mixing between $N_2$ and the SM neutrino increases as 
$\Delta_N$ is increased, we focus on a benchmark with $\Delta_N > 2$. 
The Feynman diagrams for the single $N$ and $N$-$\chi$-$\bar{\chi}$ 
channels, with Drell-Yan production and leptonic decays, are shown in 
\Cref{feynN}.

Below, we will organize our discussion by first focusing on the
regions of parameter space (which are not already excluded) that may yield 
sensitivity for single $N$ discovery in future runs of the LHC. We will 
then turn our attention to regions of parameter space that may yield 
additional sensitivity to the discovery of the 
$N$-$\chi$-$\bar{\chi}$ final state, after the discovery of $N$ has 
been established. In our discussion, the $nth-$KK mode $N_n$ may refer 
to either $(N_e)_n$ or $(N_\mu)_n$, the flavors of the KK mode coupling 
to $e$ and $\mu$ respectively.

Our estimates below for the sensitivity to the single-$N$ and $N\chi 
\bar{\chi}$ signals are based on Monte Carlo (MC) studies. The particles 
$N$, $\chi$, $\phi$ and $N_2$, and their interactions with each other as 
well as with SM fields are included in a custom \texttt{Madgraph5} (version 
\texttt{2.8.2}) \cite{madgraph} model, where the input parameters to the model 
are the scale $\Lambda$ and the mixing angle of $N$ with $\nu$. Events 
generated in this way are then passed through \texttt{Pythia 8.244} \cite{pythia} 
for showering and hadronization, and through \texttt{Delphes 3.4.2} \cite{delphes} for detector simulation. We use Delphes cards modified from \texttt{delphes\_card\_ATLAS.tcl} and 
\texttt{delphes\_card\_CMS.tcl}. For both, we adjust the lepton 
efficiency formulas to include leptons with $p_T$ down to $2$ GeV. Where 
necessary, we adopt additional lepton ID/reconstruction efficiencies by 
matching on to existing ATLAS/CMS analyses for the single-$N$ signal. 
For simplicity, we consider bounds separately on the $N_e$ coupled only 
to $e$ via the effective coupling $U_{N_ee}$ and on the $N_\mu$ coupled 
only to $\mu$ via $U_{N_\mu\mu}$. For notational compactness we use $N$ 
to represent either $N_e$ or $N_\mu$ with the $N$ flavor implied by its 
coupling to its respective charged lepton. As a benchmark, we use 
$\Delta_N = 9/4$, and we take over existing bounds for $|U_{Ne}|^2$ from 
Ref.~\cite{Chacko:2020zze}. For current bounds on $|U_{N\mu}|^2$, we 
consider the bounds adapted from \cite{Daum:1987bg, PIENU:2019usb, 
Hayano:1982wu, BNL-E949:2009dza, T2K:2019jwa, Bernardi:1985ny, 
Bernardi:1987ek, NuTeV:1999kej, DELPHI:1996qcc} and applied to the 
unparticle model in Ref.~\cite{Chacko:2020zze}.

Using the effective Lagrangian of our holographic model, we can calculate the partial widths
\begin{equation}
\Gamma(W^{\pm} \rightarrow \ell^{\pm}+N_{n})=\frac{g^{2} m_{W}}{48 \pi} \left|U_{N_{n}\ell}\right|^{2} \bigg(2+\frac{m_{N_n}^{2}}{m_{W}^{2}}\bigg)^2\bigg(1-\frac{m_{N_n}^{2}}{m_{W}^{2}}\bigg)^2
\end{equation}
for the $W$-boson to decay decay to the $n$-the KK-mode of $N$ where $\ell$ is either $\mu$ or $e$, and 
\begin{equation}
\Gamma(Z\rightarrow \nu_\ell+N_{n})=\frac{(g^{2}+g^{\prime 2}) m_{Z}}{96 \pi} \left|U_{N_{n}\ell}\right|^{2} \bigg(2+\frac{m_{N_n}^{2}}{m_{Z}^{2}}\bigg)^2\bigg(1-\frac{m_{N_n}^{2}}{m_{Z}^{2}}\bigg)^2
\end{equation}
for the $Z$-boson.

%%%%%%%%%%%%%%%%%%%%%
\subsection{Collider Searches for Single-$N$}

As mentioned above, the single-$N$ signal maps easily onto HNL searches 
parameterized in terms of $m_N$ and $|U_{N\ell}|^2$, and does not depend 
strongly on details of the UV model. Therefore, it is straightforward to 
use the results of CMS \cite{CMS} and ATLAS \cite{ATLAS} searches, as 
well as the projected sensitivity of searches for long-lived HNLs at 
MATHUSLA~\cite{Mathusla}, to identify the regions of interest in the 
parameter space of our model.

\begin{figure}
	\centering
	\includegraphics[scale=0.60]{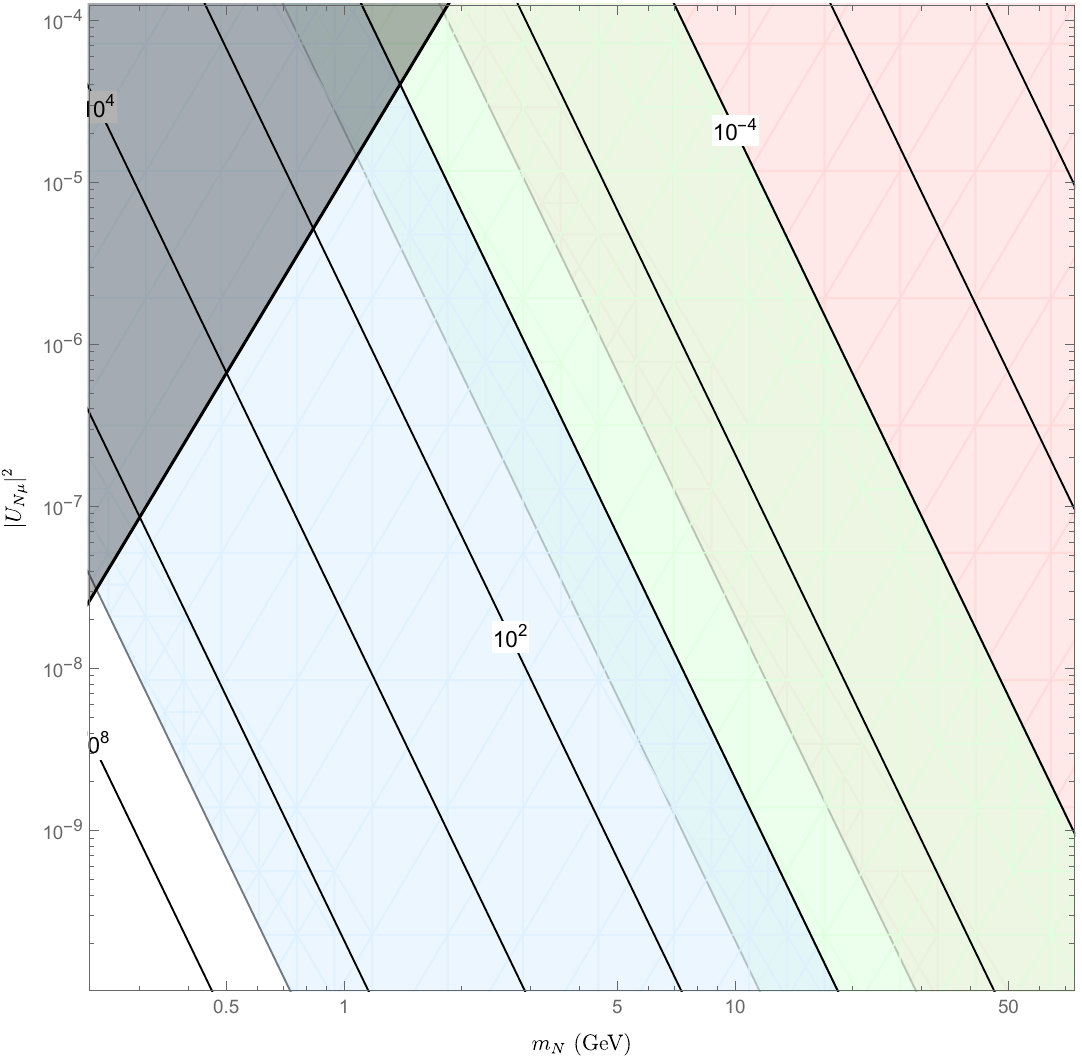}
	\caption{Contours of $c\tau$ for generic HNL particles (in meters) in the $m_N-|U_{N\ell}|^2$ plane. In region A, shaded light red, the HNL decays fast enough for prompt searches to be sensitive. In region B, shaded light green and partially overlapping regions A and C, displaced vertex searches in the tracker have sensitivity. In region C, shaded light blue, the HNL is too long-lived to be searched for by ATLAS or CMS, but dedicated long-lived particle detectors such as MATHUSLA may have sensitivity. In the grey-shaded region, the $N \to 3\nu$ decay channel dominates and the lifetime contours above are not accurate.}
	\label{figctau}
\end{figure}

We divide the parameter space into three regions according to the decay 
lifetime of $N$ particles, namely the prompt decay region (region A), 
the displaced decay region (region B), and the long-lived region (region 
C). We illustrate this in \Cref{figctau}. We assume for simplicity that 
$N$ comes in three degenerate copies, where each $N$ couples exclusively 
to a single lepton flavor, and we assume these couplings are 
flavor-universal. Since bounds on $|U_{N\tau}|^2$ are significantly 
weaker than on $|U_{N\mu,e}|^2$, we focus our attention on the electron 
and muon channels. We also assume that $N$ can be treated as a Majorana 
particle, that is, an $N$ can decay to $\ell^{\pm}$ with equal 
probabilities, resulting in charge and flavor combinations of leptons 
that have very small SM backgrounds. This is in contrast to an $N$ that 
preserves lepton number, and can therefore decay to only one of 
$\ell^{\pm}$ (while $\bar{N}$ decays to the other). The necessary 
criterion for this is $\Delta m_N \gtrsim \Gamma_N$~\cite{Das:2017hmg}, 
where $\Delta m_N$ is the splitting between the pseudo-Dirac mass 
eigenstates of $N$. This criterion is satisfied for all regions of 
parameter space that will be explored below.

In Region A, the $N$ lifetime is relatively short, and hence singlet neutrinos 
produced in colliders can be detected in prompt searches at the LHC. In 
Region B, the $N$ lifetime is $c\tau \sim \mathcal{O}(10^{-4}-10)$~m, 
which is long enough to possibly register as a displaced vertex. The 
existing ATLAS displaced search in this regime has sensitivity to HNLs 
with $m_N \sim \mathcal{O}(1) \textrm{ GeV}$ and $|U|^2 \sim 10^{-3.5} - 
10^{-5.5}$. In Region C, the $N$ is long-lived with $c\tau \sim 
\mathcal{O}(10^0 - 10^7\; m)$, and can be searched for in dedicated 
long-lived particle (LLP) detectors such as the proposed MATHUSLA 
experiment. The MATHUSLA sensitivity region for HNLs extends up to 
around $m_N \sim 5 \textrm{ GeV}$ and to about $|U|^2 \sim 10^{-9}$.

%%%%%%%%%%%%%%%%%%%%%
\subsubsection{Region A: Prompt} \label{N_A}

The leptonic $N$ decay channel, on which HNL searches are based, can be 
seen in the left panel of Fig.~\ref{feynN}. Let us consider the flavor 
and charge correlations of the three leptons in this diagram (the same 
considerations are also valid for the right panel of the same figure, 
which corresponds to the $N\chi \bar{\chi}$ signal). Note that the most 
``upstream'' lepton $\ell$ in this diagram is produced directly from the 
initial $W$. The other two leptons are produced in the $N$ decay, and 
since we assume the $N$ to couple flavor-diagonally, the next lepton has 
the same flavor as the most upstream one. We will therefore label it as 
$\ell$ as well. The third and most ``downstream'' lepton on the other 
hand is produced from an off-shell $W$, and therefore is flavor 
uncorrelated with the first two - we will label it as $\ell'$. If the 
$N$ were Dirac, then the first two $\ell$ leptons would be charge 
correlated as well as flavor correlated. However, for a Majorana $N$, 
the two leptons $\ell$ are equally likely to be same sign as to be 
opposite sign. The two leptons $\ell$ and $\ell'$ arising from the $N$ 
decay are of course always opposite sign due to charge conservation. In 
summary, when an $N$ is produced and decays through a $W$, the final state includes 
a pair of same-sign or opposite-sign leptons of the same flavor, and a 
third lepton that is uncorrelated in flavor but charge-correlated with 
the initial state. The final state also includes a neutrino, a source of 
MET in the event. We focus on the existing CMS search~\cite{CMS} in this 
region.

The CMS search focuses on the "$ee\mu$" and "$\mu\mu e$" channels, with 
35.9 fb$^{-1}$ of luminosity. The analysis considers a high mass ($m_N 
\sim \mathcal{O}(100 \textrm{ GeV})$) and a low mass ($m_N \sim 
\mathcal{O}(10 \textrm{ GeV})$) region. Only the latter is relevant for 
us, since in the high mass region the $N$ can only be produced via an 
off-shell $W$. This process has a much smaller cross section, which is 
challenging to observe above the background. The low mass region 
analysis demands that there are no opposite-sign same-flavor lepton 
pairs, which in our model requires $N$ decays to violate lepton number, 
and hence be Majorana. We list the cuts used in the low-mass analysis in 
\Cref{CMScuttable}. We check that our results based on MC estimates 
agree well with the CMS signal distributions at benchmark masses of $m_N 
= 5, 20, 30, 50 \textrm{ GeV}$ and $|U_{N\ell}|^2 = 10^{-5}$ presented 
in the appendix of Ref.~\cite{CMS}. We take this as a validation of our 
study of the $N\chi \bar{\chi}$ signal in the same search channel, 
which will be presented below.

SM backgrounds to this channel are nontrivial, and include backgrounds 
due to lepton fakes and misidentified leptons, which are difficult to 
simulate carefully. Projecting backgrounds for HNL signals at the HL-LHC 
with 14 TeV of energy is therefore beyond the scope of this work. 
However, we include the limits from Ref.~\cite{CMS} in our plots. 
Ref.~\cite{Izaguirre:2015pga} has projected a potential optimistic reach 
of searches for promptly decaying HNLs at the LHC with 300 
$\textrm{fb}^{-1}$ of data at $s=\sqrt{13} \textrm{ TeV}$. Their 
projection is shown in \Cref{fig_collider} as the solid brown line.
\begin{table}[]
	\centering
	\begin{tabular}{|l|l|}
		\hline
		\textbf{$ee\mu$ prompt signal cuts:}                                                                                                                          & \textbf{$\mu \mu e$ prompt signal cuts:}                                                                                                                          \\ \hline
		\begin{tabular}[c]{@{}l@{}}Event must contain 2 electrons\\ and 1 muon\end{tabular}                                                                    & \begin{tabular}[c]{@{}l@{}}Event must contain 2 muons and\\ 1 electron\end{tabular}                                                                   \\ \hline
		electron $p_T > 10$ GeV                                                                                                                                & electron $p_T > 10$ GeV                                                                                                                               \\ \hline
		muon $p_T > 5$ GeV                                                                                                                                     & muon $p_T > 5$ GeV                                                                                                                                    \\ \hline
		lepton $|d_0| < 0.5$ mm                                                                                                                                  & lepton $|d_0| < 0.5$ mm                                                                                                                                 \\ \hline
		lepton $|z_0| < 1.0$ mm                                                                                                                                  & lepton $|z_0| < 1.0$ mm                                                                                                                                 \\ \hline
		\begin{tabular}[c]{@{}l@{}}No opposite-sign same-flavor \\ lepton pairs\end{tabular}                                                                   & \begin{tabular}[c]{@{}l@{}}No opposite-sign same-flavor \\ lepton pairs\end{tabular}                                                                  \\ \hline
		
		$MET < 75$ GeV                                                                                                                                         & $MET < 75$ GeV                                                                                                                                        \\ \hline
		leading lepton $p_{T1} > 15$ GeV                                                                                                                       & leading lepton $p_{T1} > 15$ GeV                                                                                                                      \\ \hline
		subleading lepton $p_{T2} > 10$ GeV                                                                                                                   & subleading lepton $p_{T2} > 10$ GeV                                                                                                                  \\ \hline
		$M_{3\ell} < 80$ GeV                                                                                                                                   & $M_{3\ell} < 80$ GeV                                                                                                                                  \\ \hline
		\begin{tabular}[c]{@{}l@{}}If softest lepton is a muon \\ with $p_{T3} > 8$ GeV, \\ then either $p_{T2} > 15$ GeV \\ or $p_{T1} > 23$ GeV\end{tabular} & \begin{tabular}[c]{@{}l@{}}If softest lepton is a muon \\ with $p_{T3} < 8$ GeV, \\ then both $p_{T1} > 25$ GeV \\ and $p_{T2} > 15$ GeV\end{tabular} \\ \hline
		& \begin{tabular}[c]{@{}l@{}}If softest lepton is an electron \\ with $p_{T3} < 15$ GeV, \\ then $p_{T1} > 23$ GeV\end{tabular}                         \\ \hline
	\end{tabular}
	\caption{Summary of the cuts applied to MC events for the 
benchmark points in Search Region A, adapted from CMS's prompt HNL 
search \cite{CMS}. Here $d_0$ is the transverse impact parameter and 
$z_0$ is the longitudinal impact parameter.}
	\label{CMScuttable}
\end{table}

%%%%%%%%%%%%%%%%%%%%%
\subsubsection{Region B: Displaced} \label{N_B}

For moderately long-lived $N$, displaced vertex signatures provide a 
relatively clean channel for discovery. ATLAS has set exclusion limits for 
displaced $N$ decays in the $\mu \mu e$ signal channel ($N$ coupling to 
muons) in the range from $4-10 \textrm{ GeV}$ with $\mathcal{L} = 32.9 
\textrm{ fb}^{-1}$ of data \cite{ATLAS}. The cuts used in the analysis 
are listed in \Cref{ATLAScuttable}, and the SM background has been found 
by ATLAS to be negligible. Using MC event simulation, we 
populate the parameter region $m_N = 4-8 \textrm{ GeV}$ and 
$|U_{N\mu}|^2 = 10^{-3.5} - 10^{-6}$, and find good agreement with the 
exclusion range described in the ATLAS plots. Once again, we will take 
this as a validation of our MC methods, which we will later apply to the 
$N\chi \bar{\chi}$ signal.
\begin{table}[]
	\centering
	\begin{tabular}{|l|}
		\hline
		\textbf{$\mu\mu e$ displaced signal cuts:}                                                                                                                                        \\ \hline
		\begin{tabular}[c]{@{}l@{}}Event must contain 2 muons\\ and 1 electron\end{tabular}                                                                                               \\ \hline
		electron $p_T > 4.5$ GeV                                                                                                                                                          \\ \hline
		muon $p_T > 4$ GeV                                                                                                                                                                \\ \hline
		leading muon $p_T > 28$ GeV                                                                                                                                                       \\ \hline
		subleading muon $p_T > 5$ GeV                                                                                                                                                     \\ \hline
		\begin{tabular}[c]{@{}l@{}}Event tracks contain displaced \\ vertex within $4 < r < 300$ mm\\ made from 2 opposite-charge tracks\\ (at least one being a muon track)\end{tabular} \\ \hline
		\begin{tabular}[c]{@{}l@{}}For track pairs forming a displaced vertex, \\ $\sqrt{(\eta_1 + \eta_2)^2 + (\pi - (\phi_1 - \phi_2))^2} > 0.04$\end{tabular}                           \\ \hline
		\begin{tabular}[c]{@{}l@{}}For track pairs forming a displaced vertex,\\ $m_{inv} > 4$ GeV\end{tabular}                                                                           \\ \hline
	\end{tabular}
	\caption{List of cuts applied to MC events for 
single-$N$ samples generated in Search Region B, following from cuts applied in 
the ATLAS displaced vertex search~\cite{ATLAS}. Here $r$ is the transverse 
displacement of the vertex from the interaction point.}
	\label{ATLAScuttable}
\end{table}

\begin{figure}
	\centering
	\includegraphics[scale=0.7]{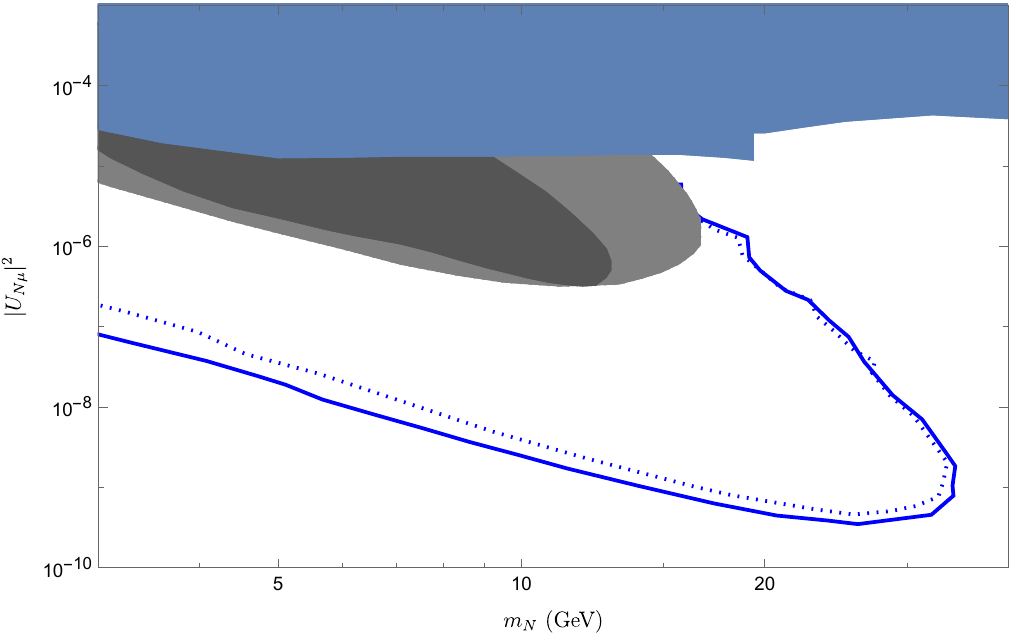}
	\caption{Existing and projected exclusion curves for displaced HNLs in the case where $N$ mixes only with $\mu$. The navy-shaded region is currently excluded by existing prompt searches. The light grey shaded region is excluded by the CMS displaced search Ref.~\cite{CMS:2022fut}; the exclusion reach of Ref.~\cite{ATLAS:2022atq} is shown in darker grey. The region where we project $\sigma_{N\chi \bar{\chi}}>5$ in an extension of the ATLAS displaced search is entirely excluded by the grey region (see \Cref{NXX_B}), and therefore not shown. The solid (dotted) blue contour shows the exclusion limit projected by Ref.~\cite{Drewes} for single $N$ in the $\mu-$ ($e$-)channel at ATLAS with $\mathcal{L} = 3 \textrm{ ab}^{-1}$.}	\label{figdisp}
\end{figure}

We note that there have been other recent studies exploring the 
potential reach of displaced HNL searches at the LHC. During the 
preparation of this work, CMS published Ref.~\cite{CMS:2022fut} placing 
limits on displaced HNLs at the LHC with $138 \textrm{ fb}^{-1}$ of 
data. The bounds reported are stronger than those of ATLAS's study in 
Ref.~\cite{ATLAS}, as CMS's search uses a larger sample of data. ATLAS 
also produced an updated search for displaced HNLs in 
Ref.~\cite{ATLAS:2022atq} with $139 \textrm{ fb}^{-1}$ of data with 
comparable reach to Ref.~\cite{CMS:2022fut}. For the purpose of 
validating our own MC procedure to project the approximate 
sensitivity to future $N\chi \bar{\chi}$ searches (see 
\Cref{NXX_B}), matching any of these studies suffices. We match our 
procedure against Ref.~\cite{ATLAS}, but display the stronger bounds 
from~\cite{CMS:2022fut,ATLAS:2022atq} in \Cref{figdisp} and 
\Cref{fig_collider}.

Additionally, Ref.~\cite{Drewes} projects the potential reach of CMS and 
ATLAS at the HL-LHC with $\mathcal{L} = 3 \textrm{ ab}^{-1}$ of data in 
the HNL parameter space using their own search developed independently 
from CMS and ATLAS. In \Cref{figdisp} we include their projected 
exclusion reach for ATLAS. This region is larger than our projection of 
the ATLAS analysis in part due to their having slightly looser $p_T$ 
cuts and utilizing both the tracker and muon chamber.

%%%%%%%%%%%%%%%%%%%%%
\subsubsection{Region C: Long-Lived} \label{N_C}

If the $N$ is long-lived, dedicated detectors for LLPs such as the 
proposed MATHUSLA facility~\cite{Mathusla}, which aims to detect 
particle decays far from the LHC interaction point, can be optimal for 
discovering the $N$. In contrast, FASER \cite{faser} has sensitivity 
mainly to highly boosted particles in the forward region, and is not as 
sensitive as MATHUSLA to HNLs in the mass range of interest to us. 
Therefore, here we focus our attention on the sensitivity of MATHUSLA.

The MATHUSLA detector is planned to form a rectangular prism which we 
take to have dimensions 200 x 200 x 20 meters and be located 100 meters 
directly above and along the beamline from the interaction point. The 
detector will be equipped with trackers with nanosecond-scale timing 
sensitivity. The MATHUSLA whitepaper includes projected limits for 
sterile right-handed neutrinos \cite{Mathusla}. MATHUSLA's sensitivity 
starts at $m_N \gtrsim 300 \textrm{ MeV}$, and the range that is of 
interest to us is up to 5~GeV. For heavier $N$, not only does the 
lifetime become smaller but the cross section also drops significantly. 
This is because below 5~GeV the cross section is dominated by heavy 
meson decays to $N$ (i.e. QCD initiated processes), while above 5~GeV, 
$N$'s are primarily produced through electroweak Drell-Yan processes. At 
the lower end of this mass range, the bounds can be as strong as 
$|U_{Ne}|^2 \sim |U_{N\mu}|^2 \sim 10^{-9}$.

To reproduce the expected sensitivity of MATHUSLA in the single-$N$ 
channel, we simulate charm and bottom production using 
Pythia~\cite{pythia}. Depending on the kinematics, we simulate the 
dominant 2-body or 3-body decay of the meson into a final state 
containing $N$ using MadGraph~\cite{madgraph}, with branching ratios 
adapted from Refs. \cite{faser,mesBR}. The MadGraph model includes the 
correct spin assignments for the meson and the final state particles, 
but the decays are approximated as fully perturbative, i.e. we do not 
include form factors. The $N$'s are then decayed using the appropriate 
$c\tau$ value at each point of the $m_N$-$|U_N|^2$ plane (neglecting 
effects of the aforementioned $N \to 3\nu$ process induced in the hidden 
sector, which is negligible in nearly all the parameter space of 
interest). MATHUSLA's efficiency to detect $N$ decays within its volume 
is taken to be 100\%, and we take the backgrounds to be negligible. Our 
projected sensitivity using this procedure agrees well with projections 
in the MATHUSLA whitepaper.

%%%%%%%%%%%%%%%%%%%%%%%%%%%%%%%%%%%
\subsection{Collider Searches for $N\chi \bar{\chi}$}
%%%%%%%%%%%%%%%%%%%%%%%%%%%%%%%%%%%

In this section, our main focus will be to evaluate whether a 
statistically significant sample of $N\chi \bar{\chi}$ events can 
be observed at the LHC. Since the only observable difference between 
single-$N$ events and $N\chi \bar{\chi}$ events is the presence of 
additional missing energy, the single-$N$ signal acts as a background in 
addition to the SM backgrounds already present. Much of the background 
involves fake or misidentified leptons and other challenging backgrounds 
to simulate, and hence careful estimation of the SM background is beyond 
the scope of this work. Instead, we focus on the prospect of seeing an 
excess $N \chi \bar{\chi}$ signal over the single-$N$ signal. We 
therefore organize the discussion in this section in the same way as 
for the single $N$ channel, both in evaluating the sensitivity of the
existing searches that we have presented in the previous section and for 
the proposed future searches to be described below. In order to reach this 
goal, we will use cuts that are optimized to take advantage of the 
differences in the kinematics of the two types of signals, such as the 
presence of additional MET and less visible energy in the $N\chi 
\bar{\chi}$ events. Our estimates of the significance of the excess 
are based on $\mathbb{N}_N / \mathbb{N}_{N\chi\bar{\chi}}$ 
statistics, where $\mathbb{N}_N$ and 
$\mathbb{N}_{N\chi\bar{\chi}}$ are the Poisson mean values of the 
two expected signals.

Note that in this subsection we will only be interested 
in whether a statistically significant number of $N \chi \bar{\chi}$ 
events (over the single-$N$ events) will pass the selection cuts of the 
existing analyses. We will not consider this as a sufficient criterion 
for discovery of the added signal component by itself, since without 
prior knowledge of the model parameters such as the $N$ mass or mixing 
angle, the combined signal may in fact fit a single-$N$ signal template 
with values of $m_N$ and $U_{N\ell}$ different from the true ones. 
Unless the experimental collaborations find other innovative methods to 
separate the two signal components, an actual discovery of this new 
channel will require first measuring the $N$ mass, for example by 
performing a search for fully visible decaying $N$'s. If the mass of $N$ 
can be thus measured, and the mixing angle determined from the cross 
section, then our results below indicate that there would be sufficient 
statistical power to resolve the new signal component. In fact, in 
section~\ref{NXX_vis} we propose a search that can be run in the visible 
$N$ channel that has a more straightforward path towards the discovery 
of the $N \chi \bar{\chi}$ signal component.

It is worth noting that, unlike the single $N$ channel, which can be 
parameterized in the same way as in HNL searches and is therefore 
largely independent of the UV model, the expected number of $N\chi 
\bar{\chi}$ events depends on a number of choices of parameters in 
the UV theory. The chief dependences are on the values of $\Delta_{N}$, 
$\Delta_{\widehat{N}}$, $\Delta_{\chi}$, $\Delta_{\widehat{\chi}}$ 
$\Delta_\Phi$, and the compositeness scale $\Lambda$. Below, we will 
work with the benchmark point $\Delta_N = \Delta_{\widehat{N}} = 2.25$, 
$\Delta_\chi= \Delta_{\widehat{\chi}} = 1.9$ and $\Delta_\Phi = 1.5$. In 
terms of IR theory parameters important for collider searches, this 
translates to $m_N/\Lambda = 1.11$, $m_{N_2}/\Lambda = 2.74$, 
$m_\chi/\Lambda = 0.65$, $m_\phi/\Lambda = 2.2$, and $|U_{N_2\ell}|^2 / 
|U_{N\ell}|^2 = 2.42\times 10^{-1}$. We take $m_N$ (or $\Lambda$) and 
$|U_{N\ell}|^2$ to be free parameters, with the rest of the mass spectra 
and couplings determined by our choices of the scaling dimensions.

%%%%%%%%%%%%%%%%%%%%%
\subsubsection{Region A: Prompt} \label{NXX_A}

We conduct a MC study to assess the signal size of the $N\chi \bar{\chi}$ component in the 
prompt $N$-decay region of the parameter space using the same procedure 
as described in \Cref{N_A}. We limit our study to $m_N \leq 25 \textrm{ 
GeV}$ since, beyond this point, $N_2$ becomes too heavy to be produced 
from on-shell $W$ decays and so the $N \chi \bar{\chi}$ event rate 
drops drastically.

We consider two kinematic variables to discriminate between single-$N$ 
and $N\chi \bar{\chi}$ events: MET and the invariant mass of the 
three final state leptons ($M_{3\ell}$). Due to the additional missing 
energy in the $N\chi \bar{\chi}$ events, the former is expected to 
be larger compared to a single-$N$ event, while $M_{3\ell}$ is expected 
to be smaller due to there being less energy left for the visible 
particles. For the same reasons, the $N\chi \bar{\chi}$ events have 
a lower efficiency under lepton $p_T$ cuts compared to the single-$N$ 
events. Therefore, we relax the lepton $p_T$ cuts of the analysis (only 
demanding that the leptons can reach the ECAL), but we leave the other 
cuts of \Cref{CMScuttable} unchanged. The reduced lepton $p_{T}$ cuts 
will result in higher SM background rates. Unfortunately, the leading 
source of these backgrounds is fake leptons, which cannot be studied 
reliably based on MC methods alone. Hence, as stated above, we focus our 
attention on the comparison of the single-$N$ and $N\chi 
\bar{\chi}$ signal components, while the search may need to be 
modified in other ways to control SM backgrounds.

\begin{figure}
	\centering
	\begin{subfigure}[b]{0.49\textwidth}
		\centering
	\includegraphics[scale=0.5]{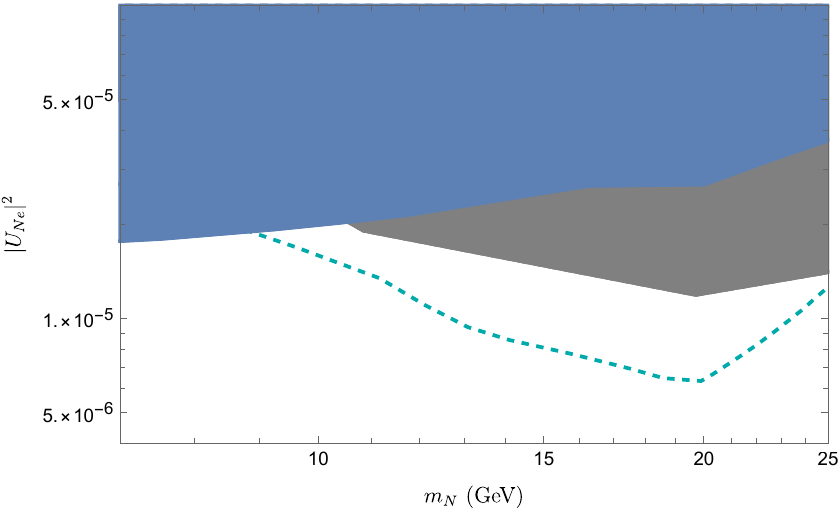}
	\caption{}
	\end{subfigure}
	\begin{subfigure}[b]{0.49\textwidth}
		\centering
	\includegraphics[scale=0.5]{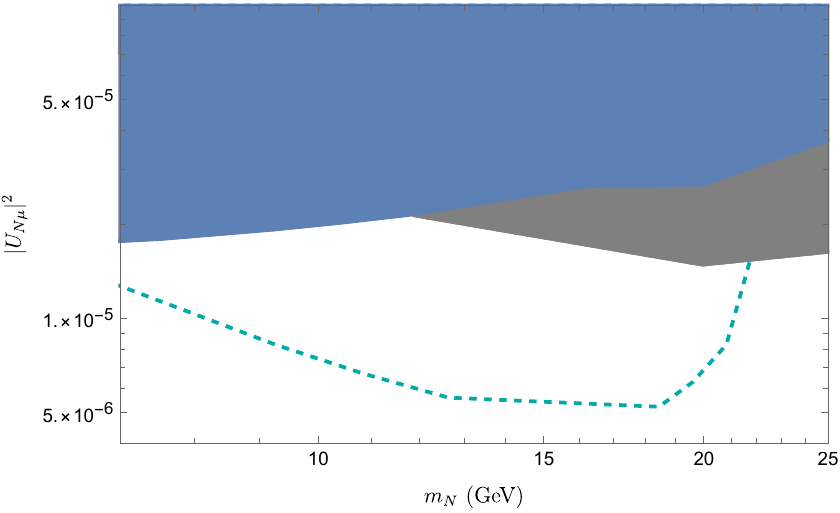}
	\caption{}
	\end{subfigure}
	\vspace*{2mm}
	\caption{Potential sensitivity reach of the HL-LHC for $N\chi 
\bar{\chi}$ signals for the electron-mixing angle (left) and 
muon-mixing angle (right). The grey regions indicate current exclusion 
limits from CMS prompt searches in \cite{CMS}, while navy regions are 
excluded by other collider and beam dump searches. Dashed lines bound 
the region where 5-$\sigma$ significance can be obtained for the $N\chi 
\bar{\chi}$ signal over the single-$N$ signal, with $\mathcal{L} = 
3000 \textrm{ fb}^{-1}$.}
\label{figprompt}
\end{figure}

We look for optimal cuts on MET and $M_{3\ell}$ to maximize the 
statistical significance of the $N\chi \bar{\chi}$ signal 
component. Due to the lower rate of $N\chi \bar{\chi}$ events, high 
statistical significance is only possible at $3~{\rm ab}^{-1}$ of 
luminosity at the HL-LHC. For both the $ee\mu$ and the $\mu\mu e$ 
channels, we find that for $m_N \sim 10-20~$GeV, 5$\sigma$ significance 
can be achieved by applying a cut $M_{3\ell, max} \sim 40-45 \textrm{ 
GeV}$, and modest MET cuts in the range $\textrm{MET}_{min} \sim 2-15 \textrm{ 
GeV}$. The region where $5\sigma$ significance can be achieved is shown 
in \Cref{figprompt}. Note that here we are not attempting to 
kinematically reconstruct the composite singlet neutrinos. A proposal 
for a search with full $N$ reconstruction will be presented below.

%%%%%%%%%%%%%%%%%%%%%
\subsubsection{Region B: Displaced} \label{NXX_B}

As in the prompt region, for the displaced region we apply the same 
methods as for our single-$N$ study (\Cref{N_B}) to the $N\chi 
\bar{\chi}$ signal. We expect this search to be most sensitive in 
the same parameter region as for the single-$N$ search, since for $m_N$ 
below this range, the $N$ is too long-lived to decay in the tracker 
region, while for $m_N$ above this range, the mixing angle must be very 
small to allow $m_N$ to decay displaced, which results in too low a 
cross section.

As with the prompt analysis, we look for cuts in the $M_{3\ell}$ and MET 
variables to maximize the significance of the $N\chi \bar{\chi}$ 
component in the signal for each parameter point. We find that for the 
full HL-LHC luminosity (${\mathcal L}=3~{\rm ab}^{-1}$), even with 
optimized MET and $M_{3\ell}$ cuts, the region in which $5\sigma$ 
significance can be obtained for the $N\chi \bar{\chi}$ events over 
single-$N$ events is entirely contained within the region already 
excluded by the ATLAS single-$N$ displaced search; therefore no contour 
for future displaced search projections for the $N\chi \bar{\chi}$ 
signal is shown. Moreover, CMS's recent displaced single-$N$ search 
Ref.~\cite{CMS:2022fut} places even stronger limits than ATLAS's search 
(the ATLAS exclusion region is entirely contained within CMS's). A 
displaced search optimized specifically for the $N\chi \bar{\chi}$ 
signal could provide more promising results. In \Cref{NXX_vis} we 
discuss the possibility of a search looking for fully visible $N$ decays.

%%%%%%%%%%%%%%%%%%%%%
\subsubsection{Region C: Long-Lived} \label{NXX_C}

Note that in the parameter region where MATHUSLA is most sensitive to 
long-lived $N$'s, the production mechanism is through heavy meson decays 
(mediated by virtual electroweak bosons). As before, in order to get 
$N\chi \bar{\chi}$ events, we need to replace the $N$ in the 
single-$N$ signal by an $N_2$, which subsequently decays. This has two 
important consequences. First, there is a range where $N_2$ is heavier 
than the meson, while $N$ is lighter than the meson. In this range, the 
$N\chi \bar{\chi}$ signal will be negligible compared to the 
single-$N$ signal because $N_2$ production is kinematically forbidden. 
Therefore, we only expect MATHUSLA to have the capability to observe an 
$N\chi \bar{\chi}$ excess over the single-$N$ signal for the lower 
$m_N$ range of its sensitivity to the single-$N$ signal. Secondly, there 
can exist ranges of parameter space where, due to chirality suppression, 
the decays of specific mesons to $N_2$ are preferred over decays to $N$. 
In this scenario, the amplitudes for two-body decays of the meson into 
$N+\ell$ or $N_2 + \ell$ are enhanced by $m_N$ or $m_{N_2}$ 
respectively~\cite{faser, mesBR}. Hence there are corners of parameter 
space where the $N\chi \bar{\chi}$ signal may be enhanced over the 
single-$N$ signal compared to naive expectations.

\begin{figure}
	\centering
	\includegraphics[scale=0.5]{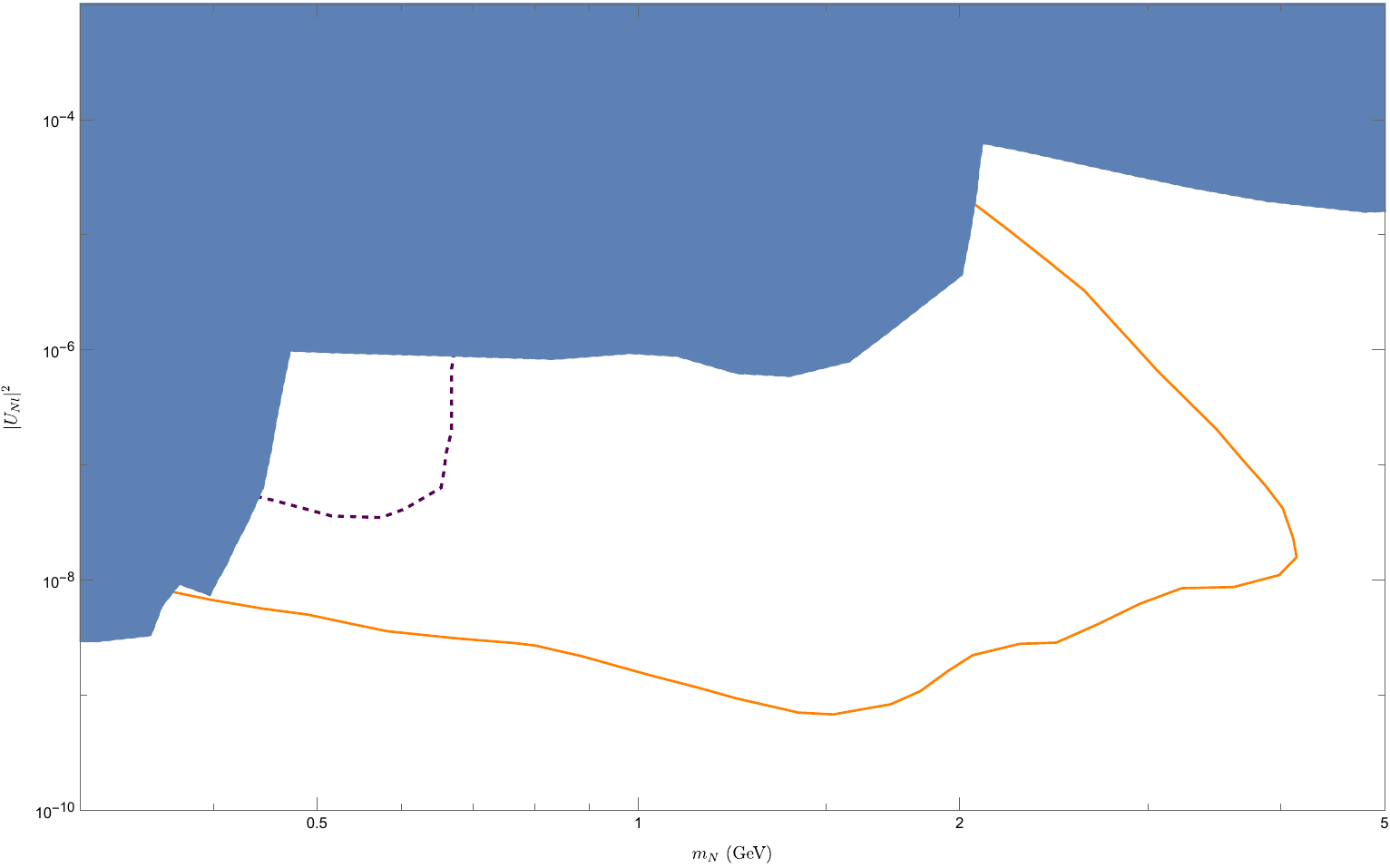}
	\caption{Projected reach for HNL decays in MATHUSLA (orange 
contour) at $\mathcal{L} = 3 \textrm{ ab}^{-1}$. The dashed purple curve 
bounds the region where an $N\chi \bar{\chi}$ excess of events 
can be observed with $5\sigma$ or greater statistical significance. The navy-shade region is excluded by existing collider and beam dump limits 
on $|U_{Ne}|^2$.}
	\label{figmathNXX}
\end{figure}

We calculate a statistical significance for the $N\chi \bar{\chi}$ 
excess by counting the total events expected to be seen in MATHUSLA for 
each point in the parameter space. In \Cref{figmathNXX}, the region 
bounded by the purple dashed curve is where the $N\chi \bar{\chi}$ 
excess has a statistical significance of 5$\sigma$ or above (and where 
the single-$N$ signal is not excluded by existing constraints). In 
comparison, the orange line represents the projected reach for the 
single-$N$ signal at the full HL-LHC luminosity of 3~ab$^{-1}$.

%%%%%%%%%%%%%%%%%%%%%
\subsection{Possibility of Searching for Fully Visible $N$ decays} \label{NXX_vis}

Until now, we have mainly focused on existing searches and projecting 
their sensitivity to single-$N$ production and $N\chi \bar{\chi}$ 
production at high luminosity. These searches are all based on the 
leptonic decays of $N$. Since the leptonic decay of $N$ contains a 
neutrino in the final state, the $N$ itself cannot be reconstructed. In 
this section, we want to consider a different type of search based on 
the fully visible decay channel $N\to \ell^{\pm}q\bar{q}'$. Since the 
$N\chi \bar{\chi}$ signal is always subdominant, we should expect 
that $N$ will be discovered first in the single-$N$ channel, probably in 
one of the fully leptonic channels. But once this happens, more 
challenging final states also become very interesting and may yield 
better sensitivity to subdominant processes such as $N\chi 
\bar{\chi}$. If $m_N$ can be fitted after the discovery of $N$ in a 
traditional channel, it will be an important variable in reducing the 
backgrounds in the fully visible search channel.

Fortunately, due to the Majorana nature of $N$, the two leptons in the 
fully visible channel (one produced in association with $N$ and the 
other arising from its decay) can be the same sign (and they will always 
have the same flavor in our model). This helps eliminate the largest SM 
backgrounds such as $Z+$jets. Unfortunately, the remaining background is 
expected to be dominated by fakes and detector effects such as lepton 
charge mismeasurement, and their reliable estimation is beyond the 
scope of this paper. In this section, we will simply compare the $N\chi 
\bar{\chi}$ signal to the single-$N$ signal in the fully visible 
channel and study the statistical significance of the excess. 
Projections for a dedicated search in this channel should be conducted 
by experimental collaborations. In other words, below we simply aim to 
demonstrate what might be possible if the backgrounds can be 
sufficiently reduced.

\begin{figure}
	\centering
	\includegraphics[scale=0.6]{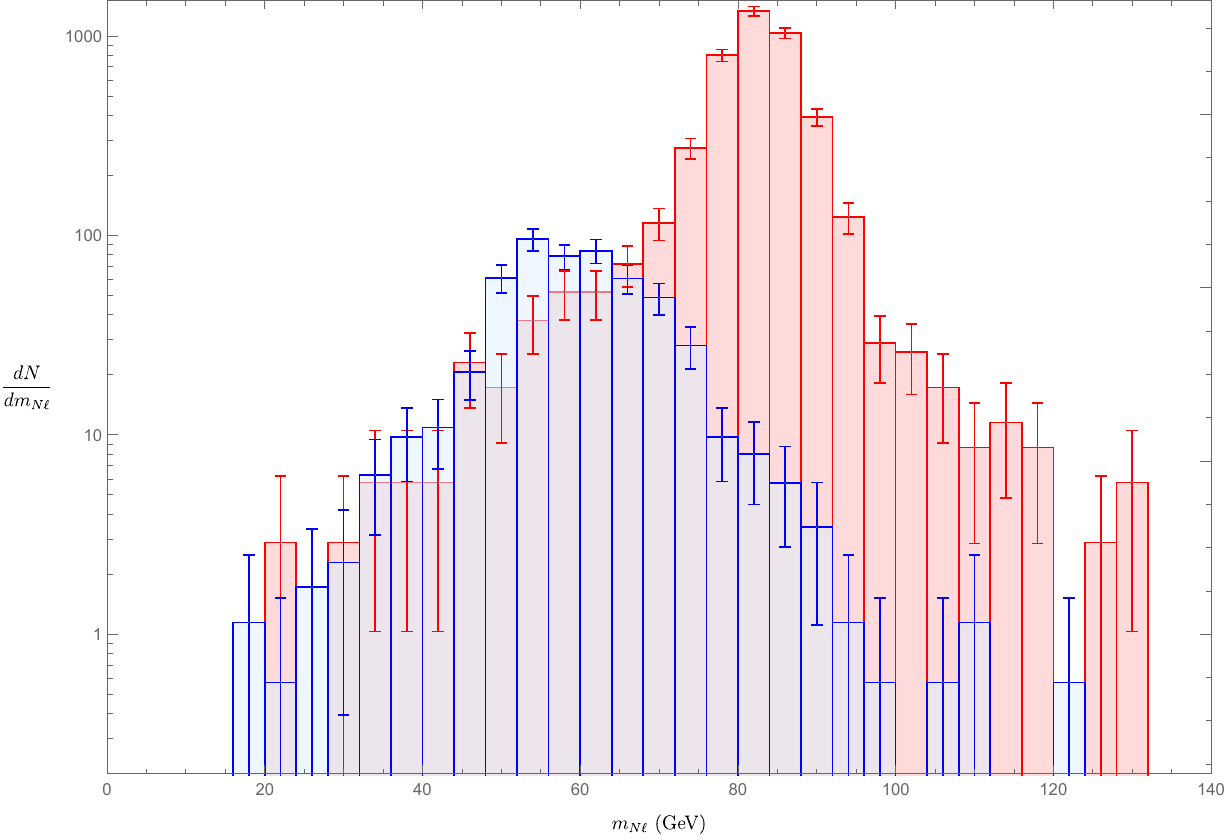}
	\caption{The distribution of $m_{N\ell}$ with $\mathcal{L} = 3 \textrm{ ab}^{-1}$ for a single-$N$ signal events (red) and $N\chi \bar{\chi}$ events (blue) at the parameter point $m_N = 10$ GeV and $|U_{N\ell}|^2 = 10^{-5}$. The error bars correspond to statistical uncertainties in our analysis, and they include uncertainties due to the finite size of our Monte Carlo event samples.}
	\label{figmNL}
\end{figure}

The fully visible decay channel of $N$ not only has a higher branching 
fraction compared to the leptonic one, but it has the very important 
advantage that the $N$ momentum can be fully reconstructed. As before, 
we take finite energy resolution into account in the study below by 
passing all events through Delphes~\cite{delphes}. In order to 
discriminate between the two signal components, we focus on the variable 
$m_{N\ell}$, the invariant mass of the $N-\ell$ system, where $\ell$ is 
the lepton produced in association with the $N$. While for single-$N$ 
events we expect the distribution of $m_{N\ell}$ to be sharply peaked 
around $m_W$, the distribution for $N\chi\bar{\chi}$ events will be 
broad and peak at $m_{N\ell} < m_W$ due to the DM particles carrying 
away energy. An example of this for the parameter point $m_N = 10 
\textrm{ GeV}$ and $|U_{N\ell}|^2 = 10^{-5}$ is shown in \Cref{figmNL}.

\begin{figure}
	\centering
	\includegraphics[scale=0.7]{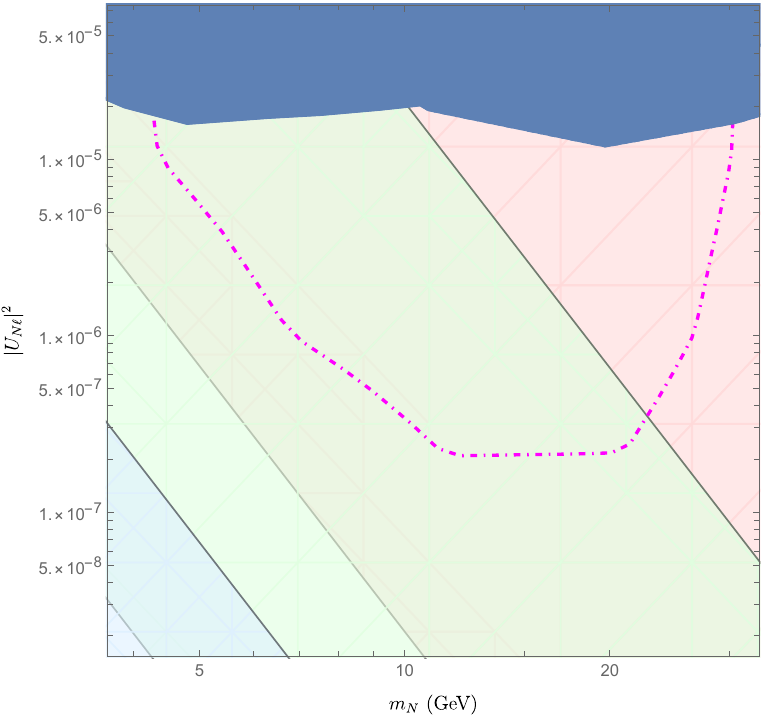}
	\caption{Above the magenta dot-dashed curve, a shape analysis of 
$m_{N\ell}$ in the visible $N$ decay channel can yield a 5$\sigma$ 
statistical significance for the $N\chi \bar{\chi}$ signal 
component in the absence of SM backgrounds. The navy-shaded region is 
currently excluded by searches for single-$N$ events. The regions shaded pink, 
green, and blue correspond to parts of parameter space optimally 
sensitive to prompt, displaced, and long-lived particle searches 
respectively.}
	\label{figVis}
\end{figure}

Accordingly, we perform a MC study for this final state (a same-sign, 
same-flavor lepton pair plus jets, $e$ and $\mu$ channels combined) with 
3~ab$^{-1}$ of luminosity, with only minimal $p_T$ requirements for the 
leptons and $p_T > 10 \textrm{ GeV}$ for jets. We require that two jets 
and one of the leptons reproduce a particle of mass $m_N$ within a 
tolerance of 20\%. We also demand that the invariant mass of the two 
leptons not be within $15 \textrm{ GeV}$ of $m_Z$, since a $Z$-veto will 
almost certainly be used in a dedicated search to reduce the $Z+$jets 
background with the charge of one lepton misidentified. By performing a 
shape analysis of the $m_{N\ell}$ distribution, we evaluate the 
statistical significance of the $\sigma_{N\chi \bar{\chi}}$ excess. 
Our results for the two combined channels assuming $U_{Ne} = U_{N\mu}$ 
are shown in \Cref{figVis}. This search can reach as far as 
$|U_{N\ell}|^2 \sim 10^{-7}$. Below $m_N \lesssim 2 \textrm{ GeV}$, $N$ 
becomes too long-lived to decay consistently in the LHC tracker system, 
but more sophisticated techniques such as searches for particles 
decaying in the muon trackers could in principle extend sensitivity 
beyond this limit. For $N$ masses above 30~GeV, the $N_2$ becomes too 
heavy to be produced via Drell-Yan processes.

\begin{figure}[t!]
	\centering
	\includegraphics[scale=0.55]{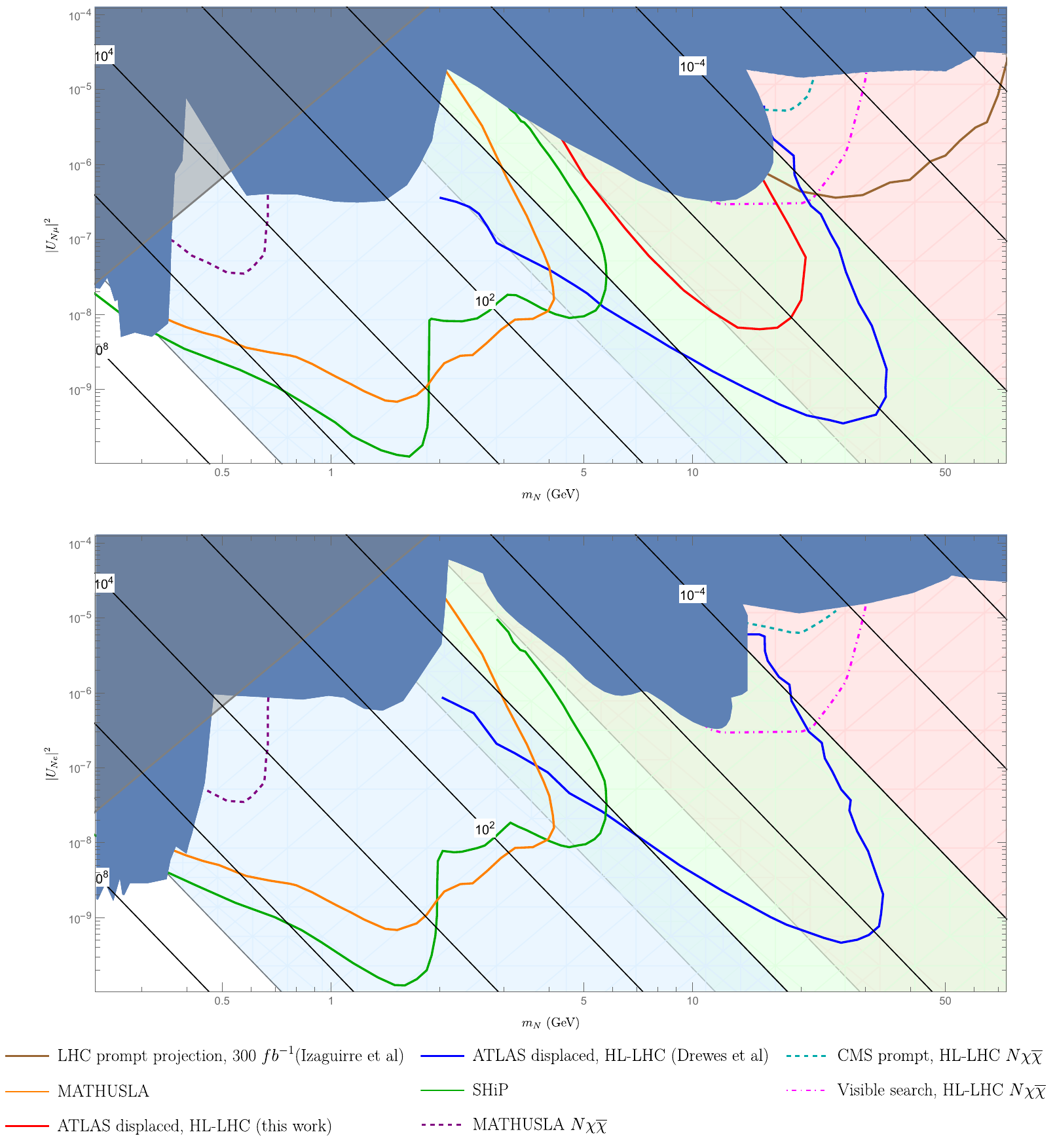}
	\caption{Parameter space of $|U_{N\ell}|^2$ plotted 
against $m_N$, with contours of the $N$ lifetime in meters overlaid (diagonal), for the muon mixing angle (above) and electron mixing angle (below). The navy-shaded regions in each plot are excluded by existing collider and beam dump constraints. Solid colored lines represent projected exclusion limits for single-$N$ sensitivity. Dashed contours represent exclusion limits for $N\chi \bar{\chi}$ sensitivity. The dot-dashed magenta contour is our optimistic $N\chi \bar{\chi}$ sensitivity for a visible collider search. In the gray-shaded region, the lifetime of $N$ in our model varies by more than a factor of 2 from that of a generic HNL due to the presence of the $N \to 3\nu$ decay channel.}
\label{fig_collider}
\end{figure}

In \Cref{fig_collider} we show our results for the $\mu$- 
and $e$-channels separately. Focusing first on the reach of collider 
searches independently from other constraints, it is easy to see that as 
the LHC continues its operation, the reach for the discovery of $N$ will 
be increased significantly both in the prompt and displaced regions via 
ongoing HNL searches, as well as in the long-lived region if the 
MATHUSLA experiment. Furthermore, a discovery of $N$ in the prompt and 
displaced regions could be followed up with a visible search along the 
lines we described, providing sensitivity for the associated production 
of a $\chi$-$\bar{\chi}$ pair, establishing their connection to the 
neutrino portal. There is a region of parameter space where MATHUSLA 
would be sensitive to $\chi$-$\bar{\chi}$ production for long-lived $N$ 
as well.

If $\chi$ constitutes all of DM, then as we saw in the 
previous section. the indirect detection constraints push the $\chi$ 
mass to $20\,$GeV and above in the $m_{\chi} > M_{N}/2$ case, which 
pushes the $N$ mass even higher. While this leaves room for the 
discovery of $N$ itself at the LHC, it eliminates all but the highest 
possible mass values for which there is sensitivity to $N 
\chi$-$\bar{\chi}$ production. As previously mentioned however, if 
$\chi$ does not constitute all of DM, then the indirect detection 
constraints are weakened, opening up the part of parameter space where 
DM production may be detectable.

%%%%%%%%%%%%%%%%%%%%%
\section{Conclusions\label{s.conclusion}}

We have explored a class of models with a strongly interacting dark 
sector coupled to the SM via the neutrino portal that can provide an 
explanation for both the origin of DM and the smallness of neutrino 
masses. We have studied the phenomenology of this scenario by modeling 
it in a warped extra dimensional framework. Within this 
higher-dimensional construction, we have discussed how neutrino masses 
can be generated via the inverse seesaw mechanism and we have shown how 
a stable state in the dark sector can play the role of a DM candidate. We 
determined the region in the parameter space of the model where the 
correct DM relic abundance can be obtained through thermal freeze-out, 
and we have studied the signals of this class of models in direct and 
indirect detection experiments and in collider searches.

Referring to the state interpolated through the neutrino portal as $N$, 
and to the DM candidate as $\chi$, the experimental signatures of the 
model depend strongly on the mass ratio $m_\chi / m_N$. When this ratio 
is smaller than $1/2$, $N$ will dominantly decay invisibly, weakening 
constraints on the model, but also making it challenging to discover. In 
this scenario, indirect detection is relatively insensitive and collider 
and beam dump searches that rely on visible decay products have no 
sensitivity. The main constraints on the model arise from DM 
self-interactions, bounds on lepton flavor violating processes (since 
the $N$ mass basis may not align with the flavor basis), direct 
detection experiments, precision electroweak measurements, and beam dump searches that do not require that 
$N$ decay in a visible channel. When the mass ratio is larger than 
$1/2$, the dominant decay of $N$ occurs through the neutrino portal into 
SM states, and therefore indirect detection experiments and collider 
searches have sensitivity. The lower bound on $m_\chi$ is significantly 
higher in this case due to constraints from the CMB and from the Fermi 
experiment. Near-future searches for lepton flavor violation will be 
able to probe a major part of the allowed parameter space. Future direct 
detection experiments will also have some level of sensitivity.

When $N$ decays visibly, for the purposes of collider searches, it falls 
into the category of a heavy neutral lepton (HNL). For HNL searches to 
have sensitivity, $N$ needs to be light enough to be produced from 
on-shell $W$-bosons (or even lighter to be produced in the decays of 
heavy mesons). Therefore, a potential discovery at the LHC is in tension 
with the existing constraints from indirect detection. The latter can 
however be weakened if $\chi$ constitutes only part of DM. In that case, 
as the luminosity of the LHC increases, additional regions in parameter space will be probed by searches looking for prompt as well as displaced 
decays of $N$, and complementary experiments such as MATHUSLA can probe 
the very long-lived $N$ region. An additional challenge is posed by 
discovering not only $N$ itself, but the production of DM particles 
along with it. We have shown that existing search strategies would not 
be sensitive to this production channel in regions of parameter space 
that are not already in tension with existing bounds. However we have 
proposed an extended search strategy based on the reconstruction of $N$ 
in a visible decay channel that would have sensitivity in the prompt 
and displaced $N$ decay regions. An interesting direction for future 
study would be the sensitivity of future experiments such as a high 
energy lepton collider to $N$ production channels such as $e^{+} 
e^{-}\to N\bar{\nu}$ which would not be limited to low $N$ masses as in 
the case of Drell-Yan production. Such a search may be sensitive to the 
region of parameter space that is compatible with indirect detection 
searches even when $\chi$ constitutes all of DM.

%%%%%%%%%%%%%%%%%%%%%%%%%%%%%%%%%%%

%%%%%%%%%%%%%%%%%%%%
\section*{Acknowledgements}
 We thank Abhish Dev and Zhen Liu for insightful discussions. ND thanks 
Taewook Youn and Vikas Aragam for helpful discussions. ZC and SD are 
supported in part by the National Science Foundation under Grant Number 
PHY-2210361. ZC is also supported in part by the US-Israeli BSF Grant 
2018236. The research of CK and ND is supported by the National Science 
Foundation Grant Number PHY-2210562. The research of SN is supported by 
the Cluster of Excellence {\it Precision Physics, Fundamental 
Interactions and Structure of Matter} (PRISMA$^+$ -- EXC 2118/1) within 
the German Excellence Strategy (project ID 39083149).

%%%%%%%%%%%%%%%%%%%

\bibliography{bib_unparticleDM}{}
\bibliographystyle{aabib}

%%%%%%%%%%%%%%%%%%
\end{document}